\def\D0{D\O~}
\ifx\nopictures Y \else \input epsf.tex \fi

\documentstyle[preprint,aps,floats]{revtex}
\begin{document}
%CsB If you do not want PACS numbers comment \draft out. 
\draft
%CsB PRL requires double space
\tighten
\baselineskip=18pt

\title{%
Soft gluon effects on lepton pairs at hadron colliders}
\author{%
C. Bal\'azs\thanks{E-mail address: balazs@pa.msu.edu} and 
C.--P. Yuan\thanks{E-mail address: yuan@pa.msu.edu}}
%\date{\today}
\address{%
Department of Physics and Astronomy, Michigan State University, \\
East Lansing, MI 48824, U.S.A.}
\maketitle

%\vspace{-8.25cm} 
\begin{flushleft}
{MSUHEP-70402 \\ CTEQ-704}
\end{flushleft}
%\vspace{5.25cm}

\begin{abstract}

With a large integrated luminosity expected at the Tevatron, a
next-to-leading order (NLO) calculation is no longer sufficient
to describe the data which yield the precision measurement of $M_W$, etc.
Thus, we extend the Collins-Soper-Sterman resummation formalism, for on-shell
vector boson production, to correctly include the effects of the polarization
and the width of the vector boson to the distributions of the decay
leptons. We show how to test the rich dynamics of the QCD multiple soft gluon
radiation, for example, by measuring the ratio $R_{CSS} \equiv \frac{\sigma 
(Q_T>Q_T^{\min})}{\sigma _{Total}}$. ($Q_T$ is the transverse momentum of
the vector boson.) We conclude that both the total rates and the
distributions of the lepton charge asymmetry predicted by the resummed
and the NLO calculations are different when kinematic cuts are applied.

\end{abstract}


\newpage

% SSSSSSSSSSSSSSSSSSSSSSSSSSSSSSSSSSSSSSSSSSSSSSSSSSSSSSSSSSSSSSSSSSSSSS
\section{Introduction}
\label{Sec:Intro}

Quantum Chromodynamics (QCD) is a field theory that is expected to 
explain all the experimental data involving strong interactions either
perturbatively or non-perturbatively~\cite{Bj}. Consider the weak boson 
($W^\pm$ and $Z^0$) production at a hadron collider, such as the Tevatron. 
In the framework of QCD the production rate of the weak bosons is calculated 
by multiplying the constituent cross section (the short-distance or the 
perturbative physics) by the parton luminosities
(the long-distance or the non-perturbative physics) \cite{qcdhand}. This
prescription of theoretical calculation was proven to the accuracy of 
${\cal O}(1/Q^2)$ and is known as the factorization theorem of 
QCD~\cite{Mueller}. ($Q$ is the mass of the vector boson.)
Since we do not yet know how to solve QCD exactly, we have to rely on the
factorization theorem to separate the perturbative part from the
non-perturbative part of the formalism for any physical observable.
The short distance contribution can be calculated perturbatively order by order 
in the strong coupling $\alpha_S$.
The long distance part has to be parametrized and fitted to the existing data 
so that it can later be used to predict the results of new experiments.
Therefore, theoretical predictions that are compared to experimental data 
always has to invoke some {\it approximation} in the calculations based upon
QCD. We refer to different prescriptions of calculations to be
different {\it models} of theory calculations which all originate from
the one and only QCD theory. For instance, to improve theory predictions
on the event shape, such as the transverse momentum ($Q_T$) distribution of
the weak boson, the commonly used theory model is the event generator, e.g.,
ISAJET \cite{isajet}, PYTHIA \cite{pythia}, or HERWIG \cite{herwig}. The
event generator can also provide information on the particle multiplicities
or the number of jets, etc. 

However, as discussed above, different models of
calculation make different approximations. Hence, a model can give more
reliable theory predictions than the others on some observables, but
may do worse for the other observables. 
A few more examples are in order. To calculate the total production rate of a
weak boson at hadron colliders, it is better to use a fixed order
perturbation calculation, and a higher order calculation is usually
found to be more reliable than a lower order calculation because it is
usually less sensitive to the choice of the scale for calculating the
parton distribution functions (PDF) or the constituent cross section 
(including the strong coupling constant $\alpha_S$).
The former scale is the factorization scale and the latter is the
renormalization scale of the process. Unfortunately, a fixed order
perturbation calculation cannot give reliable prediction of the
distribution of $Q_T$ when $Q_T$ is small. On the contrary, an event
generator can give more reliable prediction for $Q_T$ distribution in
the small $Q_T$ region, but it usually does not predict an accurate
event rate. The general feature of the above two theory models is that a
fixed order calculation is more reliable for calculating the event rate
but not the event shape, and an event generator is good for
predicting the event shape but less reliable for the event rate.

In this paper, we discuss another model of theory calculation that can give
reliable predictions on both the event rate and the shape of the 
distributions. Specifically, we are interested in the distributions of the
weak bosons and their decay products. This model of calculation is to resum
a series of large perturbative contributions due to soft gluon emission
predicted by the QCD theory. We present the QCD
resummation formalism for calculating the fully differential cross section
of the hadronically produced lepton pairs through electroweak (EW) vector
boson production and decay: $h_1h_2\rightarrow V(\rightarrow \ell _1
{\bar \ell_2})X $. 
We focus our attention on the Tevatron though our calculation is
general and applicable for any hadronic initial state $h_1h_2$ and any
colorless vector boson. For instance, the vector boson $V$ can be one of the
standard model (SM) electroweak gauge bosons $W^{\pm }$ or $Z^0$, the
virtual photon $\gamma^*$ (for producing the Drell-Yan pair), or some exotic
vector boson such as $Z'$ and $W'$ in the extended
unified gauge theories.

At the Tevatron, about ninety percent of the production cross section of the
$W^\pm$ and $Z^0$ bosons (with mass $Q$) is in the small transverse momentum 
($Q_T$) region, where $Q_T \lesssim 20$ GeV (hence $Q_T^2\ll Q^2$). 
In this region the higher order perturbative
corrections, dominated by soft and collinear gluon radiation, of the form 
$Q_T^{-2}\sum_{n=1}^\infty \sum_{m=0}^{2n-1}{}_nv_m\,
\alpha _S^n\ln^m({Q_T^2/Q^2})$, 
are substantial because of the logarithmic enhancement \cite{CSS}. 
(${}_nv_m$ are the coefficient functions for a given $n$ and $m$.)
These corrections are divergent in the $Q_T\rightarrow 0$ limit at
any fixed order of the perturbation theory. After applying the renormalization
group analysis, these singular contributions in the low $Q_T$ region can be
resummed to derive a finite prediction for the $Q_T$ distribution to compare 
with experimental data. It was proven by Collins and Soper in
Ref.~\cite{Collins} that not only the leading logs~\cite{DDT,Parisi} but 
all the large logs, including the sub-logs in the perturbative,
order-by-order calculations can be resummed for the energy
correlation in $e^+ e^-$ collisions.

For the production of vector bosons in hadron collisions two different 
formalisms were presented in the literature to resum the large contributions 
due to multiple soft gluon radiation: by Altarelli, Ellis, Greco, Martinelli
(AEGM)~\cite{AEGM}; and by Collins, Soper and Sterman (CSS) \cite{CSS}. The
detailed differences between these two formalisms were discussed in
Ref.~\cite{Arnold-Kauffman}. It was shown that the AEGM and the CSS
formalisms are equivalent up to the few highest power of $\ln
({Q_T^2/Q^2})$ at every order in $\alpha_S$ for terms proportional to
${Q}_T^{-2}$, provided $\alpha _S\,$ in the AEGM\ formalism is evaluated
at $b_0^2/b^2$ rather than at ${Q^2}$. A more noticeable difference,
except the additional contributions of order ${Q}^{-2}$ included in the
AEGM formula, is caused by different ways of parametrizing the
non-perturbative contribution in the low $Q_T$ regime. Since the CSS
formalism was proven to sum over not just the leading logs but also all the
sub-logs, and the piece including the Sudakov factor was shown to be
renormalization group invariant \cite{CSS}, we only discuss the
results of CSS formalism in the rest of this paper.

With the increasing accuracy of the experimental data on the
properties of $W^\pm$ and $Z^0$ bosons at the Tevatron, it is no longer
sufficient to only consider the effects of multiple soft gluon radiation for
an on-shell vector boson and ignore the effects coming from the decay width
and the polarization of the massive vector boson to the distributions of the
decay leptons. Hence, it is desirable to have an equivalent resummation
formalism~\cite{Balazs-Qui-Yuan} for calculating the distributions of the decay 
leptons. This formalism should correctly include the off-shellness of the 
vector boson (i.e. the effect of the width ) and the polarization information 
of the produced vector boson which determines the angular distributions of the 
decay leptons.

In the next section, we give our analytical results for such a formalism
that correctly takes into account the effects of the multiple soft gluon
radiation on the distributions of the decay leptons from the vector boson.
In Section III, we discuss the phenomenology predicted by this resummation
formalism. To illustrate the effects of multiple soft gluon radiation, we
also give results predicted from a next-to-leading order (NLO) calculation.
As expected, the observables that are directly related to the transverse
momentum of the vector boson will show large differences between the resummed 
and the NLO predictions. These observables are the transverse momentum of the
leptons from vector boson decay, the back-to-back correlations of the
leptons from $Z^0$ decay, etc. The observables that are not directly related
to the transverse momentum of the vector boson can also show noticeable
differences between the resummed and the NLO calculations if the kinematic cuts
applied to select the signal events are strongly correlated to the transverse
momentum of the vector boson. Section IV contains our detailed discussion.

Since this $Q_T$ resummation formalism only holds in the Collins-Soper
($CS$) frame~\cite{CSFrame}, we give the detailed form of the transformation 
between a four-momentum in the $CS$ frame (a special rest frame of the vector
boson) and that in the laboratory frame (the center-of-mass frame of the 
hadrons $h_1$ and $h_2$) in Appendix A. In Appendix B the analytical expression
for the NLO results, of ${\cal {O}} (\alpha _S)$, are given in $D = 4 - 2
\epsilon$ dimensions. Appendix C contains the expansion of the
resummation formula up to ${\cal O}(\alpha _S)$. Appendix D lists the
values of $A$, $B$, and $C$ functions (cf. Sec. II) used for our
numerical calculations.

We note that the resummation formalism presented in this paper can be
applied to any processes of the type $h_1 h_2 \rightarrow V(\rightarrow
\ell_1 {\bar \ell_2}) X$, where $V$ is a color neutral vector boson which 
couples to quarks and leptons via vector or axial vector currents, that is $V=e
\gamma, W^\pm, Z^0, W', Z'$, etc. Throughout this paper, we
take $V$ to be either $W^\pm$ or $Z^0$ bosons, unless specified otherwise.

% SSSSSSSSSSSSSSSSSSSSSSSSSSSSSSSSSSSSSSSSSSSSSSSSSSSSSSSSSSSSSSSSSSSSSS
\section{The Resummation Formalism}
\label{sec:Resum}

To derive the resummation formalism, we use the dimensional regularization
scheme to regulate the infrared divergencies, and adopt the 
canonical-$\gamma_5$ prescription to calculate the anti-symmetric part of 
the matrix element in $D$-dimensional space-time.\footnote{%
In this prescription, $\gamma_5$ anti-commutes with other $\gamma$'s in the
first four dimensions and commutes in the others \cite{gamma5c,twoloopf3}.}
The infrared-anomalous contribution arising from using the canonical-$%
\gamma_5$ prescription was carefully handled by applying the procedures
outlined in Ref.~\cite{Korner-Mirkes} for calculating both the virtual and the 
real diagrams.\footnote{In Ref.~\cite{Korner-Mirkes}, the authors calculated 
the anti-symmetric structure function $F_3$ for deep-inelastic scattering.}

The kinematics of the vector boson $V$ (real or virtual) can be expressed in  
terms of its mass $Q$, rapidity $y$, transverse momentum $Q_T$, and
azimuthal angle $\phi_V$, measured in the laboratory frame (the center-of-mass 
frame of hadrons $h_1$ and $h_2$). 
The kinematics of the lepton $\ell_1$ is described by $\theta$ and $\phi$, 
the polar and the azimuthal angles, defined in the Collins-Soper 
frame~\cite{CSFrame}, which is a special rest frame of the 
$V$-boson~\cite{lamtung}.
(A more detailed discussion of the kinematics can be found in Appendix A.)
The fully differential cross section for the production and decay of the
vector boson is given by the resummation formula in 
Ref.~\cite{Balazs-Qui-Yuan}: 
% EEEEEEEEEEEEEEEEEEEEEEEEEEEEEEEEEEEEEEEEEEEEEEEEEEEEEEEEEEEEEEEEEEEEEE
\begin{eqnarray}
&&\left( {\frac{d\sigma (h_1h_2\rightarrow V(\rightarrow \ell _1 {\bar \ell
_2})X)}{dQ^2\,dy\,dQ_T^2\,d\phi _V\,d\cos {\theta }\,d\phi }}\right) _{res}={%
\frac 1{96\pi ^2S}}\,{\frac{Q^2}{(Q^2-M_V^2)^2+Q^4\Gamma _V^2/M_V^2}} 
\nonumber \\
&&~~\times \left\{ {\frac 1{(2\pi )^2}}\int d^2b\,e^{i{\vec Q_T}\cdot {\vec b%
}}\,\sum_{j,k}{\widetilde{W}_{j{\bar k}}(b_{*},Q,x_1,x_2,\theta ,\phi
,C_1,C_2,C_3)}\,\widetilde{W}_{j{\bar k}}^{NP}(b,Q,x_1,x_2)\right.  \nonumber
\\
&&~~~~\left. +~Y(Q_T,Q,x_1,x_2,\theta ,\phi ,{C_4})\right\} .
\label{eq:ResFor}
\end{eqnarray}
% eeeeeeeeeeeeeeeeeeeeeeeeeeeeeeeeeeeeeeeeeeeeeeeeeeeeeeeeeeeeeeeeeeeeee
In the above equation %Eq.~\ref{eq:ResFor}, 
the parton momentum fractions are defined as $x_1=e^yQ/\sqrt{S}$ and 
$x_2=e^{-y}Q/\sqrt{S}$, where $\sqrt{S}$ is the center-of-mass (CM) energy of
the hadrons $h_1$ and $h_2$. For $V=W^{\pm }$ or $Z^0$, we adopt the
LEP line-shape prescription of the resonance behavior. The renormalization
group invariant quantity $\widetilde{W}_{j{\bar{k}}}(b)$, which sums to all
orders in $\alpha _S$ all the singular terms that behave as 
%$\alpha_S^n Q_T^{-2} \ln ^{2m -1}{(Q_T^2/Q^2)}$ ($1 \leq m \leq n$) 
$Q_T^{-2}$ $\times$ [1 or $\ln {(Q_T^2/Q^2)}$]
for $Q_T\rightarrow 0$, is 
% EEEEEEEEEEEEEEEEEEEEEEEEEEEEEEEEEEEEEEEEEEEEEEEEEEEEEEEEEEEEEEEEEEEEEE
\begin{eqnarray}
&&\widetilde{W}_{j{\bar k}}(b,Q,x_1,x_2,\theta ,\phi {,C_1,C_2,C_3} )=\exp
\left\{ -{\cal S}(b,Q{,C_1,C_2})\right\} \mid V_{jk}\mid ^2  \nonumber \\
&&~\times \left\{ \left[ \left( C_{ja}\otimes f_{a/h_1}\right) (x_1)~\left(
C_{{\bar k}b}\otimes f_{b/h_2}\right) (x_2)+\left( C_{{\bar k}a}\otimes
f_{a/h_1}\right) (x_1)~\left( C_{jb}\otimes f_{b/h_2}\right) (x_2)\right]
\right.  \nonumber \\
&&~~~~~\times (g_L^2+g_R^2)(f_L^2+f_R^2)(1+\cos ^2\theta )  \nonumber \\
&&~~~+\left[ \left( C_{ja}\otimes f_{a/h_1}\right) (x_1)~\left( C_{{\bar k}%
b}\otimes f_{b/h_2}\right) (x_2)-\left( C_{{\bar k}a}\otimes
f_{a/h_1}\right) (x_1)~\left( C_{jb}\otimes f_{b/h_2}\right) (x_2)\right] 
\nonumber \\
&&~~~~~\left. \times (g_L^2-g_R^2)(f_L^2-f_R^2)(2\cos \theta )\right\} ,
\label{eq:WTwi}
\end{eqnarray}
% eeeeeeeeeeeeeeeeeeeeeeeeeeeeeeeeeeeeeeeeeeeeeeeeeeeeeeeeeeeeeeeeeeeeee
where $\otimes $ denotes the convolution 
% EEEEEEEEEEEEEEEEEEEEEEEEEEEEEEEEEEEEEEEEEEEEEEEEEEEEEEEEEEEEEEEEEEEEEE
\begin{eqnarray}
\left( C_{ja}\otimes f_{a/h_1}\right) (x_1)=\int_{x_1}^1{\frac{d\xi _1}{\xi
_1}}\,C_{ja}\left( {\frac{x_1}{\xi _1}},b,\mu =\frac{C_3}b,C_1,C_2 \right)
\;f_{a/h_1}\left( \xi _1,\mu =\frac{C_3}b\right),  
\label{eq:Convol}
\end{eqnarray}
% eeeeeeeeeeeeeeeeeeeeeeeeeeeeeeeeeeeeeeeeeeeeeeeeeeeeeeeeeeeeeeeeeeeeee
and the $V_{jk}$ coefficients are given by 
% EEEEEEEEEEEEEEEEEEEEEEEEEEEEEEEEEEEEEEEEEEEEEEEEEEEEEEEEEEEEEEEEEEEEEE
\begin{eqnarray}
V_{jk}=\cases{ {\rm Cabibbo-Kobayashi-Maskawa~matrix~elements} & for $V =
W^\pm $ \cr $$\delta_{jk}$$ & for $V = Z^0,\gamma^*$}.
\label{eq:Vjk}
\end{eqnarray}
% eeeeeeeeeeeeeeeeeeeeeeeeeeeeeeeeeeeeeeeeeeeeeeeeeeeeeeeeeeeeeeeeeeeeee
In the above expressions $j$ represents quark flavors and $\bar k$ stands
for anti-quark flavors. The indices $a$ and $b$ are meant to sum over quarks
and anti-quarks or gluons. Summation on these double indices is implied. 
In Eq.~(\ref{eq:WTwi}) we define the couplings $f_{L,R}$ and $g_{L,R}$
through the $\ell _1 {\bar \ell_2} V$ and the 
$q{\bar q^{\prime }}V$ vertices, which are written respectively, as 
% EEEEEEEEEEEEEEEEEEEEEEEEEEEEEEEEEEEEEEEEEEEEEEEEEEEEEEEEEEEEEEEEEEEEEE
\begin{eqnarray*}
i\gamma _\mu \left[ f_L(1-\gamma _5)+f_R(1+\gamma _5)\right]
~~~{\rm and}~~~
i\gamma _\mu \left[ g_L(1-\gamma _5)+g_R(1+\gamma _5)\right] .
\end{eqnarray*}
% eeeeeeeeeeeeeeeeeeeeeeeeeeeeeeeeeeeeeeeeeeeeeeeeeeeeeeeeeeeeeeeeeeeeee
For example, for $V=W^{+},q=u$, ${\bar q^{\prime }}={\bar d}$, $\ell_1=\nu_e$, 
and ${\bar \ell_2}=e^{+}$, the couplings $g_L^2=f_L^2=G_FM_W^2/\sqrt{2}
$ and $g_R^2=f_R^2=0$, where $G_F$ is the Fermi constant. The detailed
information on the values of the parameters used in Eqs.~(\ref{eq:ResFor})
and (\ref{eq:WTwi}) is given in Table~{\ref{tbl:parameters}}. 
The Sudakov exponent ${\cal S}(b,Q{,C_1,C_2})$ in Eq.~(\ref{eq:WTwi}) is
defined as 
% EEEEEEEEEEEEEEEEEEEEEEEEEEEEEEEEEEEEEEEEEEEEEEEEEEEEEEEEEEEEEEEEEEEEEE
\begin{eqnarray}
{\cal S}(b,Q{,C_1,C_2})= 
\int_{C_1^2/b^2}^{C_2^2Q^2}{\frac{d{\bar \mu }^2}{{\bar \mu
}^2}} \left[ A\left( \alpha _S({\bar \mu }),C_1 \right) \ln \left( {\frac{
C_2^2Q^2}{{\bar \mu }^2}}\right) + B\left( \alpha _S({\bar \mu }
),C_1,C_2\right) \right] .  
\label{eq:SudExp}
\end{eqnarray}
% eeeeeeeeeeeeeeeeeeeeeeeeeeeeeeeeeeeeeeeeeeeeeeeeeeeeeeeeeeeeeeeeeeeeee
The explicit forms of the $A$, $B $ and $C$ functions and the renormalization
constants $C_i$ ($i$=1,2,3) are summarized in Appendix D. 
% TTTTTTTTTTTTTTTTTTTTTTTTTTTTTTTTTTTTTTTTTTTTTTTTTTTTTTTTTTTTTTTTTTTTTTT
\begin{table}[tbp]
\begin{center}
\begin{tabular}{crrrr}
$V$ & $M_V$ (GeV) & $\Gamma_V$ (GeV) & $g_L$ & $g_R$ \\ \hline
$\gamma$ & 0.00 & 0.00 & $g Q_f s_w/2$ & $g Q_f s_w/2$ \\ 
$W^\pm$ & 80.36 & 2.07 & $g/( 2\sqrt{2})$ & 0 \\ 
$Z^0$ & 91.19 & 2.49 & $g (T_3 - Q_f s_w^2)/(2 c_w)$ & $-g Q_f s_w^2/(2 c_w)$%
\end{tabular}
\end{center}
\caption{ Vector boson parameters and couplings to fermions. 
The $f{\bar f^{\prime}}V$
vertex is defined as $i \gamma_\mu [g_L (1 - \gamma_5) + g_R (1 + \gamma_5)]$
and $s_w = \sin\theta_w$ ($c_w = \cos\theta_w$) is the sine (cosine) of the 
weak mixing angle: $\sin^2(\theta_w(M_{Z^0}))_{\overline{MS}}= 0.2315$. $Q_f$ 
is the fermion charge ($Q_{u} = 2/3, Q_{d} = -1/3, Q_{e^-} = -1, Q_{\nu} = 0$),
and $T_3$ is the eigenvalue of the third component of the $SU(2)_L$ generator 
($T_3^{u}=1/2$, $T_3^{d} = -1/2$, $T_3^{\nu} = 1/2$, $T_3^{e^-} = -1/2$).}
\label{tbl:parameters}
\end{table}
% ttttttttttttttttttttttttttttttttttttttttttttttttttttttttttttttttttttttt

In Eq.~(\ref{eq:ResFor}) the magnitude of the impact parameter $b$ is
integrated from 0 to $\infty $. However, in the region where $b\gg 1/\Lambda
_{QCD}$, the Sudakov exponent ${\cal S}(b,Q,C_1,C_2)$ diverges as the result
of the Landau pole of the QCD coupling $\alpha _S(\mu )$ at $\mu =\Lambda
_{QCD}$, and the perturbative calculation is no longer reliable.
As discussed in the previous section, 
in this region of the impact parameter
space (i.e. large $b$), a prescription for parametrizing the
non-perturbative physics in the low $Q_T$ region is necessary. 
Following the idea of Collins and Soper~\cite{Collins}, 
the renormalization group invariant quantity 
$\widetilde{W}_{j{\bar{k}}}(b)$ is written as 
% EEEEEEEEEEEEEEEEEEEEEEEEEEEEEEEEEEEEEEEEEEEEEEEEEEEEEEEEEEEEEEEEEEEEEE
\begin{eqnarray*}
\widetilde{W}_{j{\bar k}}(b) = 
\widetilde{W}_{j{\bar k}}(b_*) \widetilde{W}
_{j{\bar k}}^{NP}(b) \,.
\end{eqnarray*}
% eeeeeeeeeeeeeeeeeeeeeeeeeeeeeeeeeeeeeeeeeeeeeeeeeeeeeeeeeeeeeeeeeeeeee
Here $\widetilde{W}_{j{\bar{k}}}(b_{*})$ is the perturbative part of 
$\widetilde{W}_{j{\bar{k}}}(b)$ and can be reliably calculated by
perturbative expansions, while $\widetilde{W}_{j{\bar{k}}}^{NP}(b)$ is the
non-perturbative part of $\widetilde{W}_{j{\bar{k}}}(b)$ that cannot be
calculated by perturbative methods and has to be determined from experimental
data. To test this assumption, one should verify that there exists a
universal functional form for this non-perturbative function $\widetilde{W}
_{j{\bar{k}}}^{NP}(b)$. This is similar to the general expectation that
there exists a universal set of parton distribution functions (PDF's) that can
be used in any perturbative QCD calculation to compare it with experimental
data. In the perturbative part of $\widetilde{W}_{j{\bar{k}}}(b)$,  
% EEEEEEEEEEEEEEEEEEEEEEEEEEEEEEEEEEEEEEEEEEEEEEEEEEEEEEEEEEEEEEEEEEEEEE
\begin{eqnarray*}
b_{*}={\frac b{\sqrt{1+(b/b_{max})^2}}} \, ,
\end{eqnarray*}
% eeeeeeeeeeeeeeeeeeeeeeeeeeeeeeeeeeeeeeeeeeeeeeeeeeeeeeeeeeeeeeeeeeeeee
and the non-perturbative function was parametrized by (cf. Ref.~\cite{CSS}) 
% EEEEEEEEEEEEEEEEEEEEEEEEEEEEEEEEEEEEEEEEEEEEEEEEEEEEEEEEEEEEEEEEEEEEEE
\begin{eqnarray}
\widetilde{W}_{j\bar k}^{NP}(b,Q,Q_0,x_1,x_2) = \exp \left[ -F_1(b) \ln \left( 
\frac{Q^2}{Q_0^2} \right) -F_{j/{h_1}}(x_1,b)-F_{{\bar k}/{h_2}
}(x_2,b)\right] ,  
\label{eq:Wnonpert}
\end{eqnarray}
% eeeeeeeeeeeeeeeeeeeeeeeeeeeeeeeeeeeeeeeeeeeeeeeeeeeeeeeeeeeeeeeeeeeeee
where $F_1$, $F_{j/{h_1}}$ and $F_{{\bar{k}}/{h_2}}$ have to be first
determined using some sets of data, and later can be used to predict the other
sets of data to test the dynamics of multiple gluon radiation predicted by
this model of the QCD theory calculation. As noted in Ref.~\cite{CSS}, $F_1$
does not depend on the momentum fraction variables $x_1$ or $x_2$, while 
$F_{j/{h_1}}$ and $F_{{\bar{k}}/{h_2}}$ in general depend on those kinematic
variables.\footnote{%
Here, and throughout this work, the flavor dependence of the non-perturbative
functions is ignored, as it is postulated in Ref.~\cite{CSS}.} The $\ln
(Q^2/Q_0^2) $ dependence associated with the $F_1$ function was predicted by
the renormalization group analysis \cite{CSS}. Furthermore, $F_1$ was shown
to be universal, and its leading behavior ($\sim b^2$) can be described by
renormalon physics \cite{Korchemsky-Sterman}. Various sets of fits to these
non-perturbative functions can be found in Refs.~\cite{Davies} and \cite
{Ladinsky-Yuan}.

In our numerical results in the next section, we use the Ladinsky-Yuan
parametrization of the non-perturbative function 
(cf. Ref.~\cite{Ladinsky-Yuan}):
% EEEEEEEEEEEEEEEEEEEEEEEEEEEEEEEEEEEEEEEEEEEEEEEEEEEEEEEEEEEEEEEEEEEEEEEE
\begin{eqnarray} \widetilde{W}_{j\bar k}^{NP}(b,Q,Q_0,x_1,x_2) = {\rm
exp} \left[- g_1 b^2 - g_2 b^2 \ln\left( {Q \over 2 Q_0} \right) - g_1
g_3 b \ln{(100 x_1 x_2)} \right], 
\label{eq:WNPLY} 
\end{eqnarray}
% eeeeeeeeeeeeeeeeeeeeeeeeeeeeeeeeeeeeeeeeeeeeeeeeeeeeeeeeeeeeeeeeeeeeeeee
where $g_1 = 0.11^{+0.04}_{-0.03}~{\rm GeV}^2$, 
$g_2 = 0.58^{+0.1}_{-0.2}~{\rm GeV}^2$, 
$g_3 =-1.5^{+0.1}_{-0.1}~{\rm GeV}^{-1}$ and $Q_0 = 1.6~{\rm GeV}$. 
(The value $b_{max}=0.5~{\rm GeV}^{-1}$ was used in determining the above 
$g_i$'s and in our numerical results.) 
These values were fit for CTEQ2M PDF with the canonical choice of
the renormalization constants, i.e. $C_1=C_3=2e^{-\gamma _E}$ 
($\gamma _E$ is the Euler constant) and $C_2=1$. 
In principle, for a calculation using a more update PDF, 
these non-perturbative parameters should be refit using a data set 
that should also include the recent high statistics $Z^0$ data from the 
Tevatron. This is however beyond the scope of this paper.

In Eq.~(\ref{eq:ResFor}), $\widetilde{W}_{j{\bar{k}}}$ sums over the soft
gluon contributions that grow as 
%$\alpha_S^n Q_T^{-2} \ln ^{2m -1}{(Q_T^2/Q^2)}$ ($1 \leq m \leq n$)
$Q_T^{-2}$ $\times$ [1 or $\ln {(Q_T^2/Q^2)}$] 
to all orders in $\alpha _S$. Contributions less singular
than those included in $\widetilde{W}_{j{\bar{k}}}$ should be calculated
order-by-order in $\alpha _S$ and included in the $Y$ term, introduced in
Eq.~(\ref{eq:ResFor}). This would, in principle, extend the applicability of
the CSS resummation formalism to all values of $Q_T$. However, 
as to be shown below, since the $A$, $B$, $C$, and $Y$ functions
are only calculated to some finite order in $\alpha _S$, the CSS resummed
formula as described above will cease to be adequate for describing data
when the value of $Q_T$ is in the vicinity of $Q$. Hence, in practice, one
has to switch from the resummed prediction to the fixed order perturbative
calculation as $Q_T \gtrsim Q$. The $Y$ term, which is defined as the 
difference between the fixed order perturbative contribution and those 
obtained by expanding the perturbative part of 
$\widetilde{W}_{j{\bar{k}}}$ to the same order, is given by 
% EEEEEEEEEEEEEEEEEEEEEEEEEEEEEEEEEEEEEEEEEEEEEEEEEEEEEEEEEEEEEEEEEEEEEE
\begin{eqnarray}
\ &&Y(Q_T,Q,x_1,x_2,\theta ,\phi ,C_4)=\int_{x_1}^1{\frac{d\xi _1}{\xi _1}}%
\int_{x_2}^1{\frac{d\xi_2}{\xi _2}}\sum_{n=1}^\infty \left[ {\frac{\alpha
_s(C_4Q)}\pi }\right] ^n  \nonumber \\
&&\ \times f_{a/h_1}(\xi _1,C_4Q)\,R_{ab}^{(n)}(Q_T,Q,\frac{x_1}{\xi _1},%
\frac{x_2}{\xi _2},\theta ,\phi )\,f_{b/h_2}(\xi _2,C_4Q),  \label{RegPc}
\end{eqnarray}
% eeeeeeeeeeeeeeeeeeeeeeeeeeeeeeeeeeeeeeeeeeeeeeeeeeeeeeeeeeeeeeeeeeeeee
where the functions $R_{ab}^{(n)}$ contain contributions less singular than 
%$\alpha_S^n Q_T^{-2} \ln ^{2m -1}{(Q_T^2/Q^2)}$ ($1 \leq m \leq n$)
$Q_T^{-2}$ $\times$ [1 or $\ln {(Q_T^2/Q^2)}$]
as $Q_T\rightarrow 0$. Their explicit expressions and the choice
of the scale $C_4$ are summarized in Appendix D.

Before closing this section, we note that the results of the usual
next-to-leading order (NLO), up to ${\cal O}(\alpha_S)$, calculation can be 
obtained by expanding the above CSS resummation formula to the $\alpha_S$ 
order, which includes both the singular piece and the $Y$ term. 
Details are given in Appendices B and D, respectively.

%SSSSSSSSSSSSSSSSSSSSSSSSSSSSSSSSSSSSSSSSSSSSSSSSSSSSSSSSSSSSSSSSSSSSSS
\section{Phenomenology}
\label{sec:Phenom}

As discussed above, due to the increasing precision of the experimental data
at hadron colliders, 
it is necessary to improve the theoretical prediction of the QCD theory by 
including the effects of the multiple soft gluon emission to all orders in 
$\alpha _S$. 
To justify the importance of such an improved QCD calculation, we compare
various distributions predicted by the resummed and the NLO calculations.
For this purpose we categorize measurables into two groups. We call an
observable to be {\it directly sensitive} to the soft gluon resummation
effect if it is sensitive to the transverse momentum of the vector boson.
The best example of such observable is the transverse momentum distribution
of the vector boson ($d\sigma /dQ_T$). Likewise, the transverse momentum
distribution of the decay lepton ($d\sigma /dp_T^\ell$) is also directly
sensitive to resummation effects. The other examples are the azimuthal angle
correlation of the two decay leptons $(\Delta \phi ^{\ell_1 {\bar \ell_2}})$, 
the balance in the transverse momentum of the two decay leptons 
$(p_T^{\ell_1}-p_T^{\bar \ell_2})$, or the correlation parameter 
$z=-{\vec p_T^{~\ell_1 }}\cdot 
{\vec p_T^{~{\bar \ell_2} }}/[\max ({{p_T^{~\ell_1 }}, 
{p_T^{~{\bar \ell_2}})}}]^2$. 
These distributions typically show large differences
between the NLO and the resummed calculations. The differences are the most
dramatic near the boundary of the kinematic phase space, such as the $Q_T$
distribution in the low $Q_T$ region and the 
$\Delta \phi ^{\ell_1 {\bar \ell_2}}$
distribution near $\pi$. Another group of observables is formed by those
which are {\it indirectly sensitive} to the resummation of the multiple soft
gluon radiation. The predicted distributions for these observables are
usually the same in either the resummed or the NLO calculations,
provided that the $Q_T$ is fully integrated out in both cases. Examples of
indirectly sensitive quantities are the total cross section $\sigma$, the
mass $Q$, the rapidity $y$, and $x_F \left( = 2 q^3/\sqrt{S} \right)$ 
of the vector boson\footnote{Here $q^3$ is the longitudinal-component of the 
vector boson momentum $q^\mu$ [cf. Eq.~(\ref{eq:QXYZ})].}, and
the rapidity $y^\ell$ of the decay lepton. However, in practice, to extract
signal events from the experimental data some kinematic cuts have to be
imposed to suppress the background events. It is important to note that 
imposing the necessary kinematic cuts usually truncate the range of the $Q_T$ 
integration, and causes different predictions from the resummed and the NLO
calculations. We demonstrate such an effect in the distributions of the
lepton charge asymmetry $A(y^\ell)$ predicted by the resummed and the NLO
calculations. We show that they are the same as long as 
there are no kinematic cuts imposed, and different
when some kinematic cuts are included. They differ the most in the large
rapidity region which is near the boundary of the phase space.

To systematically analyze the differences between the results of the NLO and
the resummed calculations we implemented the ${\cal O}(\alpha _S^0)$ (LO),
the ${\cal O}(\alpha _S)$ (NLO), and the resummed calculations in a unified
Monte Carlo package: ResBos (the acronym stands for $Res$ummed Vector $Bos$%
on production). The code calculates distributions for the hadronic
production and decay of a vector bosons via $h_1h_2\rightarrow V(\rightarrow
\ell _1 {\bar \ell_2})X$, where $h_1$ is a proton and $h_2$ can be a proton,
anti-proton, neutron, an arbitrary nucleus or a pion. Presently, $V$ can be
a virtual photon $\gamma^*$ (for Drell-Yan production), $W^{\pm }$ or $Z^0$.
The effects of the initial state soft gluon radiation are included using
the QCD soft gluon resummation formula, given in Eq.~(\ref{eq:ResFor}). 
This code also correctly takes into account the effects of the polarization 
and the decay width of the massive vector boson.

It is important to distinguish ResBos from the parton shower Monte Carlo
programs like ISAJET \cite{isajet}, PYTHIA \cite{pythia}, HERWIG \cite{herwig}, 
etc., which use the backward radiation technique~\cite{Sjostrand} to
simulate the physics of the initial state soft gluon radiation. They are
frequently shown to describe reasonably well the shape of the vector boson
distribution. On the other hand, these codes do not have the full
resummation formula implemented and include only the leading logs and
some of the sub-logs of the Sudakov factor. 
The finite part of the higher order virtual corrections
which leads to the Wilson coefficient ($C$) functions is missing from these 
event generators.
ResBos contains not only the physics from the multiple soft gluon emission, but 
also the higher order matrix elements for the production and the decay of the
vector boson with large $Q_T$, so that it can correctly predict both the event
rates and the distributions of the decay leptons.

% FFFFFFFFFFFFFFFFFFFFFFFFFFFFFFFFFFFFFFFFFFFFFFFFFFFFFFFFFFFFFFFFFFFFFFF
\begin{figure*}[t]
\begin{center}
\begin{tabular}{cc}
\ifx\nopictures Y \else{ 
\epsfysize=6.0cm 
\epsffile{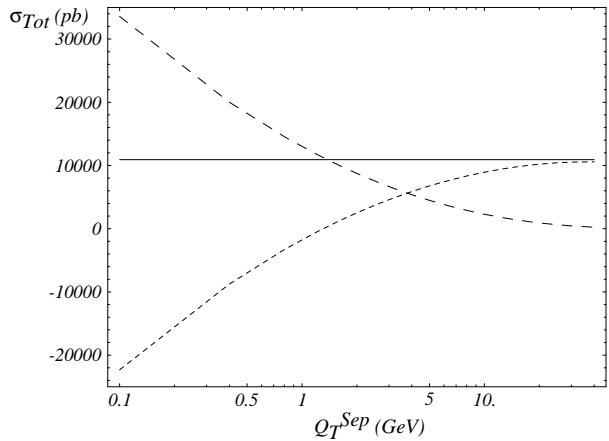}} \fi
& 
\end{tabular}
\end{center}
\caption{ Total $W^+$ production cross section as a function of the
parameter $Q_T^{Sep}$ (solid curve). The 
long dashed curve is the part of the ${\cal {O}(\alpha_S)}$ 
cross section integrated from $Q_T^{Sep}$ to the
kinematical boundary, and the short dashed 
curve is the integral from $Q_T = 0$ to $Q_T^{Sep}$ at ${\cal O}(\alpha_S)$. 
The total cross section is constant 
within $10^{-5}$ \% through more than two order of magnitude of $Q_T^{Sep}$.}
\label{fig:QTSep}
\end{figure*}
% ffffffffffffffffffffffffffffffffffffffffffffffffffffffffffffffffffffff
In a NLO Monte Carlo calculation, it is ambiguous to treat the singularity
of the vector boson transverse momentum distribution near $Q_T=0$. 
There are different ways to deal with this singularity. Usually one
separates the singular region of the phase space from the rest 
(which is calculated numerically) and handles it
analytically. We choose to divide the $Q_T$ phase space with a separation
scale $Q_T^{Sep}$. We treat the $Q_T$ singular parts of the real emission
and the virtual correction diagrams analytically, and
integrate the sum of their contributions up to $Q_T^{Sep}$. 
If $Q_T<Q_T^{Sep}$ we
assign a weight to the event based on the above integrated result and put it
into the $Q_T=0$ bin. If $Q_T>Q_T^{Sep}$, 
the event weight is given by the usual NLO calculation. 
The above procedure not only ensures a stable numerical result 
but also agrees well with the logic of the
resummation calculation. In Fig.~\ref{fig:QTSep} we demonstrate that the
total cross section, as expected, is independent of the separation scale 
$Q_T^{Sep}$ in a wide range. As explained above, 
in the $Q_T<Q_T^{Sep}$ region we approximate the $Q_T$ of the
vector boson to be zero. For this reason, we choose $Q_T^{Sep}$ as small as
possible. 
We use $Q_T^{Sep}=0.1$ GeV in our numerical calculations, unless otherwise 
indicated.
This division of the transverse momentum phase space gives us practically
the same results as the invariant mass phase space slicing technique. This
was precisely checked by the lepton charge asymmetry results predicted by
DYRAD \cite{Giele}, and the NLO [up to ${\cal O}(\alpha_S)$] calculation within
the ResBos Monte Carlo package. 

To facilitate our comparison, we calculate the NLO and the resummed
distributions using the same parton luminosities and parton distribution
functions, EW and QCD parameters, and renormalization and factorization scales
so that any difference found in the distributions is clearly due to the
different QCD physics included in the theoretical calculations.
(Recall that they are different models
of calculations based upon the same QCD theory, and the resummed calculation 
contains the dynamics of the multiple soft gluon radiation.) 
This way we compare the resummed and the NLO 
results on completely equal footing. The parton distributions used in the 
different order calculations are listed in Table~\ref{tbl:PDF}. 
In Table~\ref{tbl:PDF} and the rest of this work, we denote 
by Resummed ${\cal O}(\alpha _S^2)$ the result of the resummed calculation
with $A^{(1,2)}$, $B^{(1,2)}$ and $C^{(0,1)}$ included [cf. Appendix D];
by Resummed ${\cal O}(\alpha _S)$
with $A^{(1)}$, $B^{(1)}$ and $C^{(0,1)}$;
by Resummed ${\cal O}(\alpha _S^0)$
with $A^{(1)}$, $B^{(1)}$ and $C^{(0)}$.
(Similarly, later in Table~\ref{tbl:Total}, CSS ${\cal O}(\alpha _S^2)$
implies that $A^{(1,2)}$, $B^{(1,2)}$ and $C^{(0,1)}$ included in the 
resummation calculation, etc.)
In the following, we discuss the relevant experimental observables 
predicted by these models of calculations using the ResBos code.
Our numerical results are given for the Tevatron, a $p {\bar p}$
collider with $\sqrt{S} = 1.8$ TeV, and CTEQ4 PDF's unless specified 
otherwise.

% TTTTTTTTTTTTTTTTTTTTTTTTTTTTTTTTTTTTTTTTTTTTTTTTTTTTTTTTTTTTTTTTTTTTTTT
\begin{table}[tbp]
\begin{center}
\begin{tabular}{l|ccc|ccc}
& \multicolumn{3}{c|}{Fixed order} & \multicolumn{3}{c}{Resummed} \\ 
& ${\cal O}(\alpha _S^0)$ & ${\cal O}(\alpha _S)$ & ${\cal O}(\alpha _S^2)$
& ${\cal O}(\alpha _S^0)$ & ${\cal O}(\alpha _S)$ & ${\cal O}(\alpha _S^2)$
\\ \hline
PDF & CTEQ4L & CTEQ4M & CTEQ4M & CTEQ4L & CTEQ4M & CTEQ4M %\\ 
%$\alpha _S$ (order) & 1 & 2 & 2 & 1 & 2 & 2
\end{tabular}
\end{center}
\caption{List of PDF's used at the different models of calculations. 
The values of the strong coupling constants used with the CTEQ4L and CTEQ4M 
PDF's are $\alpha _S^{(1)}(M_{Z^0}) = 0.132$ and 
$\alpha _S^{(2)}(M_{Z^0}) = 0.116$ respectively.}
\label{tbl:PDF}
\end{table}
% ttttttttttttttttttttttttttttttttttttttttttttttttttttttttttttttttttttttt

% sssssssssssssssssssssssssssssssssssssssssssssssssssssssssssssssssssssss

\subsection{Vector Boson Transverse Momentum Distribution}

\begin{figure*}[t]
\begin{center}
\begin{tabular}{cc}
\ifx\nopictures Y \else{ \epsfysize=6.0cm $\!\!\!\!\!\!\!\!\!\!\!\!$
\epsffile{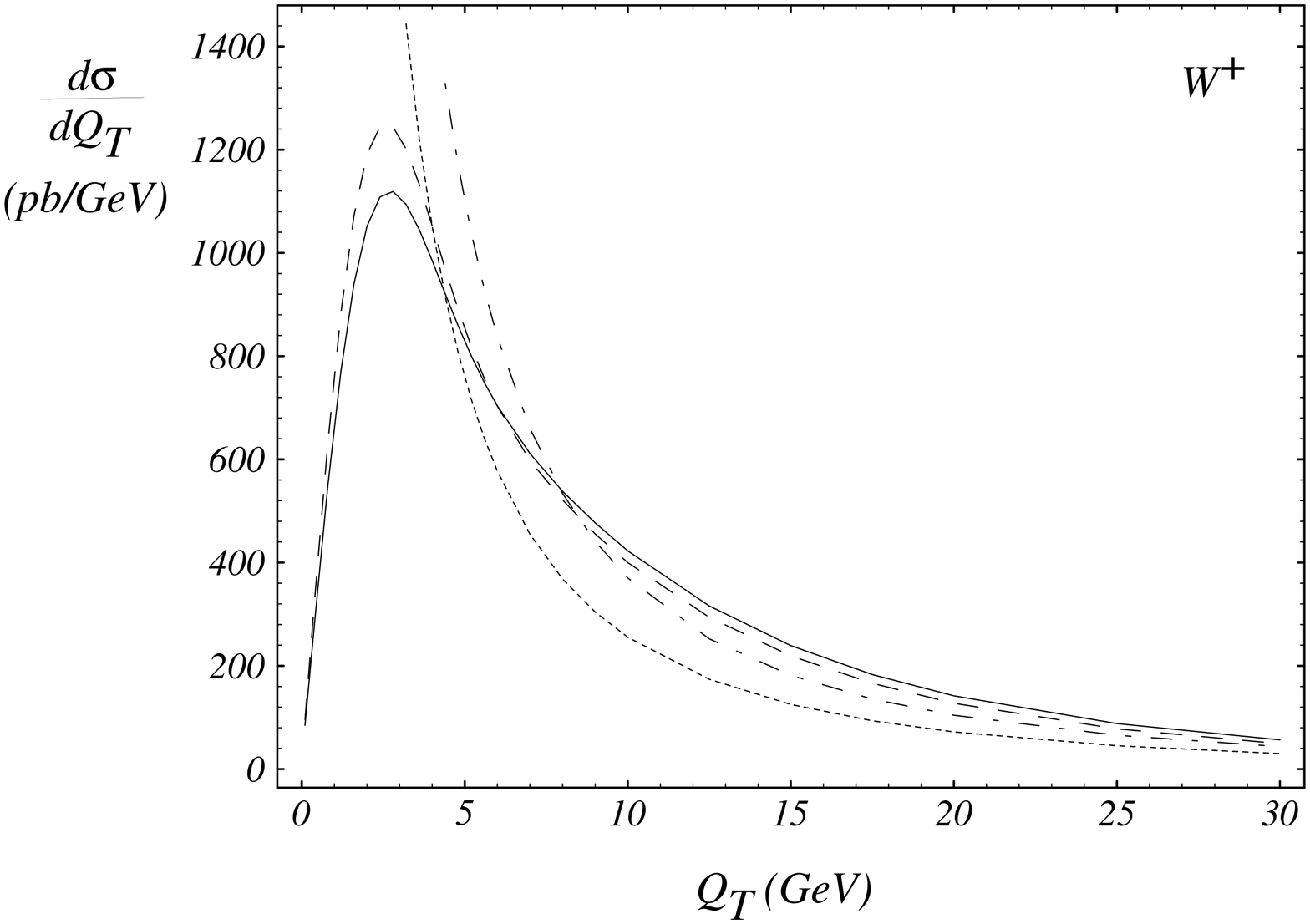} 
\epsfysize=6.0cm
\epsffile{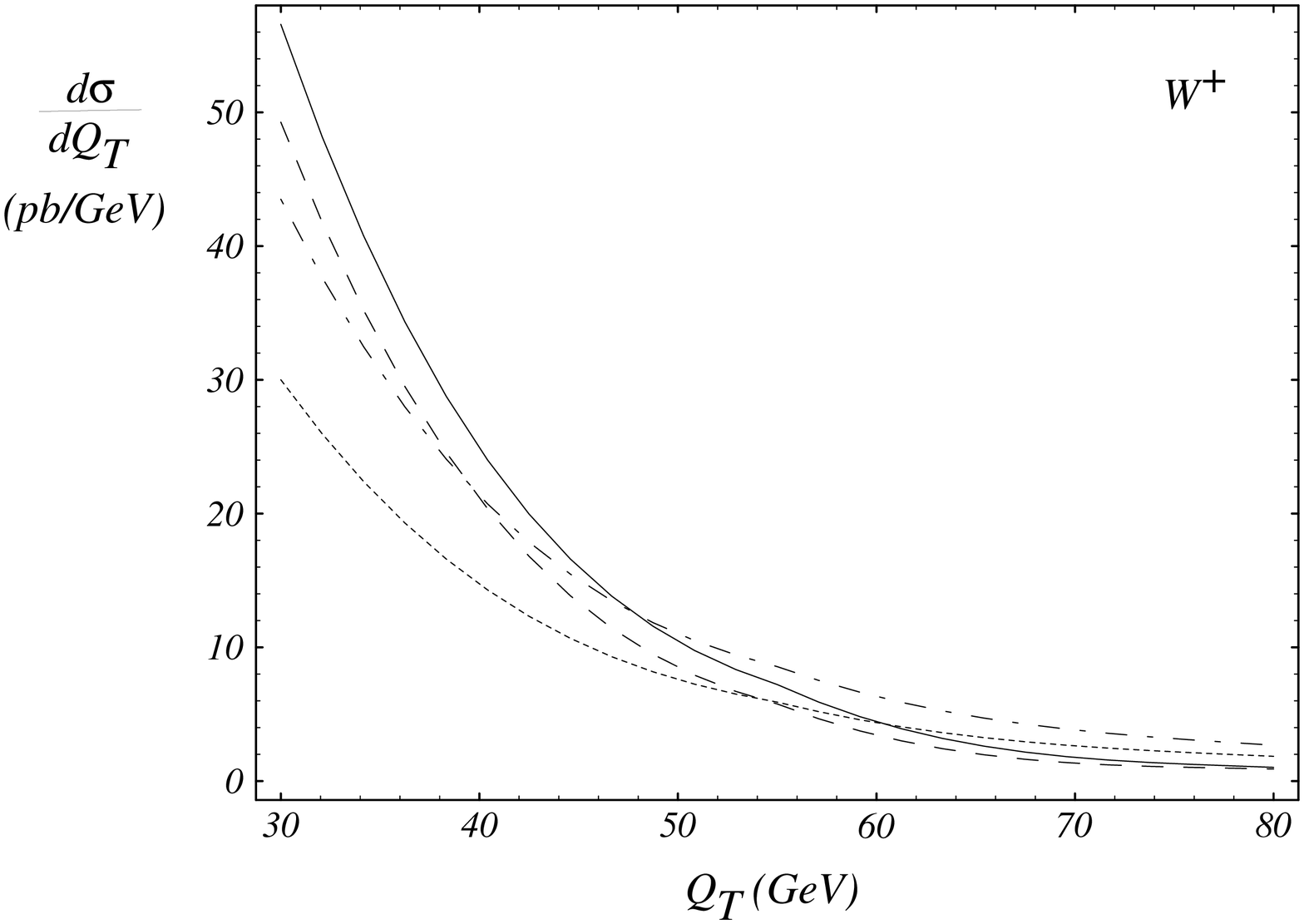} 
} \fi &  \\ 
\ifx\nopictures Y \else{ \epsfysize=6.0cm $\!\!\!\!\!\!\!\!\!\!\!\!$
\epsffile{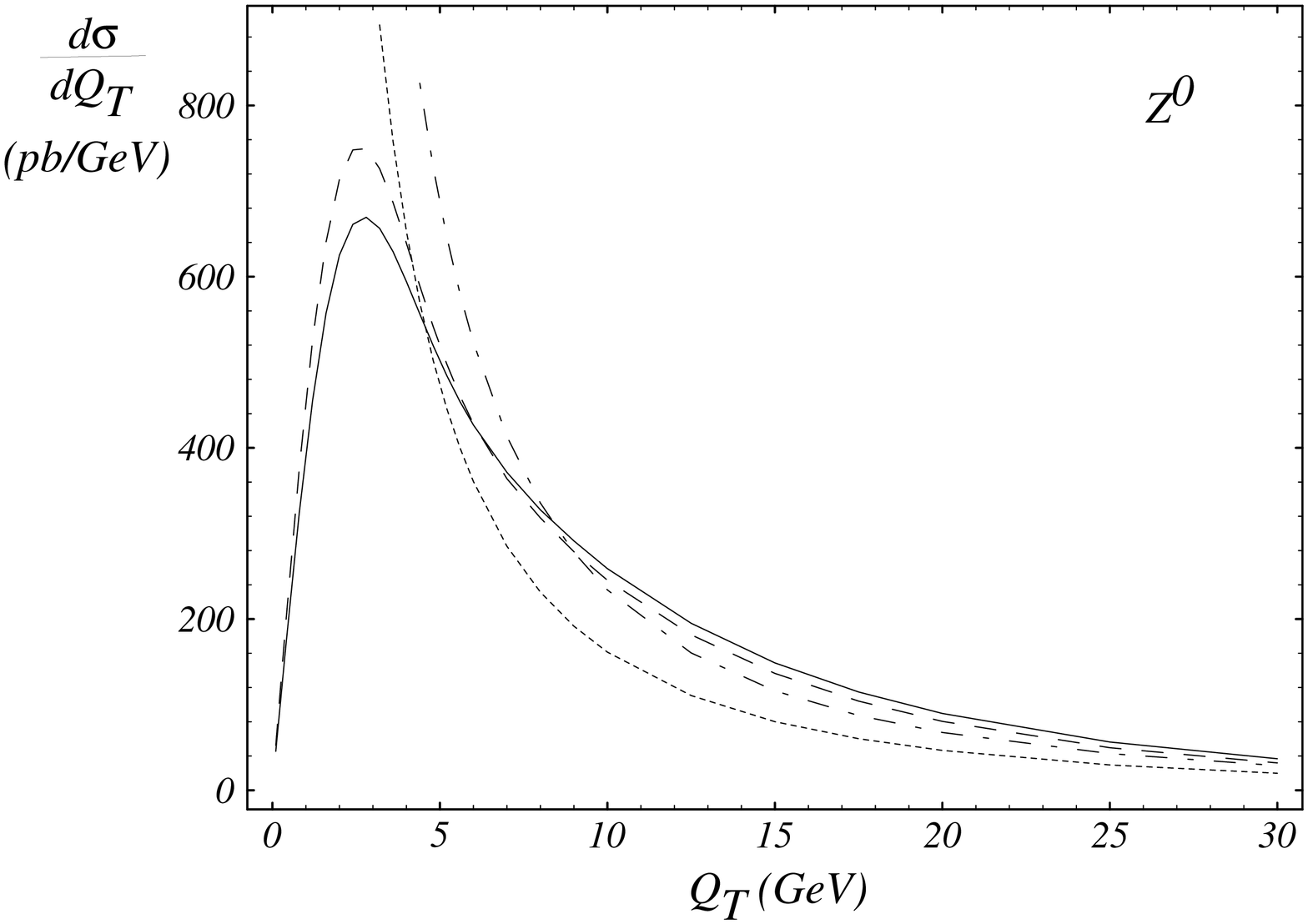} 
\epsfysize=6.0cm
\epsffile{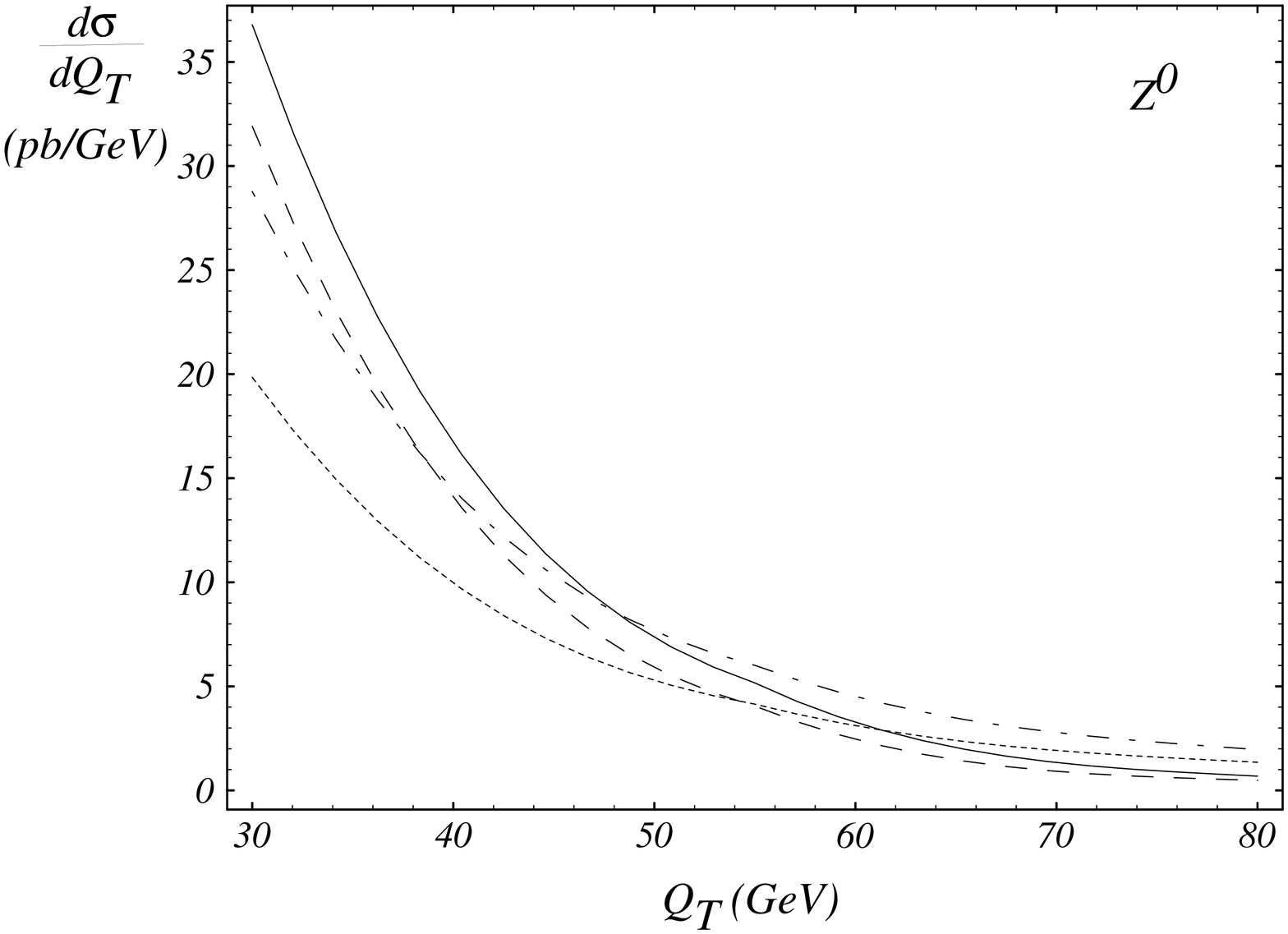} 
} \fi & 
\end{tabular}
\end{center}
\caption{The low and intermediate $Q_T$ regions of the $W^+$ and $Z^0$ 
distributions at the
Tevatron, calculated in fixed order ${\cal {O}(\alpha_S)}$ (dotted)
and ${\cal O}(\alpha_S^2)$ (dash-dotted), and resummed ${\cal {O}
(\alpha_S)}$ (dashed) and ${\cal O}(\alpha_S^2)$ (solid) 
%[cf. Table~\ref{tbl:Total}]. 
[cf. Table~III].
The cross-over occurs at 54 GeV for the ${\cal O}(\alpha_S)$, 
and at 49 GeV for the ${\cal O}(\alpha_S^2)$ $W^\pm$ distributions. 
The situation is very similar for the $Z^0$ boson.
}
\label{fig:Matching}
\end{figure*}

According to the parton model the primordial transverse momenta of partons
entering into the hard scattering are zero. 
This implies that a $\gamma ^{*}$, $W^{\pm }$ or $Z^0$ boson produced 
in the Born level process has no
transverse momentum, so that the LO $Q_T$ distribution is a Dirac-delta
function peaking at $Q_T=0$. In order to have a vector boson produced with a 
non-zero $Q_T$, an additional parton has to be emitted from the initial
state partons. This happens in the QCD process. However the singularity at 
$Q_T=0$ prevails up to {\it any fixed order} in $\alpha _S$ of the
perturbation theory, and the transverse momentum distribution $d\sigma
/dQ_T^2 $ is proportional to $Q_T^{-2} \times [1~{\rm {or}~}\ln (Q_T^2/Q^2)]$
at small enough transverse momenta. The most important feature of the
transverse momentum resummation formalism is to correct this unphysical
behavior and render $d\sigma /dQ_T^2$ finite at zero $Q_T$. The 
$Q_T$ distributions of the $W^{+}$ and $Z^0$ bosons predicted 
by the NLO and the resummed calculations are shown in Fig.~\ref{fig:Matching}. 
% FFFFFFFFFFFFFFFFFFFFFFFFFFFFFFFFFFFFFFFFFFFFFFFFFFFFFFFFFFFFFFFFFFFFFFF
% ffffffffffffffffffffffffffffffffffffffffffffffffffffffffffffffffffffff

\begin{figure*}[t]
\begin{center}
\begin{tabular}{cc}
\ifx\nopictures Y \else{ 
\epsfysize=6.0cm 
\epsffile{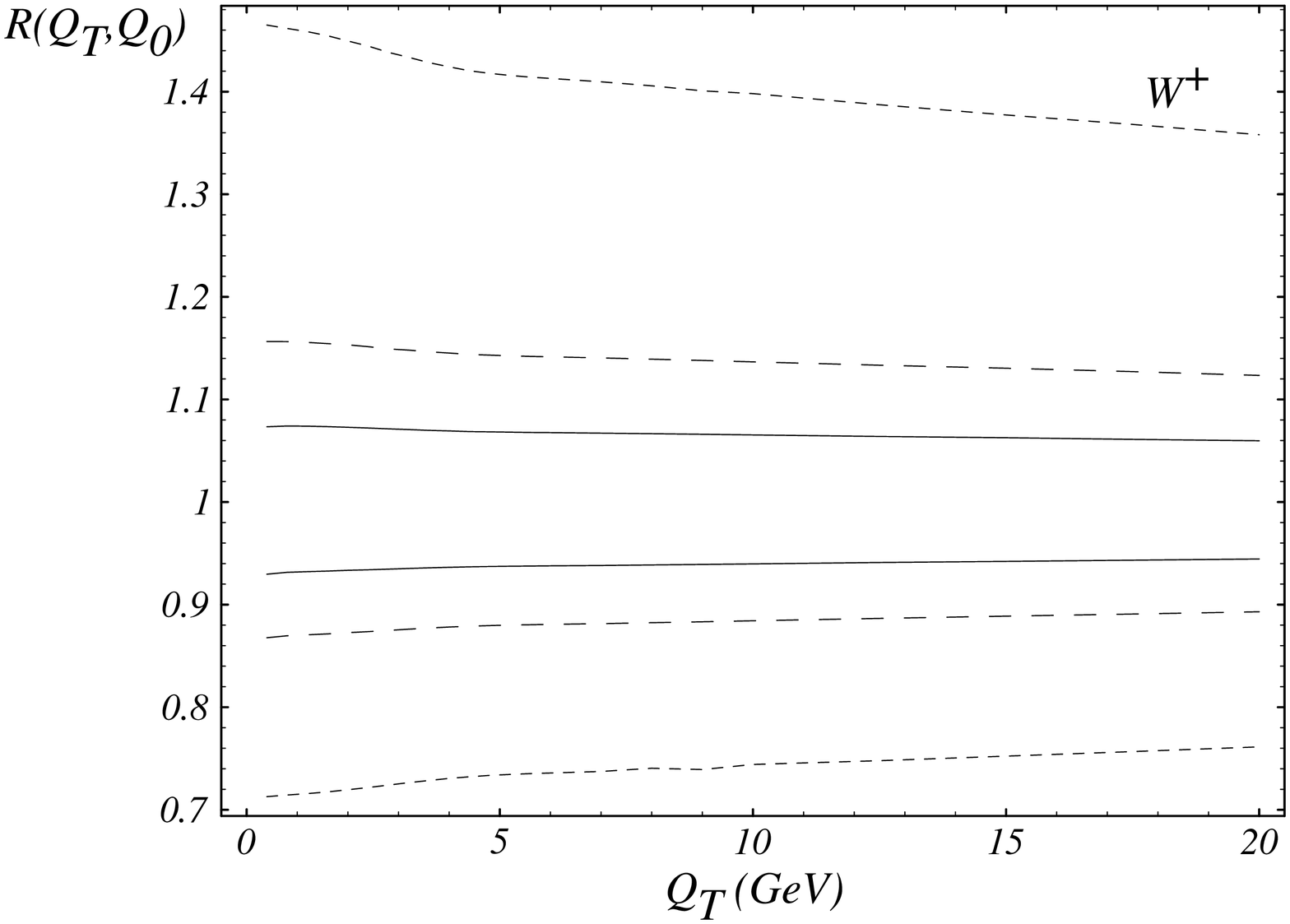}} \fi \\
\ifx\nopictures Y \else{ 
\epsfysize=6.0cm 
\epsffile{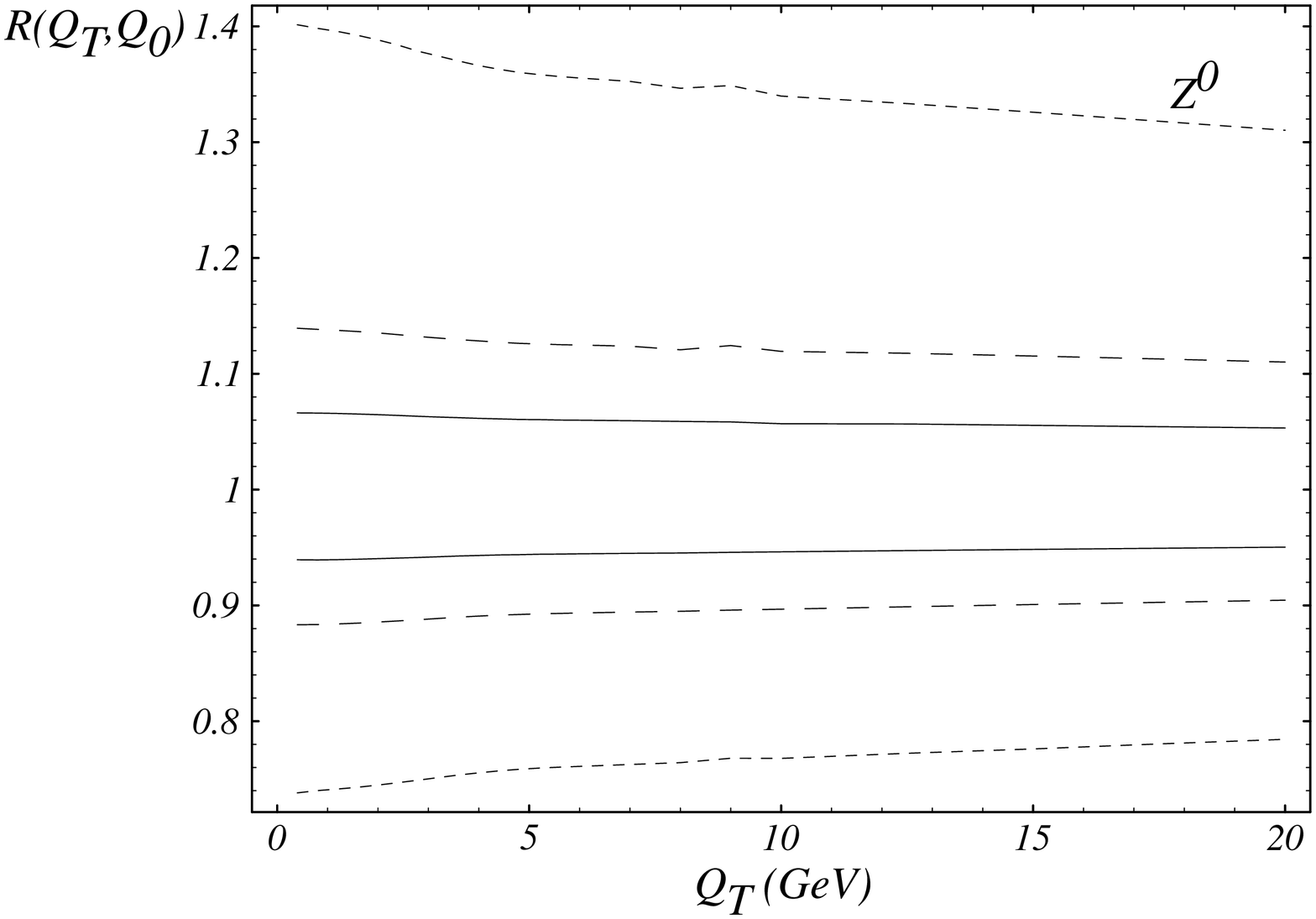}} \fi & 
\end{tabular}
\end{center}
\caption{ The ratio $R(Q_T,Q_0)$, with $y_0 = 0$, for $W^+$ and $Z^0$ 
bosons as a function of $Q_T$. For $W^+$, solid lines are: 
$Q_0 = 78$ GeV (upper) and 82 GeV (lower), dashed: 
$Q_0 = 76$ GeV (upper) and 84 GeV (lower), dotted:
$Q_0 = 70$ GeV (upper) and 90 GeV (lower). 
For $Z^0$ bosons, solid lines: 
$Q_0 = 88$ GeV (upper) and 92 GeV (lower), dashed: 
$Q_0 = 86$ GeV (upper) and 94 GeV (lower), dotted:
$Q_0 = 80$ GeV (upper) and 100 GeV (lower).}
\label{fig:RqTQ}
\end{figure*}

\begin{figure*}[t]
\begin{center}
\begin{tabular}{cc}
\ifx\nopictures Y \else{ 
\epsfysize=6.0cm 
\epsffile{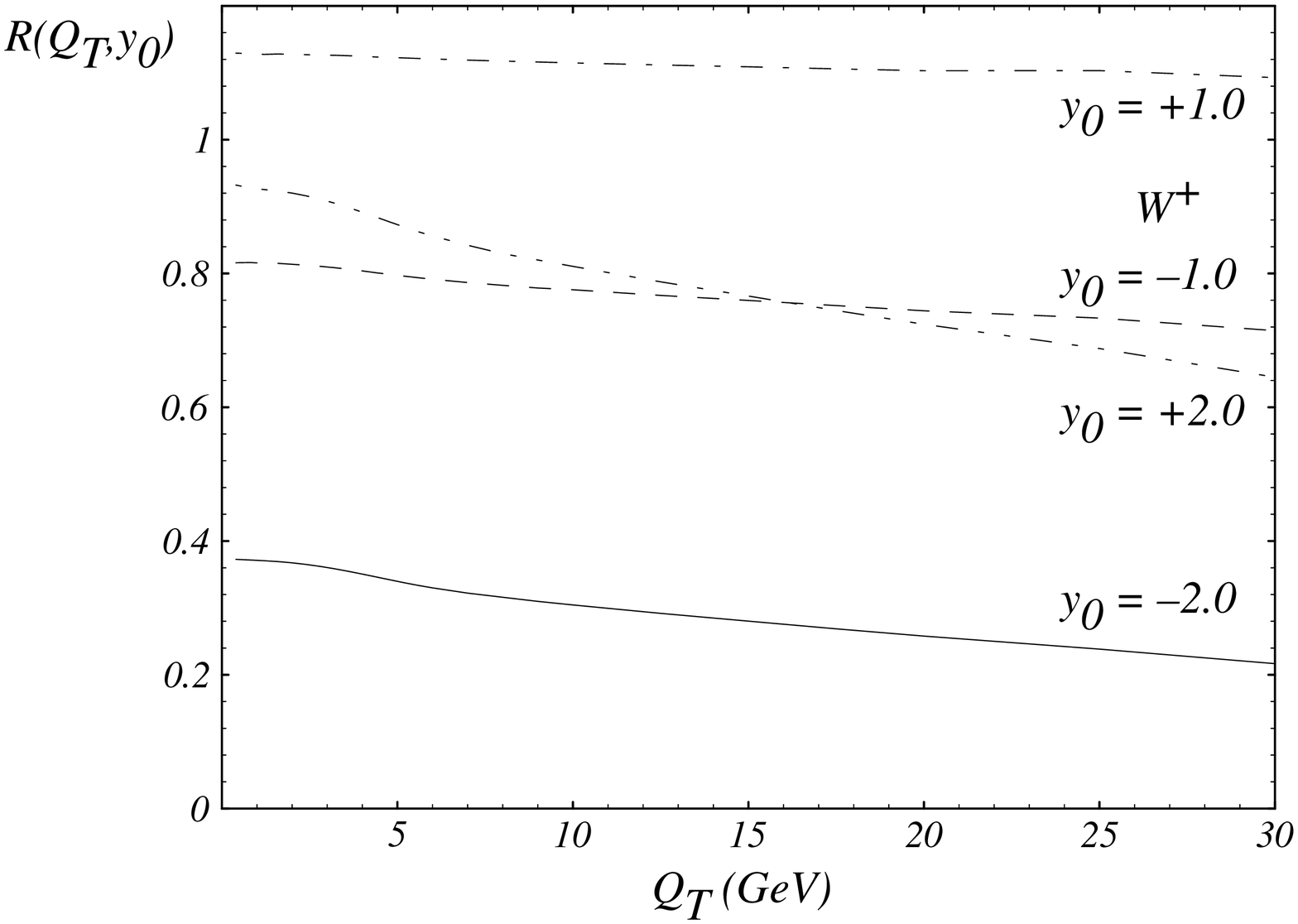}} \fi \\
\ifx\nopictures Y \else{ 
\epsfysize=6.0cm 
\epsffile{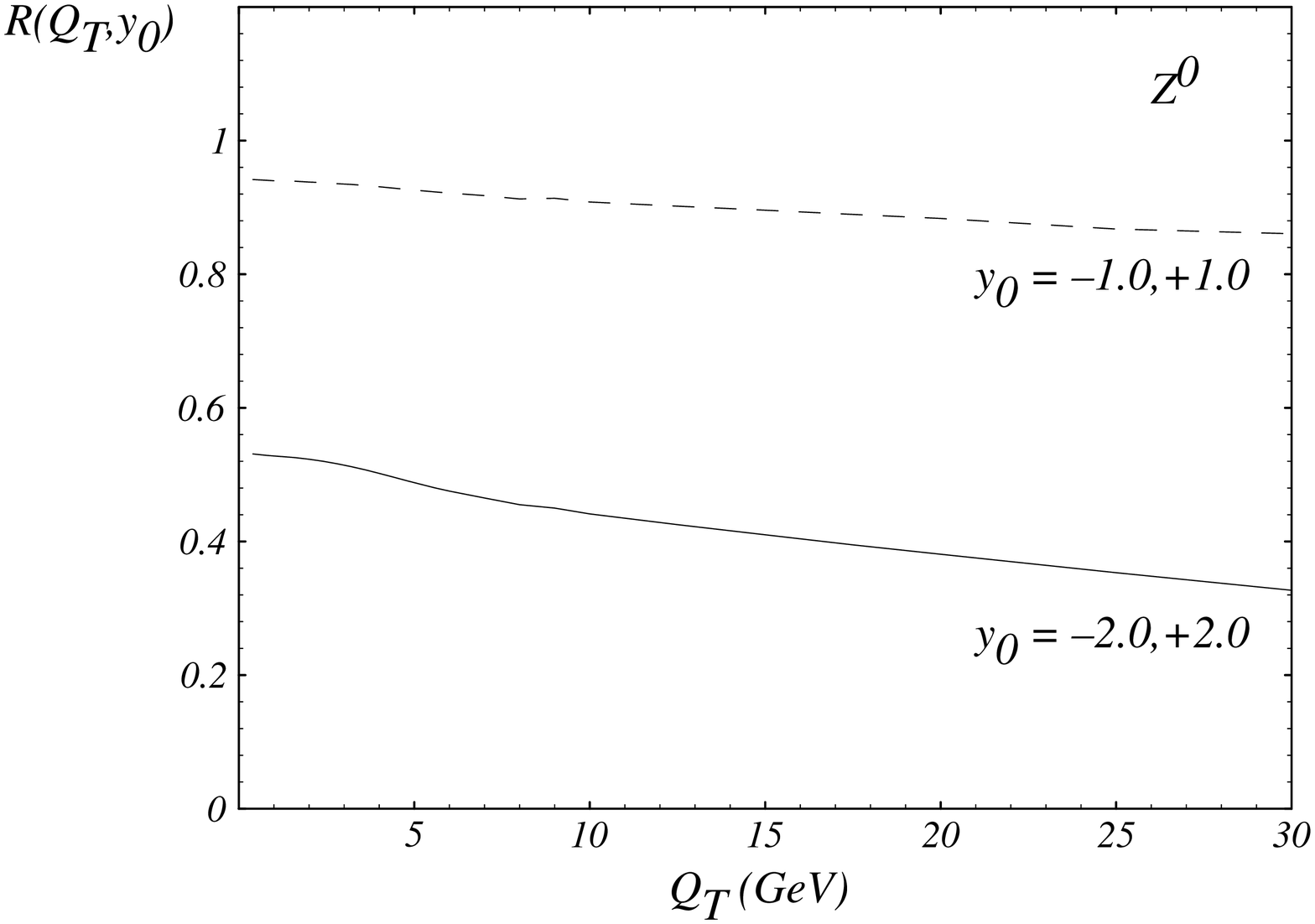}} \fi & 
\end{tabular}
\end{center}
\caption{The ratio $R(Q_T,y_0)$, with $Q_0 = M_V$, for $W^+$ and $Z^0$ bosons 
as a function of $Q_T$.}
\label{fig:RqTy}
\end{figure*}

We find that in the resummed calculation, after taking out the resonance
weighting factor $Q^2/((Q^2-M_V^2)^2+Q^4\Gamma _V^2/M_V^2)$ in Eq.~(\ref
{eq:ResFor}), the shape of the transverse momentum distribution of the 
vector boson $V$ between $Q_T=0$ and $20$ GeV is remarkably constant for $%
Q$ being in the vicinity of $M_V$. Fixing the rapidity $y$ of the vector boson
$V$ at some value $y_0$ and taking the ratio 
% EEEEEEEEEEEEEEEEEEEEEEEEEEEEEEEEEEEEEEEEEEEEEEEEEEEEEEEEEEEEEEEEEEEEEE
%\begin{eqnarray*}
%R(Q_T,Q_0)={\frac{\left. d\sigma /dQ^2dQ_T^2dy\right| _{fixed\;y}}{\left.
%d\sigma /dQ^2dQ_T^2dy\right| _{Q=M_V,\;fixed\;y}}},
%\end{eqnarray*}
% eeeeeeeeeeeeeeeeeeeeeeeeeeeeeeeeeeeeeeeeeeeeeeeeeeeeeeeeeeeeeeeeeeeeee
% EEEEEEEEEEEEEEEEEEEEEEEEEEEEEEEEEEEEEEEEEEEEEEEEEEEEEEEEEEEEEEEEEEEEEE
\begin{eqnarray*}
R(Q_T,Q_0)=
{\frac
{\left. \displaystyle \frac{d\sigma}{d Q^2d Q_T^2dy}\right| _{Q=Q_0,y=y_0}}
{\left. \displaystyle \frac{d\sigma}{d Q^2d Q_T^2dy}\right| _{Q=M_V,y=y_0}}},
\end{eqnarray*}
% eeeeeeeeeeeeeeeeeeeeeeeeeeeeeeeeeeeeeeeeeeeeeeeeeeeeeeeeeeeeeeeeeeeeee
we obtain almost constant curves (within 3 percent) for $Q = M_V \pm 10$ GeV 
(cf. Fig.~\ref{fig:RqTQ}) for $V = W^+$ and $Z^0$. 
%??? Why there's hardly any $Q^2$ and y dependence of the shape??? 
The fact that the shape of the transverse momentum distribution shows such 
a weak dependence on the invariant mass $Q$ in the vicinity of the vector boson
mass can be used to make the Monte Carlo implementation of the resummation
calculation faster. This weak dependence was also used in the \D0 $W$ mass 
analysis when assuming that the mass dependence of the fully differential 
$W$ boson production cross section factorizes as a multiplicative 
term~\cite{Flattum}. 
%??? Why else is this important ???
% FFFFFFFFFFFFFFFFFFFFFFFFFFFFFFFFFFFFFFFFFFFFFFFFFFFFFFFFFFFFFFFFFFFFFFF
% ffffffffffffffffffffffffffffffffffffffffffffffffffffffffffffffffffffff
Similarly, we define the ratio
% EEEEEEEEEEEEEEEEEEEEEEEEEEEEEEEEEEEEEEEEEEEEEEEEEEEEEEEEEEEEEEEEEEEEEE
\begin{eqnarray*}
R(Q_T,y_0)=
{\frac{\left. \displaystyle \frac{d\sigma}{dQ^2dQ_T^2dy}\right| _{Q=M_V,y=y_0}}
      {\left. \displaystyle \frac{d\sigma}{dQ^2dQ_T^2dy}\right| _{Q=M_V,y=0}}},
\end{eqnarray*}
% eeeeeeeeeeeeeeeeeeeeeeeeeeeeeeeeeeeeeeeeeeeeeeeeeeeeeeeeeeeeeeeeeeeeee
to study the $Q_T$ shape variation as a function of the vector boson rapidity. 
Our results are shown in Fig.~\ref{fig:RqTy}. Unlike the ratio $R(Q_T,Q_0)$
shown in Fig.~\ref{fig:RqTQ}, the distributions of $R(Q_T,y_0)$ for the 
$W^\pm$ and $Z^0$ bosons are clearly different for any value of the rapidity
$y_0$. 

To utilize the information on the transverse momentum of the $W^+$ boson in
Monte Carlo simulations to reconstruct the mass of the $W^+$ boson, it was
suggested in Ref.~\cite{Reno} to predict $Q_T(W^+)$ distribution from the
measured $Q_T(Z^0)$ distribution and the calculated ratio of 
$Q_T(W^+)$ and $Q_T(Z^0)$
predicted by the resummation calculations \cite{CSS,Arnold-Kauffman}, in
which the vector boson is assumed to be on its mass-shell. 
%Besides of the fact that the number of lepton pairs from $Z^0$'s is about
%one order of magnitude smaller due to the 
%differences in couplings and branching ratios. 
Unfortunately, this idea will not work with a good precision because, 
as clearly shown in Fig.~\ref{fig:RqTy}, 
the ratio of the $W^+$ and $Z^0$ transverse momentum distributions 
depends on the rapidities of the vector bosons. 
Since the rapidity of
the $W^+$ boson cannot be accurately reconstructed without knowing the
longitudinal momentum (along the beam pipe direction) of the neutrino, 
which is in the form of missing energy carried away by the neutrino, 
this dependence cannot be incorporated in data analysis and the above 
ansatz cannot be realized in practice for a precision measurement 
of $M_W$.\footnote{%
If a high precision measurement were not required, 
then one could choose from the two-fold solutions for the neutrino longitudinal 
momentum to calculate the longitudinal momentum of the $W^\pm$ boson.} 
Only the Monte Carlo implementation of the exact 
matrix element calculation (ResBos) can correctly predict the distributions 
of the decay leptons, such as the transverse mass of the $W^\pm$ boson, 
and the transverse momentum of the charged lepton, so that 
they can be directly compared with experimental data to extract the value of 
$M_W$. We comment on these results later in this section. 
% FFFFFFFFFFFFFFFFFFFFFFFFFFFFFFFFFFFFFFFFFFFFFFFFFFFFFFFFFFFFFFFFFFFFFFF
% ffffffffffffffffffffffffffffffffffffffffffffffffffffffffffffffffffffff

\begin{figure*}[t]
\begin{center}
\begin{tabular}{cc}
\ifx\nopictures Y \else{ 
\epsfysize=6.0cm 
\epsffile{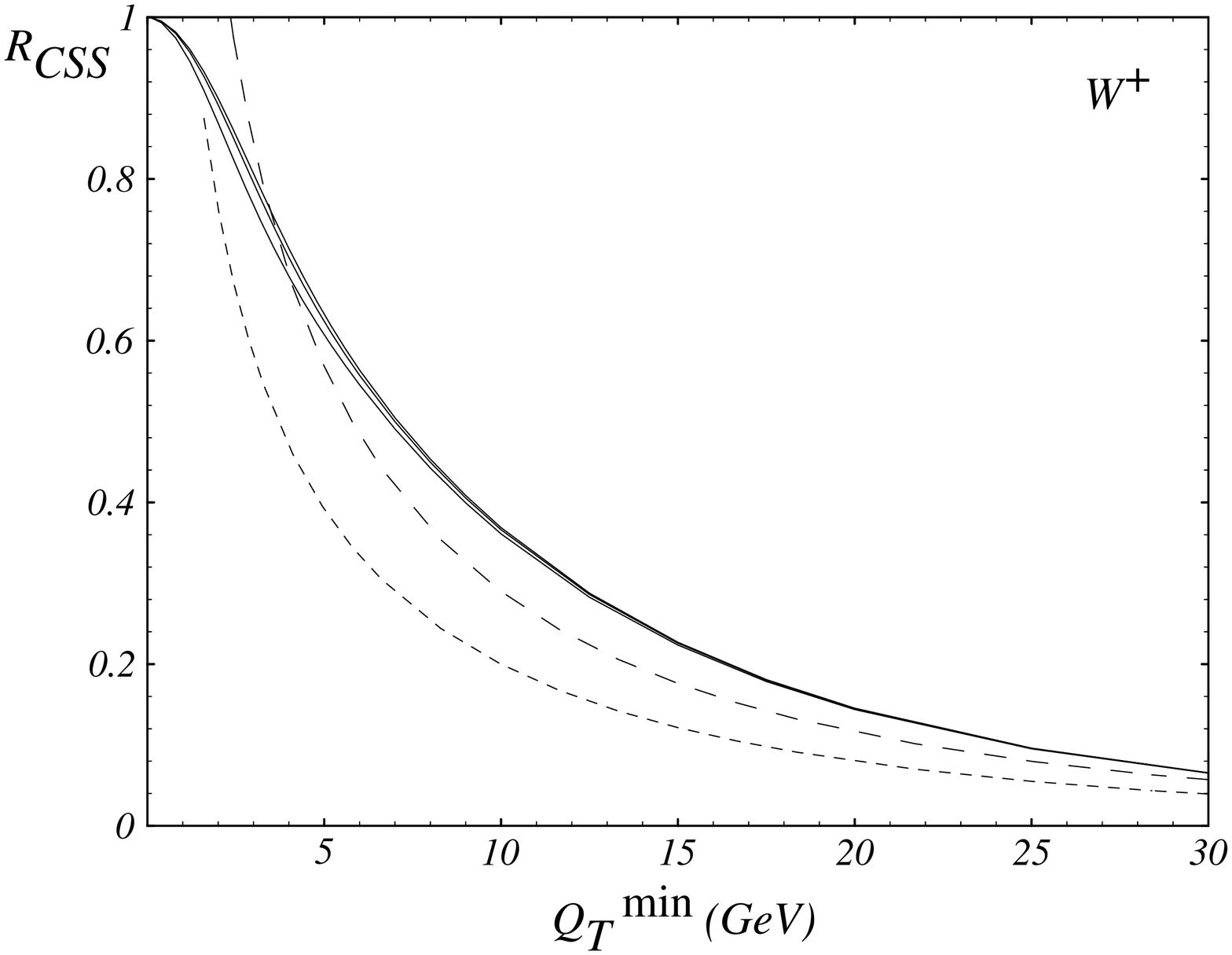} } 
\fi & \\ 
\ifx\nopictures Y 
\else{ \epsfysize=6.0cm 
\epsffile{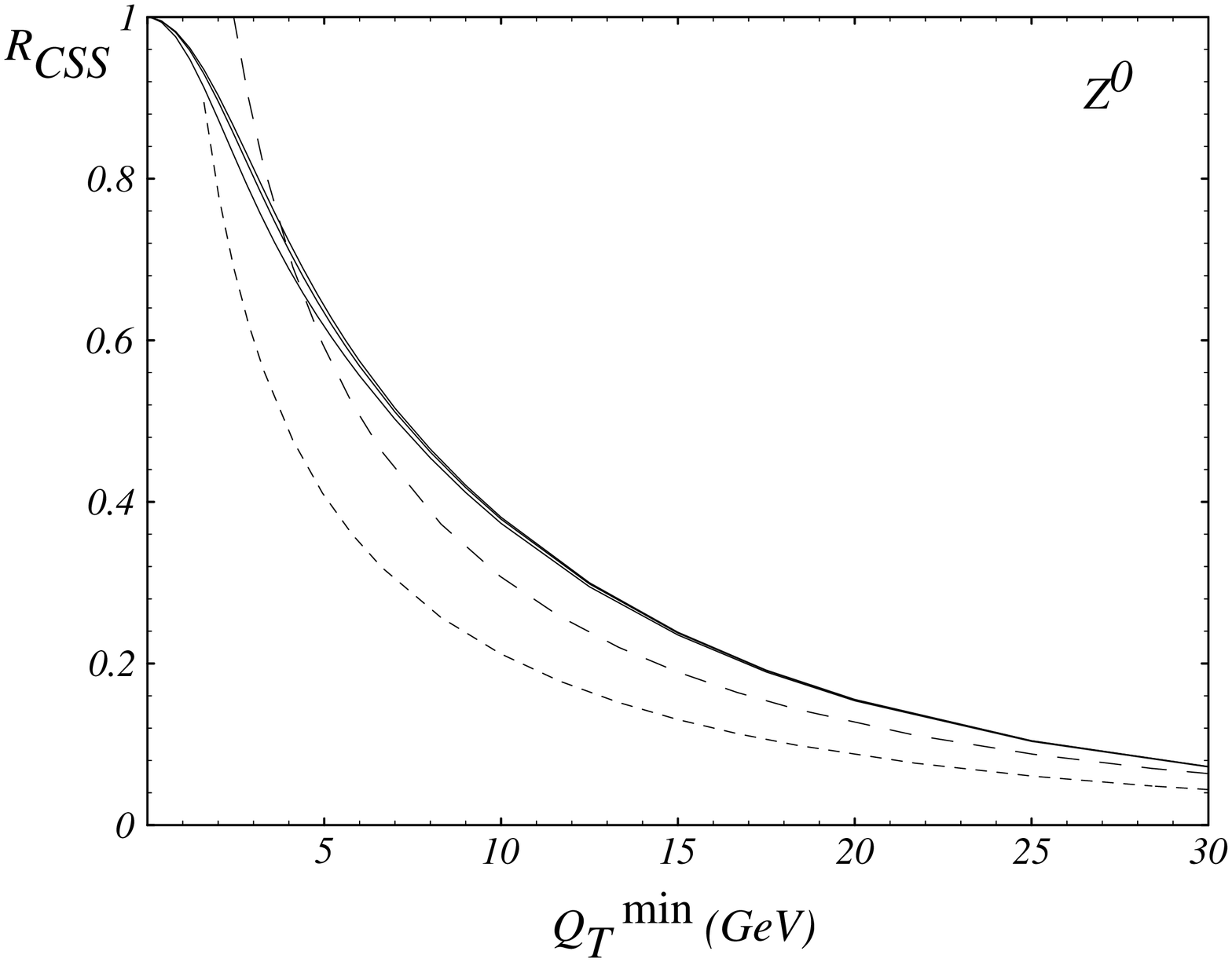} } 
\fi & 
\end{tabular}
\end{center}
\caption{ The ratio $R_{CSS}$ as a function of $Q_T^{\min}$ for $W^+$ and 
$Z^0$ bosons. The fixed order (${\cal O}(\alpha_S)$ short dashed, 
${\cal O}(\alpha_S^2)$ dashed) curves are ill-defined in the low $Q_T$ 
region. The resummed (solid) curves are calculated for $g_2$ = 0.38 (low), 
0.58 (middle) and 0.68 (high) GeV$^2$ values. }
\label{fig:Integrated}
\end{figure*}

Another way to compare the results of the resummed and the NLO calculations
is given by the distributions of $\sigma (Q_T>Q_T^{\min })/\sigma _{Total}$, 
as shown in Fig.~\ref{fig:Integrated}.
We defined the ratio as 
% EEEEEEEEEEEEEEEEEEEEEEEEEEEEEEEEEEEEEEEEEEEEEEEEEEEEEEEEEEEEEEEEEEEEEE
\[
R_{CSS} \equiv \frac{\sigma (Q_T>Q_T^{\min })}{\sigma _{Total}}=\frac 1{%
\sigma _{Total}}\int_{Q_T^{\min }}^{Q_T^{\max}}dQ_T\;\frac{d\sigma
(h_1h_2\rightarrow V)}{dQ_T}, 
\]
% eeeeeeeeeeeeeeeeeeeeeeeeeeeeeeeeeeeeeeeeeeeeeeeeeeeeeeeeeeeeeeeeeeeeee
where $Q_T^{\max}$ is the largest $Q_T$ allowed by the phase space. 
In the NLO calculation, $\sigma(Q_T>Q_T^{\min })$ grows
without bound near $Q_T^{\min }=0$, as the result of the singular behavior 
$1/Q_T^2$ in the matrix element. 
The NLO curve runs well under the resummed one in the 2 GeV $<Q_T^{\min }<$ 
30 GeV region, and the $Q_T$ distributions from the NLO and the resummed 
calculations have different shapes even in the region where $Q_T$ is of the 
order 15 GeV. 

With large number of
fully reconstructed $Z^0$ events at the Tevatron, one should be able to use
data to discriminate these two theory calculations. In view of this result it 
is not surprising that the \D0 analysis of the $\alpha _S$ measurement \cite
{D0alphaS} based on the measurement of $\sigma (W+1\;jet)/\sigma (W+0\;jet)$ 
does not support the NLO\ calculation in which the effects of the multiple
gluon radiation are not included. We expect that if this measurement was
performed by demanding the transverse momentum of the jet to be larger than
about 50 GeV, at which scale the resummed and the NLO distributions in 
Fig.~\ref{fig:Matching} cross, the NLO calculation would adequately describe 
the data. 
% FFFFFFFFFFFFFFFFFFFFFFFFFFFFFFFFFFFFFFFFFFFFFFFFFFFFFFFFFFFFFFFFFFFFFFF
% ffffffffffffffffffffffffffffffffffffffffffffffffffffffffffffffffffffff

To show that for $Q_T$ below 30 GeV, the QCD multiple soft gluon radiation 
is important to explain the \D0 data \cite{D0alphaS}, we also include
in Fig.~\ref{fig:Integrated} the prediction for the $Q_T$ distribution at
the order of $\alpha^2_S$. As shown in the figure, the $\alpha^2_S$ curve is 
closer to the resummed curve which proves that for this range of $Q_T$ the
soft gluon effect included in the $\alpha^2_S$ calculation is important for
predicting the vector boson $Q_T$ distribution. In other words, in this
range of $Q_T$, it is more likely that soft gluons accompany the $W^\pm$ 
boson than just a single hard jet associated with the vector boson
production.
For large $Q_T$, it becomes more likely to have hard jet(s) produced with
the vector boson.

\begin{figure*}[t]
\begin{center}
\begin{tabular}{cc}
\ifx\nopictures Y \else{ 
\epsfysize=6.0cm 
\epsffile{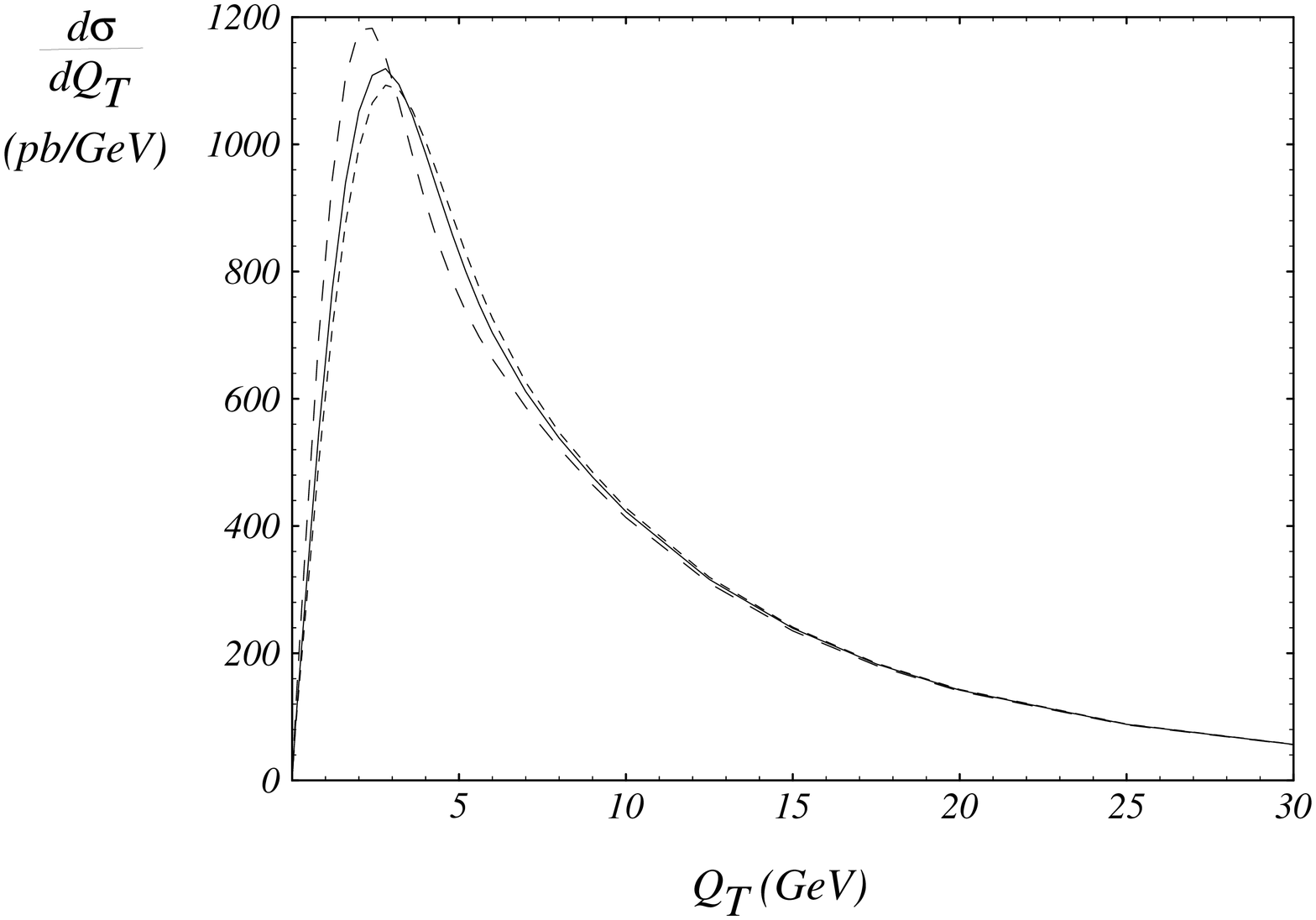} } 
\fi & \\
\ifx\nopictures Y \else{ 
\epsfysize=6.0cm 
\epsffile{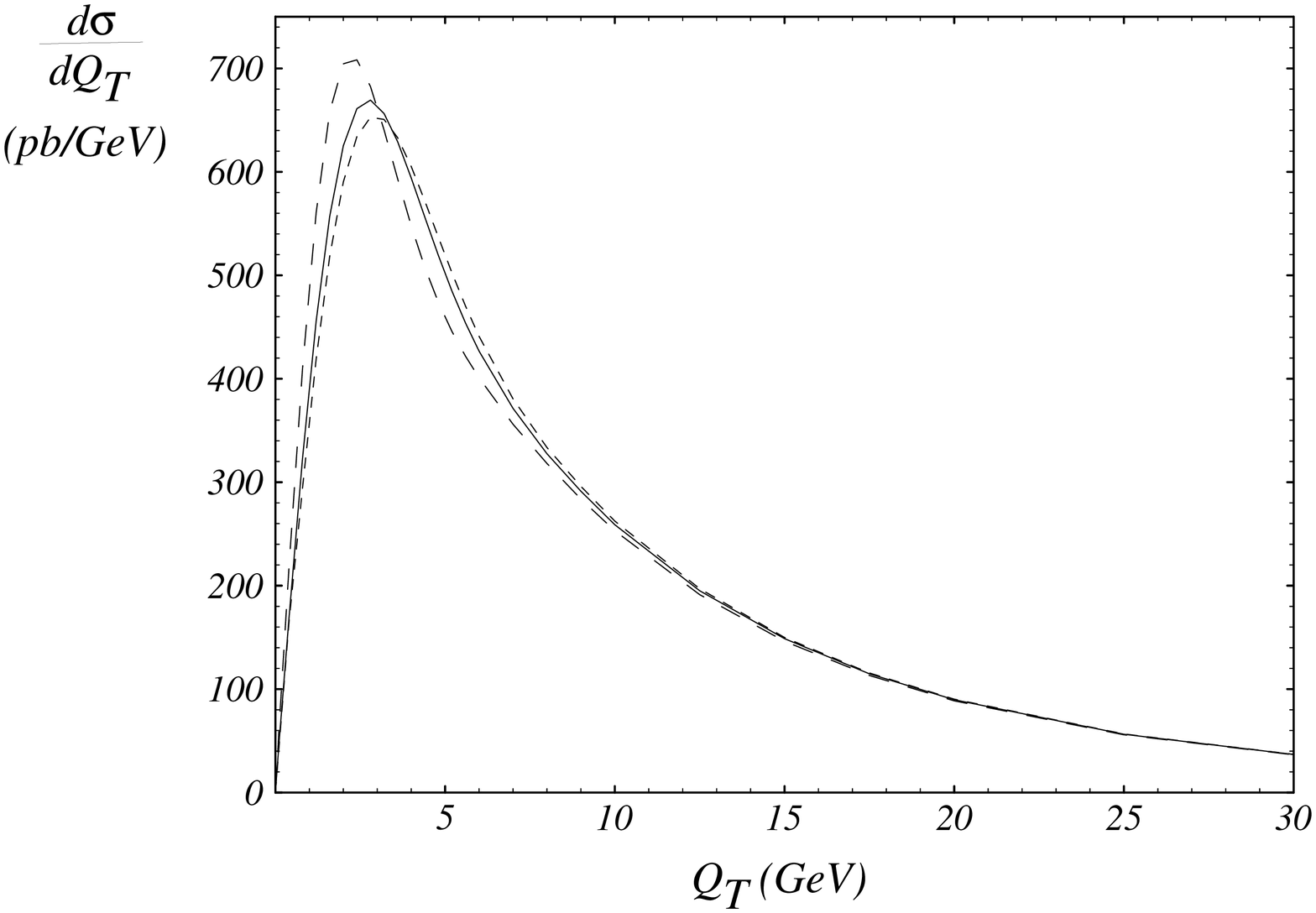} } 
\fi &
\end{tabular}
\end{center}
\caption{ Transverse momentum distributions of $W^+$ and $Z^0$ bosons 
calculated with low (long dash, $g_2$ = 0.38 GeV$^2$), nominal (solid, 
$g_2$ = 0.58 GeV$^2$) and high (short dash, $g_2$ = 0.68 GeV$^2$) $g_2$ 
non-perturbative parameter values. The low and high excursions in $g_2$ 
are the present one standard deviations from the nominal value in the 
Ladinsky-Yuan parametrization. }
\label{fig:W_qT_W_g2}
\end{figure*}

Measuring $R_{CSS}$ in the low $Q_T$ region (for $Q_T \lesssim Q/2$) provides a
stringent test of the dynamics of the multiple soft gluon radiation
predicted by the QCD theory. The same measurement of $R_{CSS}$
can also provide information about some part of the non-perturbative physics
associated with the initial state hadrons. As shown in Fig.~\ref{fig:W_qT_W_g2}
and in Ref.~\cite {Ladinsky-Yuan}, the effect of the non-perturbative physics 
on the $Q_T$ distributions of the 
$W^\pm$ and $Z^0$ bosons produced at the Tevatron is
important for $Q_T$ less than about 10 GeV. This is evident by observing
that different parametrizations of the non-perturbative functions do not
change the $Q_T$ distribution for $Q_T > 10$ GeV, although they do 
dramatically change the shape of $Q_T$ for $Q_T < 10$ GeV. 
Since for $W^\pm$ and $Z^0$ production, the $\ln(Q^2/Q^2_0)$ term is large, the
non-perturbative function, as defined in Eq.~(\ref{eq:Wnonpert}), is dominated
by the $F_1(b)$ term (or the $g_2$ parameter) which is supposed to be universal 
for all Drell-Yan type processes and related to the physics of the renormalon 
\cite{Korchemsky-Sterman}. Hence, the measurement of $R_{CSS}$ cannot only be 
used to test the dynamics of the QCD multiple soft gluon radiation,  
in the $10 \, {\rm GeV} \, < Q_T < 40$ GeV region, but may also be used to 
probe this part of non-perturbative physics for $Q_T < 10$ GeV. 
It is therefore important to
measure $R_{CSS}$ at the Tevatron. With a large sample of $Z^0$ data at Run 2, 
it is possible to determine the
dominant non-perturbative function which can then be used to calculate the
$W^\pm$ boson $Q_T$ distribution to improve the accuracy of the $M_W$ and the 
charged lepton rapidity asymmetry measurements.
% FFFFFFFFFFFFFFFFFFFFFFFFFFFFFFFFFFFFFFFFFFFFFFFFFFFFFFFFFFFFFFFFFFFFFFF
% ffffffffffffffffffffffffffffffffffffffffffffffffffffffffffffffffffffff

% sssssssssssssssssssssssssssssssssssssssssssssssssssssssssssssssssssssss

\subsection{The Total Cross Section}
\label{subsec:Total}

Before we compare the distributions of the decay leptons, we examine the 
question whether or not the $Q_T$ resummation formalism changes the prediction 
for the total cross section. In Ref.~\cite{AEM85} it was shown that in the 
AEGM formalism, which differs from the CSS formalism, the ${\cal O}(\alpha _S)$
total cross section is obtained after integrating their resummation formula
over the whole range of the phase space.

In the CSS formalism, without including the $C$ and $%
Y$ functions, the fully integrated resummed result recovers the ${\cal O}
(\alpha _S^0)$ cross section, provided that $Q_T$ is integrated from zero to 
$Q$. This can be easily verified by expanding the resummation formula up
to ${\cal O}(\alpha _S)$, dropping the $C^{(1)}$ and the $Y$ pieces (which are
of order ${\cal O}(\alpha _S)$), and integrating over the lepton variables. 
It yields 
% EEEEEEEEEEEEEEEEEEEEEEEEEEEEEEEEEEEEEEEEEEEEEEEEEEEEEEEEEEEEEEEEEEEEEE
\begin{eqnarray}
\ &&\int_0^{P_T^2}dQ_T^2\;\frac{d\sigma }{dQ^2dydQ_T^2}=\frac{\sigma _0}S 
\delta(Q^2 - M_V^2) \nonumber \\
&&\ \ \ \ \ \times \left\{ \left( 1-\frac{\alpha _S(Q)}\pi \left[ \frac 12
A^{(1)}\ln ^2\left( \frac{Q^2}{P_T^2}\right) +B^{(1)}\ln \left( \frac{Q^2}{
P_T^2}\right) \right] \right) \;f_{j/h_1}(x_1,Q^2)\;f_{\overline{k}
/h_2}(x_2,Q^2)\right.  \nonumber \\
&&\ \ \ \ \ -\frac{\alpha _S(Q)}{2\pi }\ln \left( \frac{Q^2}{P_T^2}\right) 
\left[ \left( P_{j\leftarrow a}\otimes
f_{a/h_1}\right) (x_1,Q^2)\;f_{\overline{k}/h_2}(x_2,Q^2) \right.
\nonumber \\
&& ~~~~~~~~~~~~~~~~~~~~~~~~~~~ \left. \left. +
\;f_{j/h_1}(x_1,Q^2)
\left( P_{\overline{k}\leftarrow b}\otimes f_{b/h_2}\right) (x_2,Q^2)\right] + 
j\leftrightarrow \overline{k}\right\} ,  
\label{eq:Total}
\end{eqnarray}
% eeeeeeeeeeeeeeeeeeeeeeeeeeeeeeeeeeeeeeeeeeeeeeeeeeeeeeeeeeeeeeeeeeeeee
where $P_T$ is the upper limit of the $Q_T$ integral and we fixed the mass 
of the vector boson for simplicity.
To derive the above result we have used the
canonical set of the $C_i$ ($i=1,2,3$) coefficients (cf. Appendix C and D).
When the upper limit $P_T$ is taken to be $Q$, all the logs in the above
equation vanish and Eq.~(\ref{eq:Total}) reproduces the Born level 
(${\cal O}(\alpha_S^0)$) cross section. Similar conclusion holds for higher
order terms from the expansion of the resummed piece when $C$ and $Y$ are not
included. This is evident because the singular pieces from the expansion
are given by
% EEEEEEEEEEEEEEEEEEEEEEEEEEEEEEEEEEEEEEEEEEEEEEEEEEEEEEEEEEEEEEEEEEEEEE
\begin{eqnarray*}
\left. \frac{d\sigma }{dQ_T^2}\right| _{singular}=\frac 1{Q_T^2}%
\sum_{n=1}^\infty \sum_{m=0}^{2n-1}{}_nv_m\alpha _S^n\ln ^m\left( \frac{Q_T^2}
{Q^2}\right)
\end{eqnarray*}
% eeeeeeeeeeeeeeeeeeeeeeeeeeeeeeeeeeeeeeeeeeeeeeeeeeeeeeeeeeeeeeeeeeeeee
The integral of these singular terms will be proportional to $\ln
(Q^2/P_T^2) $ raised to some power. 
Again, for $P_T=Q$ all the logs vanish and the tree level result is
obtained.

The inclusion of $C^{(1)}$ and $Y$ functions will change the above conclusion 
and lead to a different total cross section, because $C^{(1)}$ contains
the hard part virtual corrections and $Y$ contains the hard gluon
radiation. 
In the $Q_T < Q$ region the resummed piece dominates, because it resums
the singular pieces: $Q_T^{-2}\times [1$ or $\ln (Q^2/Q_T^2)]$, to all
order. However, in the $Q_T>Q$ region the perturbative result should be
used, because the singular pieces do not dominate (the logs are small 
in this case) and the other
contributions in the fixed order perturbative contributions can be
important. In order to define the total cross section a prescription of
smooth transition between the resummed and the fixed order perturbative
results is necessary. In Fig.~\ref{fig:Matching}, we show the resummed
(1,1) (resummation with $A^{(1)},B^{(1)}$ and $C^{(0,1)}$ included) 
and the fixed order ${\cal O}(\alpha _S)$ $Q_T$ distributions 
for $W^+$ and $Z^0$ bosons.
As shown, the resummed (1,1) and the fixed order curves are 
close to each other for $Q/2<Q_T<Q$, and they cross
near $Q_T\sim Q/2$. Henceforth, we adopt the following matching procedure.
For $Q_T$ values below the crossing point we use the resummed curve, and above 
it the ${\cal O}(\alpha _S)$ curve, to define a resummed ${\cal O}(\alpha _S)$
distribution. This distribution is smooth (although not differentiable at
the matching point), and most importantly it
does not alter either the resummed or the ${\cal O}(\alpha _S)$ distributions 
in the kinematic regions where they are proven to be valid. 
This simple matching
prescription provides us with an ${\cal O}(\alpha _S)$ resummed total cross
section with an error of ${\cal O}(\alpha_s^2)$, as shown in Ref.~\cite
{Arnold-Kauffman}. In practice this translates into less than a percent
deviation between the fixed order and the resummed
${\cal O}(\alpha _S)$ total cross sections. 
This can be understood from the earlier discussion that
if the matching were done at $Q_T$ equal to $Q$, then the total cross section 
calculated from the CSS resummation formalism should be the same as that
predicted by the NLO calculation, provided that $C^{(1)}$ and $Y^{(1)}$ are 
included.
However, this matching prescription would not result in a smooth curve for the 
$Q_T$ distribution at $Q_T=Q$. The small difference between the 
resummed ${\cal O}(\alpha_S)$ and the NLO [fixed order, ${\cal O}(\alpha_S)$] 
total cross sections comes from the matching procedure described
above. This difference indicates the size of the higher order corrections not 
included in the NLO calculation.

In order to compare the $Q_T$ distribution with experimental data we also 
include the effect of some known higher order corrections to the Sudakov 
factor and plot the resummed (2,1) (with $A^{(1,2)},B^{(1,2)}$ and 
$C^{(0,1)}$ included) and the fixed order ${\cal O}(\alpha _S^2)$ 
distributions~\cite{Arnold-Reno}.\footnote{%
The $K$\ factor, which is defined to be the ratio of the cross sections at 
${\cal O}(\alpha _S^2)$ and ${\cal O}(\alpha _S)$, as the function of $Q_T$, 
changes within about 5\% in the plotted $Q_T$ region. 
For $W^+$: $K(Q_T=30$ GeV$)=1.48,~K(50\;$GeV$)=1.43,~K(70\;$GeV$)=1.39$; 
and for $Z^0$: $K(Q_T=30$ GeV$)=1.47,~K(50\;$GeV$)=1.43,~K(70\;$GeV$)=1.39$.} 
We match the two distributions at the value of $Q_T$, about 48 GeV, 
where they cross over for $W^+$ production, that defines the resummed 
${\cal O}(\alpha_S^2)$ distribution in the whole $Q_T$ region. 
The total cross section predicted from various theory calculations are 
listed in Table \ref{tbl:Total}. 
% TTTTTTTTTTTTTTTTTTTTTTTTTTTTTTTTTTTTTTTTTTTTTTTTTTTTTTTTTTTTTTTTTTTTTTT
\begin{table}[tbp]
\begin{center}
\begin{tabular}{l|c|cc|c|cc|r}
$V$ & $E_{cm}$ & \multicolumn{2}{c|}{Fixed Order} 
& CSS (1,1) + Y & \multicolumn{2}{c|}{CSS (2,1) + Y}
& Experiment \\ 
& (TeV) & ${\cal O}(\alpha _S^0)$ & ${{\cal O} (\alpha _S)}$ & $\oplus$
Pert. ${\cal O}(\alpha _S )$ & $\oplus$ Pert. ${\cal O}(\alpha _S )$ & $%
\oplus$ Pert. ${\cal O}(\alpha _S^2)$ & (Ref.~\cite{CDFTotal}) \\ \hline
W$^{+}$ & 1.8 & 8.81 & 11.1 & 11.3 & 11.3 & 11.4 & 11.5 $\pm $ 0.7 \\ 
W$^{+}$ & 2.0 & 9.71 & 12.5 & 12.6 & 12.6 & 12.7 &  \\ 
Z$^0$ & 1.8 & 5.23 & 6.69 & 6.79 & 6.79 & 6.82 & 6.86 $\pm $ 0.36 \\ 
Z$^0$ & 2.0 & 6.11 & 7.47 & 7.52 & 7.52 & 7.57 & 
\end{tabular}
\end{center}
\par
\caption{Total cross sections of $p {\bar p} \rightarrow (W^+ {\rm or}~ Z^0)
X$ at the present and upgraded Tevatron, calculated in different
prescriptions, in units of nb. 
The finite order total cross section results are based on the calculations 
in Ref.~[27]. 
%\cite{AEM85}.
The Pert. ${\cal O}(\alpha _S^2)$ results were obtained from Ref.~[28].
The ``$+$'' signs refer to the inclusion of the $Y$ piece and the ``$\oplus$''
signs to the switch from the resummed to the fixed order calculations.}
\label{tbl:Total}
\end{table}
% ttttttttttttttttttttttttttttttttttttttttttttttttttttttttttttttttttttttt

We note that kinematic cuts affect the total cross section in a subtle manner. 
It is obvious from our matching prescription that the resummed 
${\cal O}(\alpha _S^2)$ and the fixed order ${\cal O}(\alpha _S)$ curves in 
Fig.~\ref{fig:Matching} will never cross. On the
other hand, the resummed ${\cal O}(\alpha _S^2)$ total cross section 
is about the same as the fixed order ${\cal O}(\alpha _S)$ cross
section when integrating $Q_T$ from 0 to $Q$. These two facts imply that
when kinematic cuts are made on the $Q_T$ distribution with $Q_T < Q$, we will 
obtain a higher total cross section in the fixed order ${\cal O}(\alpha _S)$ 
than in the resummed ${\cal O}(\alpha _S^2)$ calculation. 
In this paper we follow the CDF cuts (for the $W^+$ boson mass analysis)
and demand $Q_T < 30$ GeV~\cite{CDFMW}. 
Consequently, in many of our figures, to be shown
below, the fixed order ${\cal O}(\alpha_S)$ curves give about 3\% higher
total cross section than the resummed ones.

% sssssssssssssssssssssssssssssssssssssssssssssssssssssssssssssssssssssss

\subsection{Lepton Charge Asymmetry}
\label{subsec:LCA}

The CDF lepton charge asymmetry measurement~\cite{CDFLCA} played a crucial
role in constraining the slope of the $u/d$ ratio in recent parton
distribution functions. It was shown that one of the largest theoretical
uncertainty in the $W^\pm$ mass measurement comes from the parton 
distributions~\cite{TEV2000}, and the lepton charge asymmetry was shown 
to be correlated
with the transverse mass distribution~\cite{Stirling-Martin}. Among others,
the lepton charge asymmetry is studied to decrease the errors in the
measurement of $M_W$ coming from the parton distributions. 
Here we investigate the effect of the resummation on the lepton
rapidity distribution, although it is not one of those observables which are
most sensitive to the $Q_T$ resummation, i.e. to the effect of multiple soft
gluon radiation.

\begin{figure*}[t]
\begin{center}
\begin{tabular}{cc}
\ifx\nopictures Y \else{ 
\epsfysize=6.0cm 
\epsffile{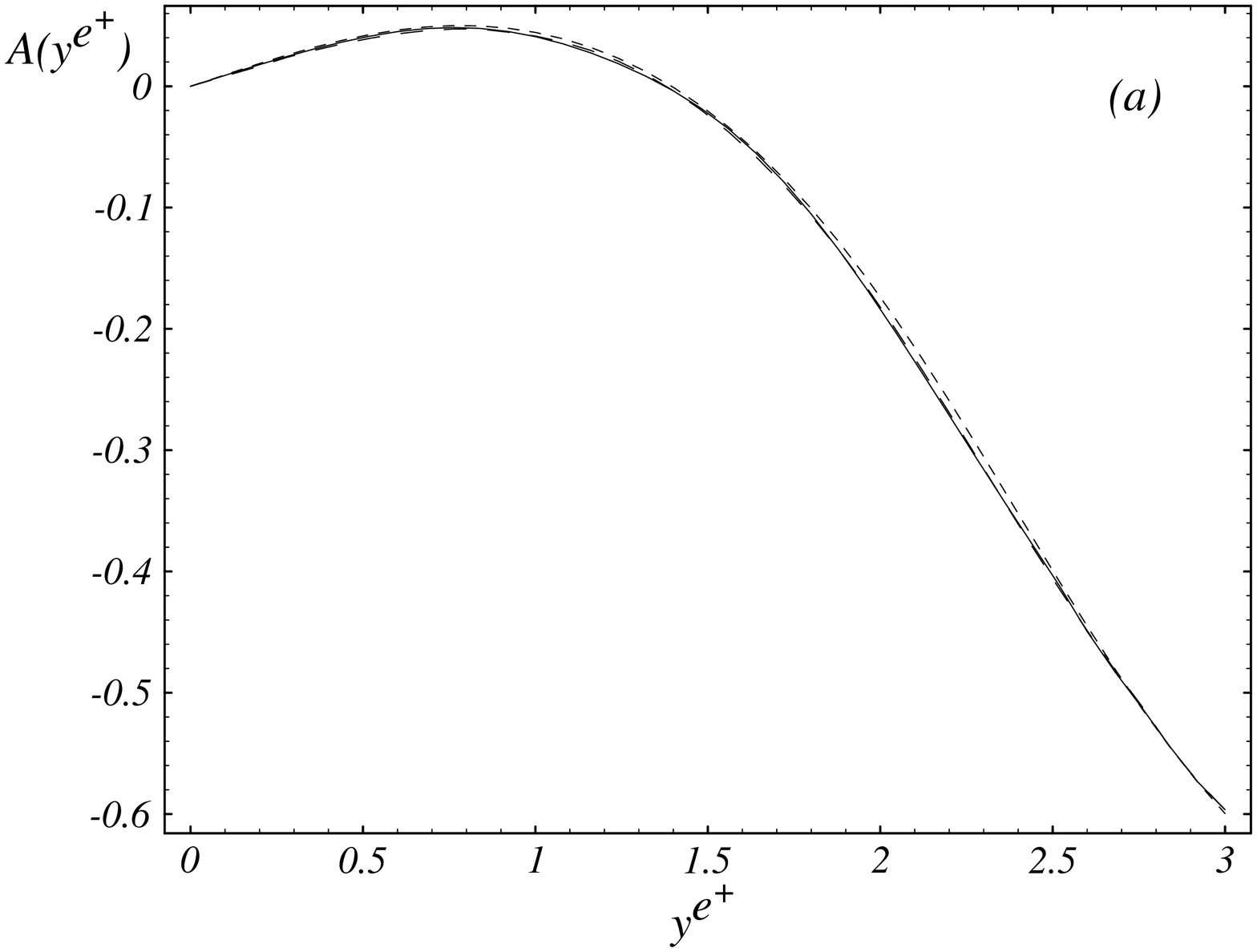} } 
\fi &  \\ 
\ifx\nopictures Y 
\else{ \epsfysize=6.0cm 
\epsffile{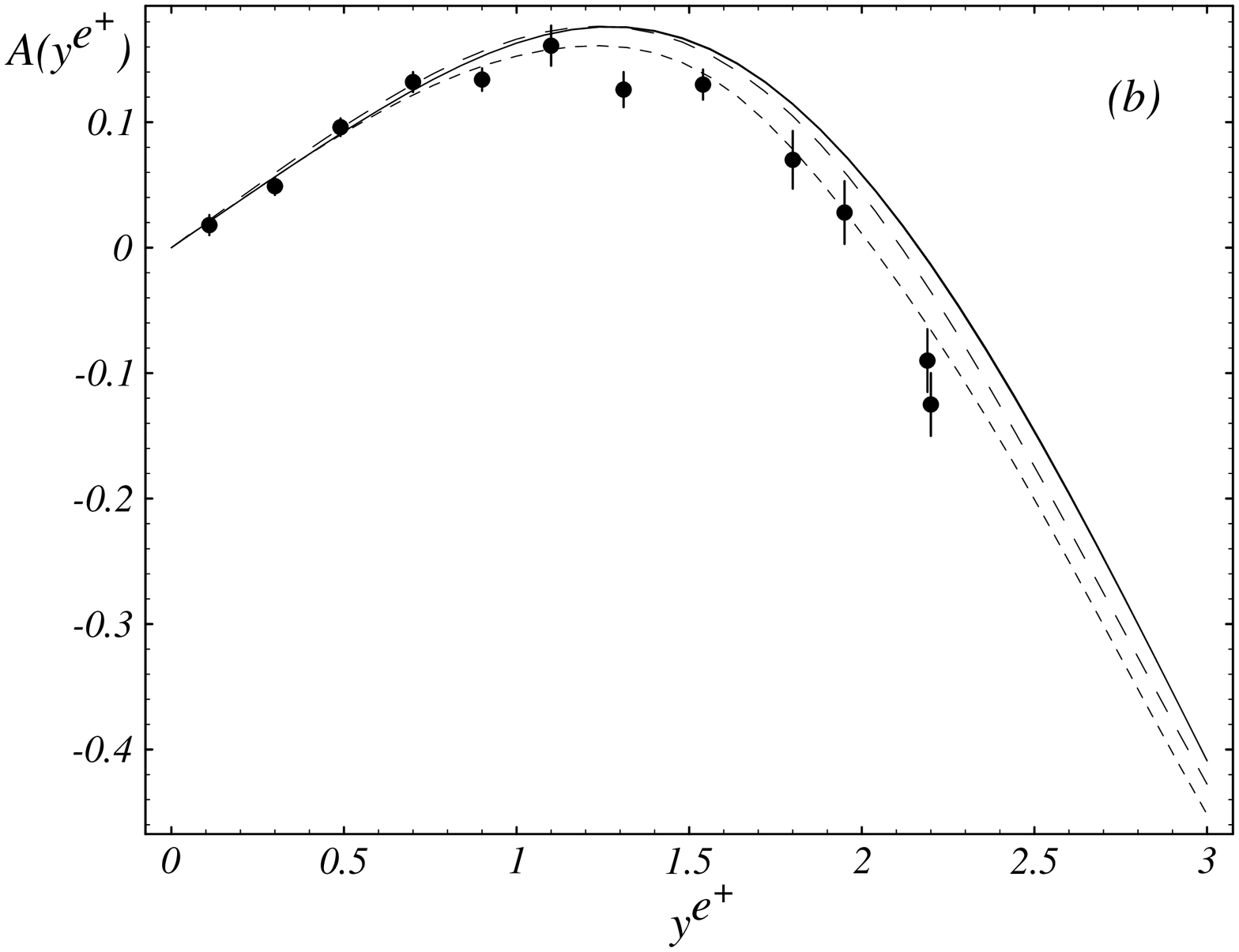} } 
\fi & 
\end{tabular}
\end{center}
\caption{Lepton charge asymmetry distributions. 
(a) Without any kinematic cuts, the NLO (long dashed) and the 
resummed ${\cal O}(\alpha_S)$ (solid) curves overlap and the LO (short dashed)
curve differs somewhat from them. 
(b) With cuts ($Q_T<30~{\rm GeV}, p_T^{e^+,\nu}>25~{\rm GeV}$),
the effect of the different $Q_T$ distributions renders the lepton rapidity 
asymmetry distributions different. The two resummed curves calculated with 
$g_2$ = 0.58 and 0.78 GeV$^{-2}$ cannot be distinguished on this plot.}
\label{fig:W_y_e_Asy}
\end{figure*}

The definition of the charge asymmetry is 
% EEEEEEEEEEEEEEEEEEEEEEEEEEEEEEEEEEEEEEEEEEEEEEEEEEEEEEEEEEEEEEEEEEEEEE
\begin{eqnarray*}
A(y)=\frac{\displaystyle \frac{d\sigma}{dy_+}-\frac{d\sigma}{dy_-}}
          {\displaystyle \frac{d\sigma}{dy_+}+\frac{d\sigma}{dy_-}},
\end{eqnarray*}
% eeeeeeeeeeeeeeeeeeeeeeeeeeeeeeeeeeeeeeeeeeeeeeeeeeeeeeeeeeeeeeeeeeeeee
where $y_{+}$ ($y_{-}$) is the rapidity of the positively (negatively)
charged particle (either vector boson or decay lepton). 
Assuming CP invariance,\footnote{Here we ignore the small CP violating effect 
due to the CKM matrix elements in the SM.} the following relation holds:
% EEEEEEEEEEEEEEEEEEEEEEEEEEEEEEEEEEEEEEEEEEEEEEEEEEEEEEEEEEEEEEEEEEEEEE
\begin{eqnarray*}
{\frac{d\sigma }{dy_{+}}}(y)={\frac{d\sigma }{dy_{-}}}(-y).
\end{eqnarray*}
% eeeeeeeeeeeeeeeeeeeeeeeeeeeeeeeeeeeeeeeeeeeeeeeeeeeeeeeeeeeeeeeeeeeeee
Hence, the charge asymmetry is frequently written as 
% EEEEEEEEEEEEEEEEEEEEEEEEEEEEEEEEEEEEEEEEEEEEEEEEEEEEEEEEEEEEEEEEEEEEEE
\begin{eqnarray*}
A(y)=\frac{\displaystyle \frac{d\sigma}{dy}(y>0)-\frac{d\sigma}{dy}(y<0)}
          {\displaystyle \frac{d\sigma}{dy}(y>0)+\frac{d\sigma}{dy}(y<0)}.
\end{eqnarray*}
% eeeeeeeeeeeeeeeeeeeeeeeeeeeeeeeeeeeeeeeeeeeeeeeeeeeeeeeeeeeeeeeeeeeeee
For the charge asymmetry 
of the vector boson ($W^{\pm }$) or the charged decay lepton ($\ell ^{\pm }$), 
the fixed order and the resummed ${\cal O}(\alpha _S)$ 
[or ${\cal O}(\alpha _S^2)$]
results are the same, provided that there are no kinematic cuts imposed.
This is because the shape difference in the vector boson transverse
momentum has been integrated out and the total cross sections are the
same up to higher order corrections in $(\alpha _S)$.
In Fig.~\ref{fig:W_y_e_Asy}(a) we show the lepton charge asymmetry without
cuts for CTEQ4M PDF. The NLO and the resummed curves overlap,
although they differ from the ${\cal O}(\alpha _S^0)$ prediction.
% FFFFFFFFFFFFFFFFFFFFFFFFFFFFFFFFFFFFFFFFFFFFFFFFFFFFFFFFFFFFFFFFFFFFFFF
% ffffffffffffffffffffffffffffffffffffffffffffffffffffffffffffffffffffff

On the other hand, when kinematic cuts are applied to the decay leptons,
the rapidity distributions of the vector bosons or the leptons
in the fixed order and the resummed calculations are different.
Restriction of the phase space implies
that only part of the vector boson transverse momentum distribution is
sampled. The difference in the resummed and the fixed order $Q_T$
distributions will prevail as a difference in the rapidity distributions of
the charged lepton. We can view this phenomenon in a different (a Monte
Carlo) way. In the rest frame of the $W^\pm$, the decay kinematics is the same,
whether it is calculated up to ${\cal O}(\alpha _S)$ or within the
resummation formalism. On the other hand, the $W^\pm$ rest frame is different
for each individual Monte Carlo (MC) event depending on the order of the 
calculation. 
This difference is caused by the fact that the $Q_T$ distribution of the 
$W^\pm$ is different in the ${\cal O}(\alpha _S)$ and the resummed 
calculations, and the kinematic cuts select different events in these two 
cases. 
Hence, even though the $Q_T$ distribution of the $W^\pm$ is integrated out, 
when calculating the lepton rapidity distribution,
we obtain slightly different predictions in the two calculations. 
% >>> ??? Why deviate at large y ???
The difference is larger for larger $|y^{\ell}|$, being
closer to the edge of the phase space, because the
soft gluon radiation gives high corrections there and this effect up to all
order in $\alpha _S$ is contained in the resummed but only up to order of 
$\alpha _S$ in the NLO calculation. 
Because the rapidity of the lepton and that of the vector boson are highly 
correlated, large rapidity leptons
mostly come from large rapidity vector bosons. 
Also, a vector boson with large rapidity tends to have low transverse momentum,
because the available phase space is limited to low $Q_T$ for a $W^\pm$ boson 
with large $|y|$. Hence, the difference in the low $Q_T$ distributions 
of the NLO and the resummed calculations yields the difference in the 
$y^{\ell}$ distribution for leptons with high rapidities.

Asymmetry distributions of the charged lepton with cuts using the CTEQ4M
PDF are shown in Fig.~\ref{fig:W_y_e_Asy}(b). 
The applied kinematic cuts are: 
$Q_T<30$~GeV, $p_T^{e^+,\nu}>25$. 
These are the cuts that CDF used when extracted the lepton rapidity 
distribution from their data~\cite{CDFLCA}.
We have checked that the ResBos fixed order ${\cal O}(\alpha _S)$ curve 
agrees well with the DYRAD~\cite{Giele} result. 
As anticipated, the ${\cal O}(\alpha _S^0)$, ${\cal O}(\alpha _S)$ and
resummed results deviate at higher rapidities ($|y^e|>1.5$).\footnote{Here
and henceforth, unless specified otherwise, by a resummed calculation we 
mean our resummed ${\cal O}(\alpha _S^2)$ result.}
The deviation between the NLO and the resummed curves indicates that to 
extract information on the PDF in the large rapidity region, the resummed 
calculation, in principle, has to be used if the precision of the data is 
high enough to distinguish these predictions. 
Fig.~\ref{fig:W_y_e_Asy}(b) also shows the negligible dependence
of the resummed curves on the non-perturbative parameter $g_2$. We plot the 
result of the resummed calculations with the nominal $g_2$ = 0.58 GeV$^{-2}$,
and with $g_2$ = 0.78 GeV$^{-2}$ which is two standard deviations higher. 
The deviation between these two curves (which is hardly observable on the 
figure) is much smaller than the deviation between the resummed and the NLO 
ones.

There is yet another reason why the lepton charge asymmetry 
can be reliably predicted only by the resummed calculation.
When calculating the lepton
distributions in a numerical ${\cal O}(\alpha _S)$ code, one has to
artificially divide the vector boson phase space into hard and soft regions,
depending on -- for example -- the energy or the $Q_T$ of the emitted gluon
(e.g. $q\overline{q} \rightarrow W+\;$hard or soft gluon). The observables
calculated with this phase space slicing technique acquire a dependence on
the scale which separates the hard from the soft regions. To emphasize this
dependence as an example, we show that when the phase space is divided by the $
Q_T$ separation, the dependence of the asymmetry on the scale $Q_T^{Sep}$
can be comparable to the difference in the ${\cal O}(\alpha _S)$ and the
resummed results. This means that there is no definite prediction 
from the NLO calculation for the lepton rapidity distribution.
Only the resummed calculation can give an unambiguous prediction for the 
lepton charge asymmetry.

\begin{figure*}[t]
\begin{center}
\begin{tabular}{cc}
\ifx\nopictures Y \else{ 
\epsfysize=6.0cm 
\epsffile{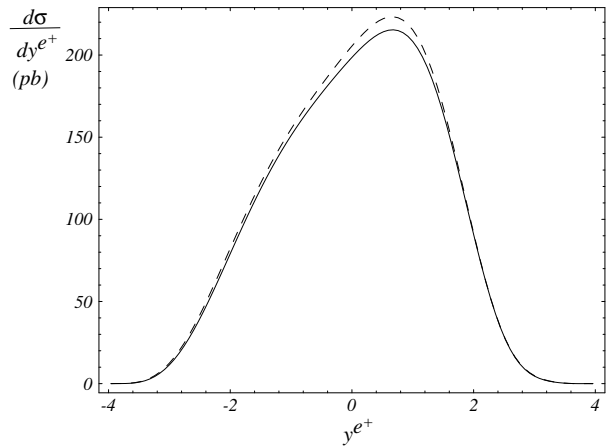} } 
\fi & 
\end{tabular}
\end{center}
\caption{Distributions of positron rapidities from the decays of $W^+$'s 
produced at the Tevatron, predicted by the resummed (solid) and the NLO
(dashed) calculations with the same kinematic cuts as for the asymmetry plot.}
\label{fig:W_y_e}
\end{figure*}

Before closing this section, we also note that although in the lepton 
asymmetry distribution the NLO and resummed results are about the same for 
$|y^{e^+}| < 1$, it does not imply that the rapidity distributions of the 
leptons predicted by those two theory models are the same. As shown in 
Fig.~\ref{fig:W_y_e}, this difference can in principle be observable with a 
large statistics data sample and a good knowledge of the luminosity of the 
colliding beams.
% FFFFFFFFFFFFFFFFFFFFFFFFFFFFFFFFFFFFFFFFFFFFFFFFFFFFFFFFFFFFFFFFFFFFFFF
% ffffffffffffffffffffffffffffffffffffffffffffffffffffffffffffffffffffff

% sssssssssssssssssssssssssssssssssssssssssssssssssssssssssssssssssssssss

\subsection{Transverse Mass Distribution}
%\label{subsec:TM}

Since the invariant mass of the $W^{\pm }$ boson cannot be reconstructed 
without knowing the longitudinal momentum of the neutrino, one has to find a
quantity that allows an indirect determination of the mass of the $W^\pm$ 
boson.
In the discovery stage of the $W^{\pm }$ bosons at the CERN\ ${\rm S p 
\overline{p} S}$ collider, the mass and width were measured
using the transverse mass distribution of the charged lepton-neutrino pair 
from the $W^\pm$ boson decay. Ever since the early eighties, the transverse 
mass distribution, $m_T = \sqrt{2p_T^{e}p_T^\nu (1-\cos \Delta \phi_{e
\nu })}$, has been known as the best measurable for the extraction of both 
$M_W$ and $\Gamma _W$, for it is insensitive to the transverse momentum of the 
$W^\pm$ boson. 
The effect of the non-vanishing vector boson transverse momentum on
the $m_T$ distribution was analyzed ~\cite{Barger,Smith} 
well before the $Q_T$ distribution of the $W^\pm$ boson was correctly 
calculated
by taking into account the multiple soft gluon radiation. Giving an average
transverse boost to the vector boson, the authors of Ref.~\cite{Barger}
concluded that for the fictive case of $\Gamma _W$ = 0, the end points of the
transverse mass distribution are fixed at zero: $d\sigma
/dm_T^2(m_T^2=0)=d\sigma /dm_T^2(m_T^2=M_W^2)=0$. The sensitivity of the
$m_T$ shape to a non-zero $Q_T$ is in the order of $\left\langle
(Q_T/M_W)^2\right\rangle \approx $1\% without affecting the end points
of the $m_T$ distribution. 
%cpy
%This is contrasted with the fact that the $p_T^{\ell}$ distribution
%sensitivity is an order of magnitude larger $\left\langle
%Q_T/M_W\right\rangle \approx $10\% both in shape and the location of the
%upper end point. 
Including the effect of the finite width of the $W^\pm$ boson, the authors in
Ref.~\cite{Smith} showed that the shape and the location of the Jacobian
peak are not sensitive to the $Q_T$ of the $W^\pm$ boson either.
The non-vanishing
transverse momentum of the $W^{\pm }$ boson only significantly modifies the
$m_T$ distribution around $m_T=0$. 
%Among the conlusions of the above paper one of the most important is 
%that the shape of the $m_T$ distribution around the Jacobian 
%peak is indeed very sensitive
%to the width of the $W^{\pm }$.

\begin{figure*}[t]
\begin{center}
\begin{tabular}{cc}
\ifx\nopictures Y \else{ 
\epsfysize=6.0cm 
\epsffile{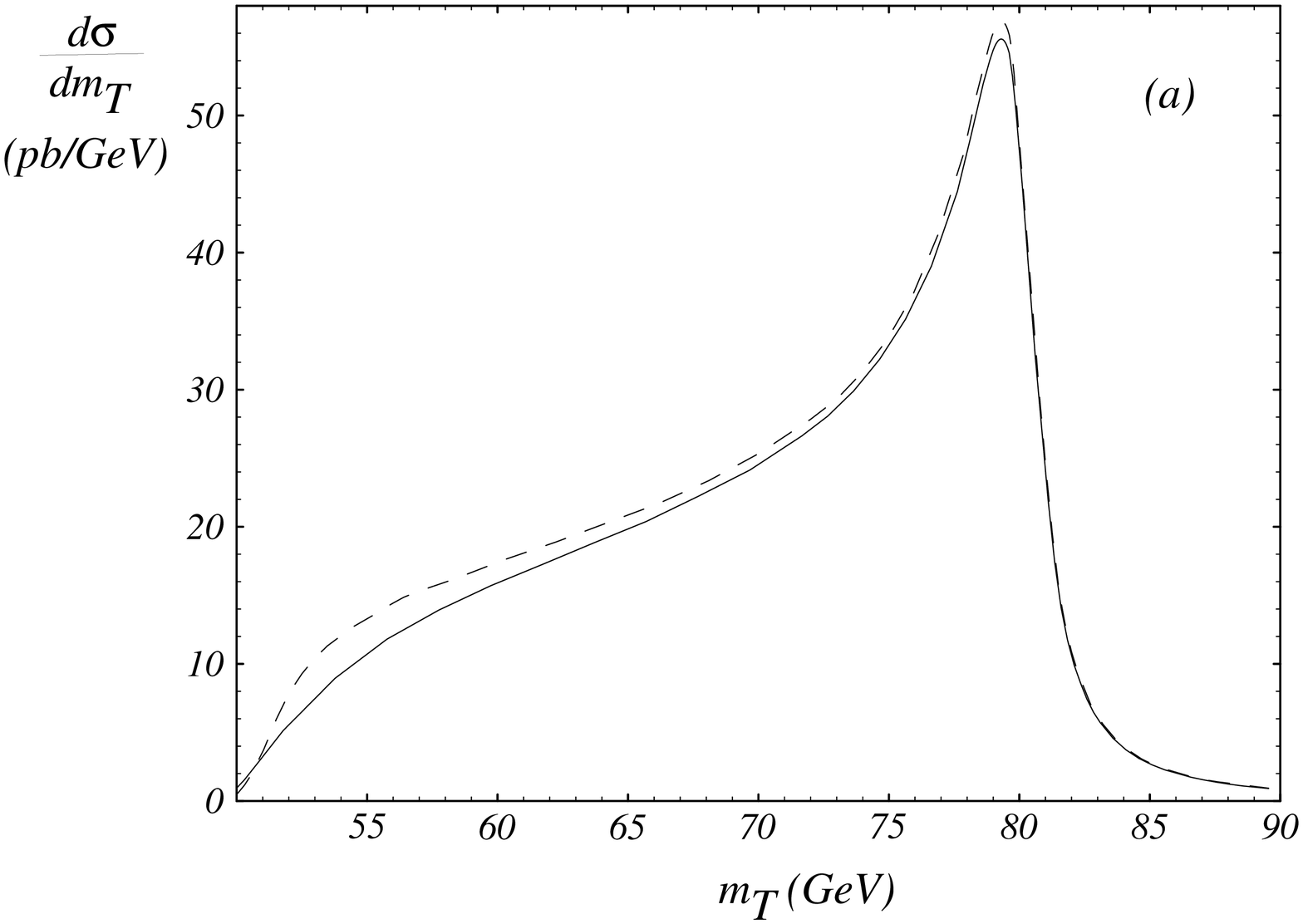} } 
\fi &  \\ 
\ifx\nopictures Y 
\else{ \epsfysize=6.0cm 
\epsffile{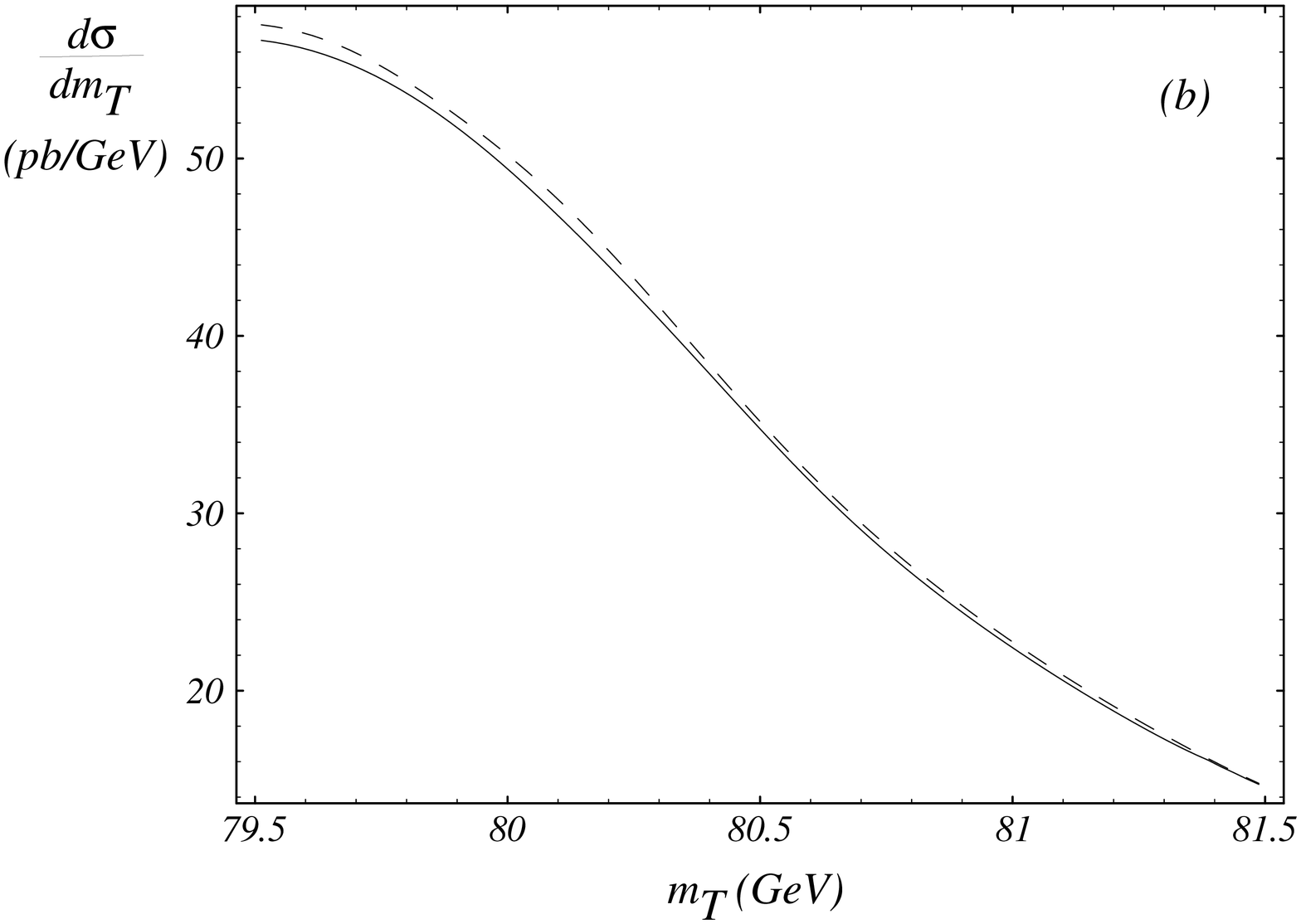} } 
\fi & 
\end{tabular}
\end{center}
\caption{
Transverse mass distribution for $W^{+}$ production and decay at
the 1.8 TeV Tevatron.
}
\label{fig:W_mt}
\end{figure*}

Our results confirm that the shape of the Jacobian peak is quite insensitive 
to the order of the calculation. 
We show the NLO and the resummed transverse mass distributions in 
Fig.~\ref{fig:W_mt} for $W^\pm$ bosons produced 
at the Tevatron with the kinematic cuts: 
$Q_T < 30$ GeV, $p_T^{e^+,\nu} > 25$ GeV and $|y^{e^+}|<3.0$.
% and $|y^{\nu}|<3.5$.
Fig.~\ref{fig:W_mt}(a) covers the full (experimentally interesting) $m_T$ 
range while Fig.~\ref{fig:W_mt}(b) focuses on the $m_T$ range which 
contains most of the information about the $W^\pm$ mass. 
There is little visible difference between the $shapes$ of the NLO 
and the resummed $m_T$ distributions. 
On the other hand, the right shoulder of the curve appears to be ``shifted'' 
by about 50 MeV, 
because, as noted in Section~\ref{subsec:Total}, the total cross sections 
are different after the above cuts imposed in the NLO and the resummed 
calculations. 
At Run 2 of the Tevatron, with large integrated luminosity 
($\sim 2~{\rm fb}^{-1}$),
the goal is to extract the $W^\pm$ boson mass with a precision of 30-50 MeV 
from the $m_T$ distribution \cite{TEV2000}. 
%If $M_W$ is to be measured with this precision 
%the transverse momentum distribution must be known very accurately. 
Since $M_W$ is sensitive to the position of the Jacobian peak \cite{Smith},
the high precision measurement of the $W^\pm$ mass has to rely on the resummed 
calculations.
% FFFFFFFFFFFFFFFFFFFFFFFFFFFFFFFFFFFFFFFFFFFFFFFFFFFFFFFFFFFFFFFFFFFFFFF
% ffffffffffffffffffffffffffffffffffffffffffffffffffffffffffffffffffffff

% --- Limitations
Extraction of $M_W$ from the transverse mass distribution has some drawbacks
though. The reconstruction of the transverse momentum $p_T^\nu $ of the
neutrino involves the measurement of the underlying event transverse
momenta: $\vec{p}_T^{~\nu }=-\vec{p}_T^{~\ell }-\vec{p}_T^{~recoil}-\vec{p}%
_T^{~underlying~event}$. This resolution degrades by the number of
interactions per crossing ($N_{I_c}$)~\cite{TEV2000}. 
%The $p_T^\nu$ (and so $m_T$) resolution degreeds 
%with increasing luminosity due
%to multiple interactions ( uncertainty~$\Sqrt{N_{ia}}$) <- Tev2000.
With a high luminosity ($\sim 100\,{\rm fb}^{-1}$) at the 2 TeV Tevatron 
(TEV33)~\cite{TEV33}, $N_{I_c}$ can be as large as 10, so that the Jacobian 
peak is badly smeared. 
%Since $\langle m_T \rangle$ is inversly correlated with the error of the
%extracted mass $\Delta M_W$ this error will be large.
This will lead to a large uncertainty in the measurement of $M_W$. For this
reason the systematic precision of the $m_T$ reconstruction will be less at
the high luminosity Tevatron, and an $M_W$ measurement that relies on the
lepton transverse momentum distribution alone could be more promising. 
We discuss this further in the next section. 
%The ratio of $m_T^{W^\pm}/m_T^{Z^0}$
%may not suffer so much from the above effect. At the high luminosity
%Tevatron the shape comparison, based on a statistical methods (like the
%Kolmogorov-Smirnov method), may utilize the $m_T$ distributions in the
%extraction of $M_W$.

The theoretical limitation on the $M_W$ measurement using the $m_T$
distribution comes from the dependence on the
non-perturbative sector, i.e. from the PDF's and the non-perturbative 
parameters in the resummed formalism.
Assuming the PDF's and these
non-perturbative parameters to be independent variables, the
uncertainties introduced are estimated to be less than 50 MeV and 10
MeV, respectively, at the TEV33 \cite{Flattum,SM96EW}. It is clear 
that the main theoretical uncertainty comes from the PDF's. As
to the uncertainty due to the non-perturbative parameters (e.g. $g_2$)
in the CSS resummation formalism, it can be greatly reduced by carefully
study the $Q_T$ distribution of the $Z^0$ boson which is expected to
be copiously produced at Run 2 and beyond.

% --- $M_W$ @ LHC:
The $M_W$ measurement at the LHC may also be promising. Both ATLAS and CMS
detectors are well optimized for measuring the leptons and the missing 
$E_T$~\cite{SM96EW}. 
The cross section of the $W^+$ boson production is about four times larger 
than that at the Tevatron, and
in one year of running with 20 fb$^{-1}$ luminosity yields a few
times $10^7$ $W\rightarrow \ell \nu $ events after imposing similar cuts to
those made at the Tevatron. 
%m_T shape diff betw LO & NLO ~ 3\% (10\% @ Tev)
Since the number of interactions per crossing may be significantly lower 
(in average $N_{I_c}$ = 2) 
at the same or higher luminosity than that at the TEV33~\cite{SM96EW},
the Jacobian peak in the $m_T$ distribution 
will be less smeared at the LHC than at the TEV33. 
%m_T is more peaked =>
%  since m_T is inversly correlated with \Delta M_W -> \Delta M_W is smaller at
%  LHC than at Tev
Furthermore, the non-perturbative effects are relatively smaller 
at the LHC because the perturbative Sudakov factor dominates. 
On the other hand, the probed region
of the PDF's at the LHC has a lower value of the average $x~(\sim 10^{-3})$
than that at the Tevatron $(\sim 10^{-2})$, hence the uncertainty from the
PDF's might be somewhat larger. 
A more detailed study of this subject is desirable.

\subsection{Lepton Transverse Momentum}

%The measurement of $M_W$ has a crucial importance in testing the symmetry
%breaking sector of the Standard Model~\cite{Rosner}
%LEP will measure the W mass with about th esame precision then TEV33
%it is desirable to have as many independent measurement as possible.  
Due to the limitations mentioned above, the transverse mass ($m_T$) method
may not be the only and the most promising way for the precision
measurement of $M_W$ at some future hadron colliders. 
As discussed above, the observable $m_T$ was used because of its insensitivity 
to the high order QCD corrections. 
In contrast, the lepton transverse momentum ($p_T^\ell$) distribution receives 
a large, ${\cal O}\left( \left\langle Q_T/M_W\right\rangle \right) \sim 10\%$%,
perturbative QCD correction at the order $\alpha _S$, as compared to the
Born process. With the resummed results in hand it
becomes possible to calculate the $p_T^\ell $ distribution precisely within
the perturbative framework, and to extract the $W^\pm$ mass 
straightly from the transverse momentum distributions of the decay leptons.

Just like in the $m_T$ distribution, the mass of the $W^\pm$ boson is mainly 
determined by the shape of the distribution near the Jacobian peak. 
The location of the maximum of the peak
is directly related to the $W^\pm$ boson mass, while the theoretical width of
the peak varies with its decay width $\Gamma _W$. Since the Jacobian peak is
modified by effects of both $Q_T$ and $\Gamma _W$, it is important to take
into account both of these effects correctly. In our calculation (and in
ResBos) we have properly included both effects. 
%Actual details of the $W^\pm$ mass extraction from the $p_T^{\ell}$ 
%distribution require further analysis.

\begin{figure*}[t]
\begin{center}
\begin{tabular}{cc}
\ifx\nopictures Y \else{ 
\epsfysize=6.0cm 
\epsffile{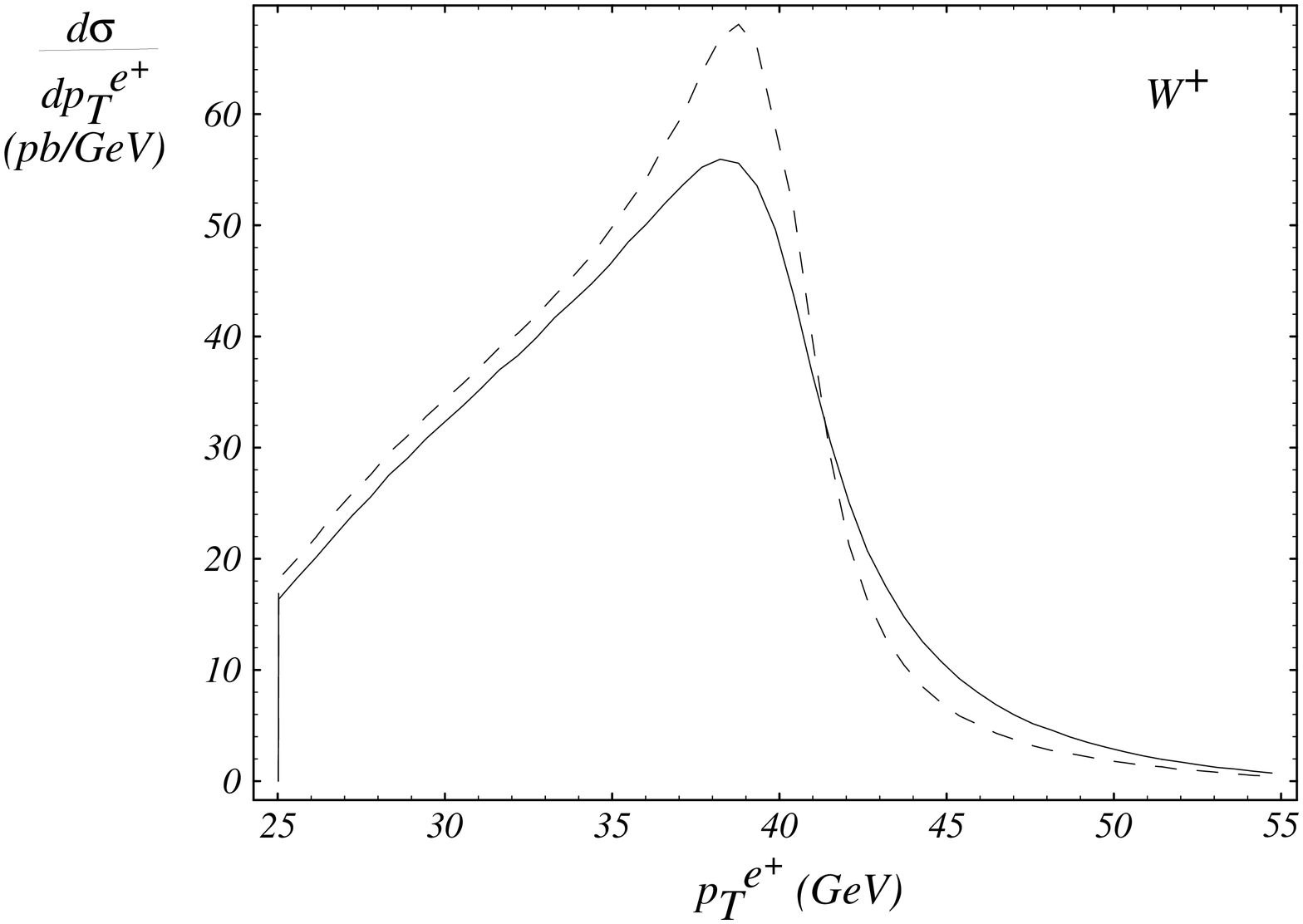} } 
\fi &  \\ 
\ifx\nopictures Y 
\else{ \epsfysize=6.0cm 
\epsffile{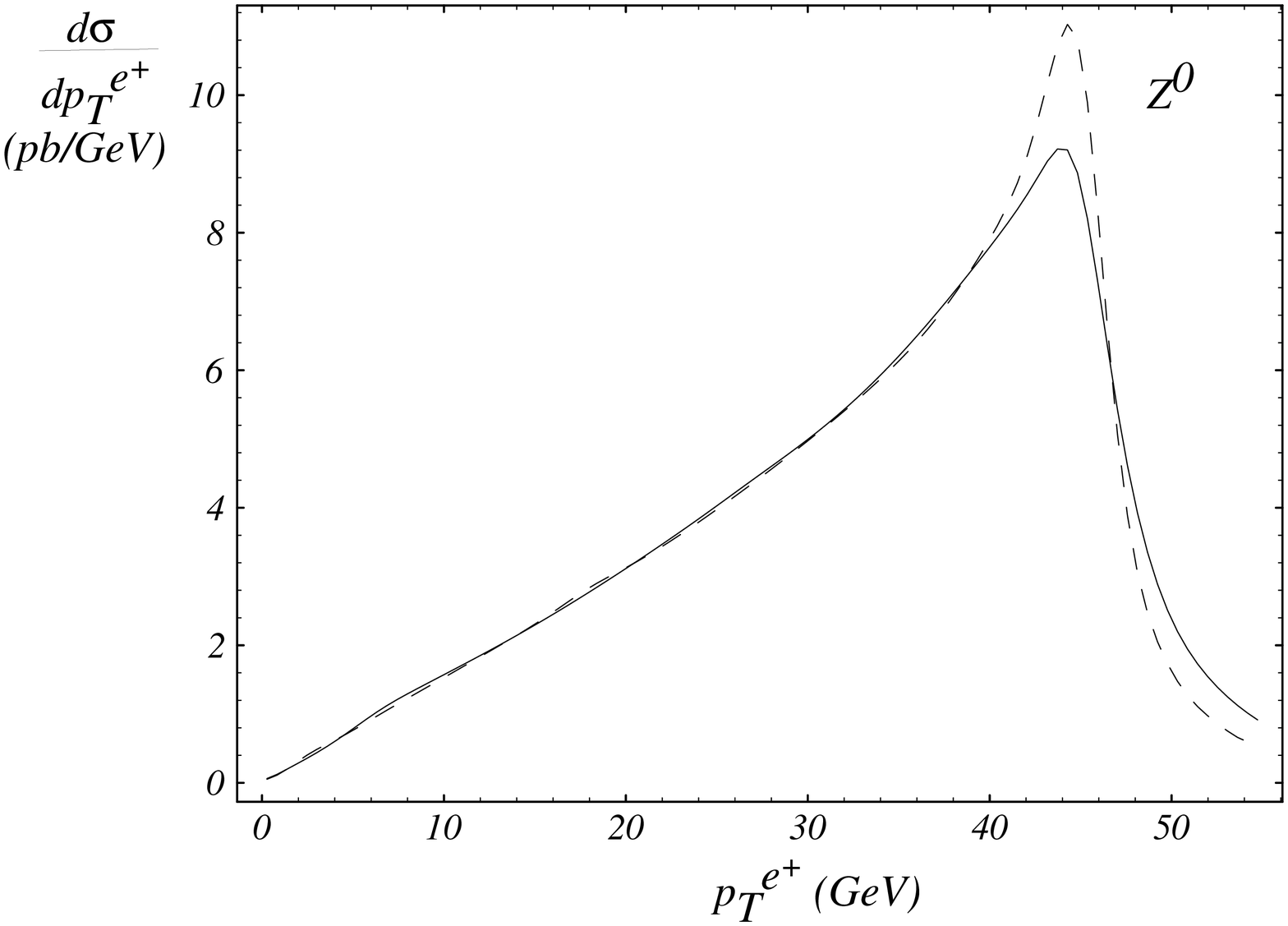} } 
\fi & 
\end{tabular}
\end{center}
\caption{Transverse momentum distributions of $p_T^{e^{+}}$ from $W^{+}$
and $Z^0$ decays for the NLO (dashed) and the resummed ${\cal O}(\alpha _S)$ 
(solid) calculations. 
Resumming the initial state multiple soft-gluon emission has 
the typical effect of smoothening and broadening the Jacobian peak 
(at $p_T^{e^{+}}=M_V/2$). 
%The NLO distribution is singular and ill-defined near $M_V/2$. 
%Preliminary experimental data are also shown from the Tevatron.
The CDF cuts are imposed on the $W^+$ distributions, 
but there are no cuts on the $Z^0$ distributions.}
\label{fig:V_pT_e}
\end{figure*}

The effect of resummation on the transverse momentum distribution of the 
charged lepton from $W^{+}$ and $Z^0$ decays is shown in Fig.~\ref{fig:V_pT_e}. 
The NLO and the resummed distributions differ a great amount 
even without imposing any kinematic cuts. 
The clear and sharp Jacobian peak of the NLO
distribution is strongly smeared by the finite transverse momentum of the 
vector boson introduced by multiple gluon radiation. This higher order effect 
cannot be correctly calculated in any finite order of the perturbation theory
and the resummation formalism has to be used.
% FFFFFFFFFFFFFFFFFFFFFFFFFFFFFFFFFFFFFFFFFFFFFFFFFFFFFFFFFFFFFFFFFFFFFFF
% ffffffffffffffffffffffffffffffffffffffffffffffffffffffffffffffffffffff

% --- MW from pT
%clean, do not measure underlying event.
%We examine the possibility of extracting $M_W$ from the $p_T^e$ distribution.
%The great advantage lies in the simplicity of this measurement. 
%$p_T^e$ does not depend on $p_T^\nu$. 
One of the advantages of using the $p_T^\ell $ distribution to determine $M_W
$ is that there is no need to reconstruct the $p_T^\nu $ distribution which
potentially limits the precision of the $m_T$ method. 
From the theoretical side, the limitation is in the knowledge of the
non-perturbative sector. 
%the PDF's and the non-perturbative parameters $g_i$. 
Studies at \D0 \cite{Adam} show that the $p_T^\ell $ distribution is most
sensitive to the PDF's and the value of the non-perturbative parameter $g_2$. 
The $p_T^\ell $ distribution is more sensitive to the PDF choice, 
than the $m_T$ distribution is. The uncertainty in the
PDF causes an uncertainty in $M_W$ of about 150 MeV, which is about three 
times as large as that using the $m_T$ method~\cite{Adam}. A 0.1
GeV$^2$ uncertainty in $g_2$ leads to about $\Delta M_W=30$ MeV uncertainty
from the $p_T^{\ell}$ fit, which is about five times worse than that 
from the $m_T$ measurement~\cite{Adam}. 
%While better global fit is not available 
%for the non-perturbative parameters $g_i$, 
Therefore, to improve the $M_W$ measurement, it is necessary to include the 
$Z^0$ data sample at the high luminosity Tevatron to refit the $g_i$'s and 
obtain a tighter constrain on them from the $Q_T$ distribution of the $Z^0$ 
boson.
The \D0 study showed that an accuracy of $\Delta g_2=0.01$ GeV$^2$ 
can be achieved with Run 2 and TeV33 data, %with $10^33$ cm$^{-2}$ luminosity, 
%100 times the statistics of Run 1B, 
which would contribute an error of $\Delta M_W<5$ MeV from the $p_T^\ell $ 
\cite{Adam}. In this case the uncertainty coming from the PDF's remains to be 
the major theoretical limitation. 
% --- @ LHC 
At the LHC, the $p_T^\ell$ distribution
can be predicted with an even smaller theoretical error coming from the 
non-perturbative part, because at higher
energies the perturbative Sudakov factor dominates over the non-perturbative
function.

% --- MW from pT Ratios
It was recently suggested to extract $M_W$ from the ratios of the transverse
momenta of leptons produced in $W^{\pm }$ and $Z^0$ decay~\cite{Giele-Keller}.
The theoretical advantage is that the non-perturbative uncertainties are
decreased in such a ratio. On the other hand, it is not enough that the
ratio of cross sections is calculated with small theoretical errors. For a
precision extraction of the $W^\pm$ mass the theoretical calculation must be
capable of reproducing the individually observed transverse momentum
distributions themselves. The $W^\pm$ mass measurement
requires a detailed event modeling, understanding of
detector resolution, kinematical acceptance and efficiency effects, 
which are different for the $W^{\pm }$ and $Z^0$ events, as illustrated
above. Therefore, the ratio of cross sections can only
provide a useful check for the $W^\pm$ mass measurement.

\begin{figure*}[t]
\begin{center}
\begin{tabular}{cc}
\ifx\nopictures Y \else{ 
\epsfysize=6.0cm 
\epsffile{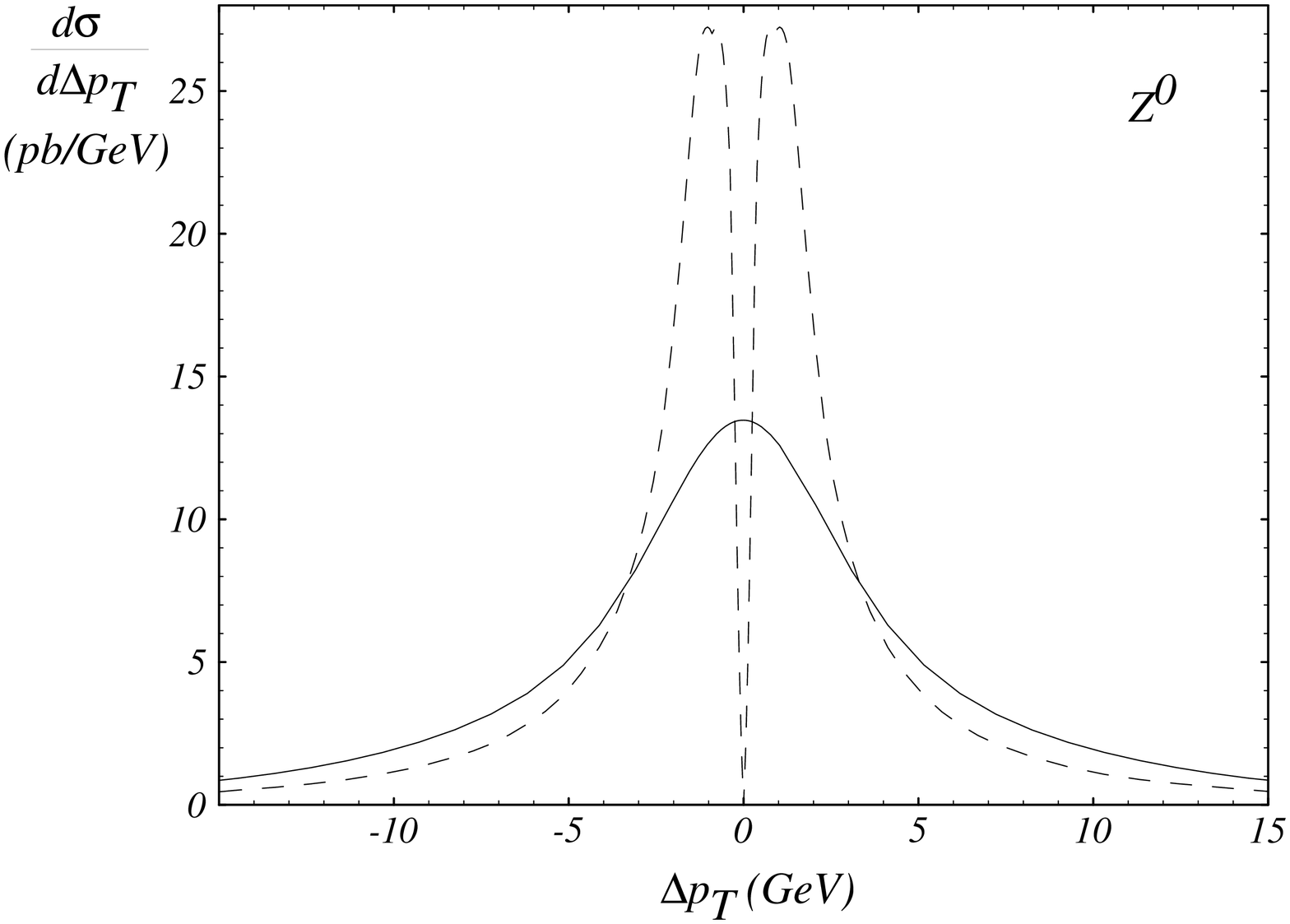} } 
\fi &  \\ 
\ifx\nopictures Y 
\else{ \epsfysize=6.0cm 
\epsffile{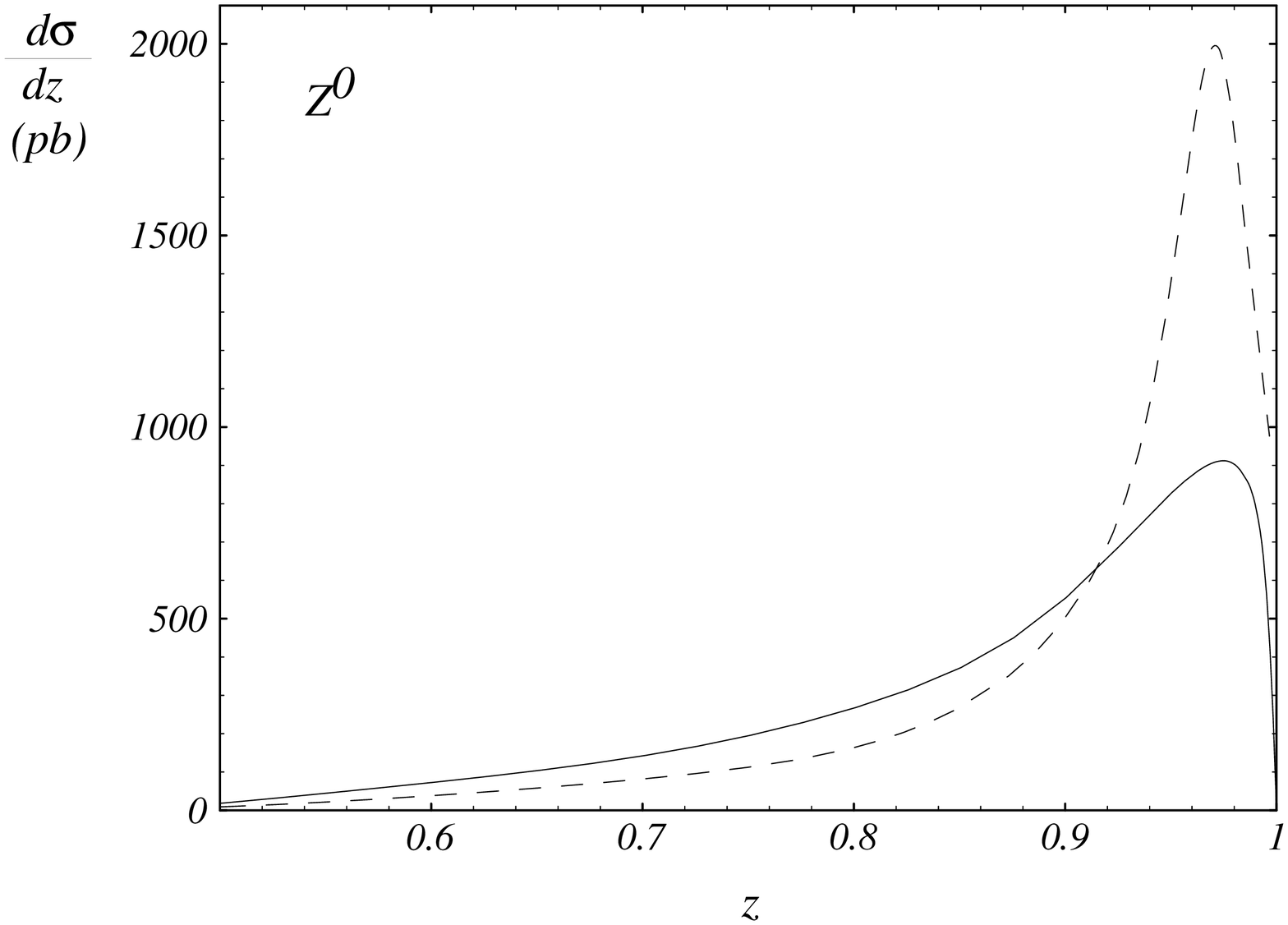} } 
\fi & 
\end{tabular}
\end{center}
\caption{Balance in transverse momentum $\Delta {p}_T=|{\vec p}_T^{\,\ell
_1}|-|{\vec p}_T^{\,{\bar \ell_2}}|$ and angular correlation ${z=-\vec p}
_T^{\,\ell _1}\cdot 
{\vec p}_T^{\,{\bar \ell_2}}/[\max 
(p_T^{\ell_1},p_T^{{\bar \ell_2}})]^2$ 
of the decay leptons from $Z^0$ bosons produced at the Tevatron.}
\label{fig:Z_pTCorr}
\end{figure*}

% --- pT^1 - pT^2 
For Drell-Yan events or lepton pairs from $Z^0$ decays, additional
measurable quantities can be constructed from the lepton transverse
momenta. They are the distributions in the balance of the transverse momenta 
$(\Delta {p}_T=|{\vec{p}}_T^{\,\ell _1}|-|{\vec{p}}_T^{\,{\bar \ell_2}}|)$ 
and the angular correlation of the two lepton momenta 
$({z=-\vec{p}}_T^{\,\ell_1}\cdot 
{\vec{p}}_T^{\,{\bar \ell_2}}/[\max 
(p_T^{\,\ell _1},p_T^{\,{\bar \ell_2}})]^2)$. 
It is expected that these quantities are also sensitive to the effects of
the multiple soft gluon radiation. These distributions are shown in 
Figure~\ref{fig:Z_pTCorr}. 
As shown, the resummed distributions significantly differ from the NLO ones. 
In these, and the following figures for $Z^0$ decay distributions, it is 
understood that the following kinematic cuts are imposed: 
$Q_T^{Z^0}<30$~GeV, $p_T^{e^{+},e^{-}}>25$ GeV and $|y^{e^{+},e^{-}}|<3.0$, 
unless indicated otherwise. 
% FFFFFFFFFFFFFFFFFFFFFFFFFFFFFFFFFFFFFFFFFFFFFFFFFFFFFFFFFFFFFFFFFFFFFFF
% ffffffffffffffffffffffffffffffffffffffffffffffffffffffffffffffffffffff

% sssssssssssssssssssssssssssssssssssssssssssssssssssssssssssssssssssss

\subsection{Lepton Angular Correlations}

\begin{figure*}[t]
\begin{center}
\begin{tabular}{cc}
\ifx\nopictures Y \else{ \epsfysize=6.0cm
\epsffile{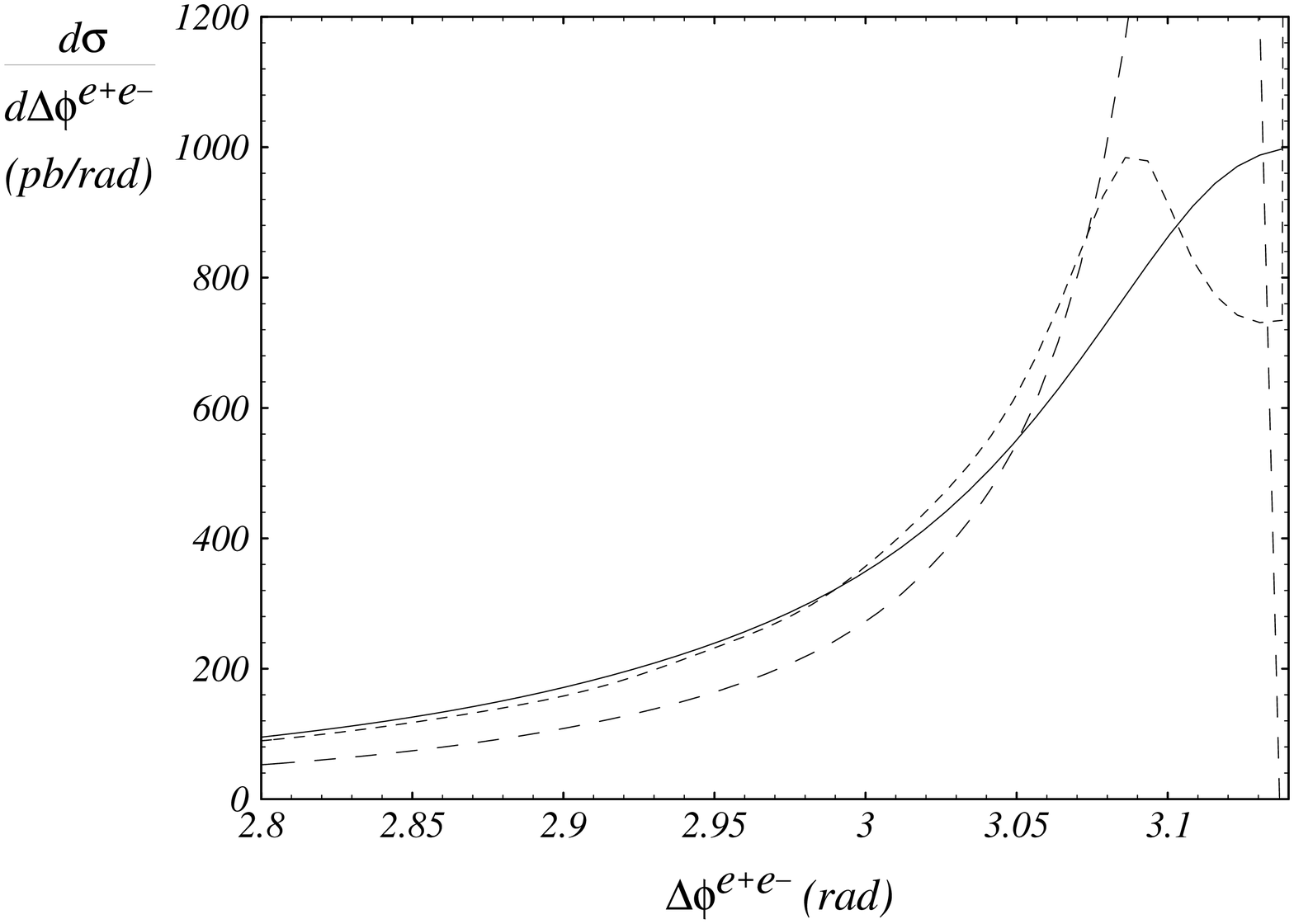}} 
\fi & 
\end{tabular}
\end{center}
\caption{The correlation between the lepton azimuthal angles near the region 
$\Delta \phi \sim \pi$ for $p{\bar p}\rightarrow 
(Z^0 \rightarrow e^{+}e^{-})X$. 
The resummed (solid) distribution gives the correct angular correlation of the 
lepton pair. The NLO (dashed lines) distribution near $\Delta \phi =\pi $
is ill-defined and depends on $Q_T^{Sep}$ (the scale for separating 
soft and hard gluons in the NLO calculation).
The two NLO distributions were calculated with $Q_T^{Sep} = 1.2$ GeV 
(long dash) and $Q_T^{Sep} = 2.0$ GeV
(short dash). }
\label{fig:Z_DelPhi_ee}
\end{figure*}

% >>> $\Delta \phi $
Another observable that can serve to test the QCD theory beyond the
fixed-order perturbative calculation is the difference in the azimuthal
angles of the leptons $\ell _1$ and ${\bar \ell_2}$ from the decay of a vector
boson $V$. In practice, this can be measured for $\gamma ^{*}\text{ or }%
Z^0\rightarrow \ell_1 {\bar \ell_2}$. We show in Fig.~\ref{fig:Z_DelPhi_ee} the
difference in the azimuthal angles of $e^{+}$ and $e^{-}$ ($\Delta \phi
^{e^{+}e^{-}}$), measured in the laboratory frame for $Z^0\rightarrow
e^{+}e^{-}$, calculated in the NLO and the resummed approaches. As
indicated, the NLO result is ill-defined in the vicinity of $\Delta \phi
\sim \pi $, where the multiple soft-gluon radiation has to be resummed to
obtain physical predictions. 
% FFFFFFFFFFFFFFFFFFFFFFFFFFFFFFFFFFFFFFFFFFFFFFFFFFFFFFFFFFFFFFFFFFFFFF
% ffffffffffffffffffffffffffffffffffffffffffffffffffffffffffffffffffffff

\begin{figure*}[t]
\begin{center}
\begin{tabular}{cc}
\ifx\nopictures Y 
\else{ \epsfysize=6.0cm 
\epsffile{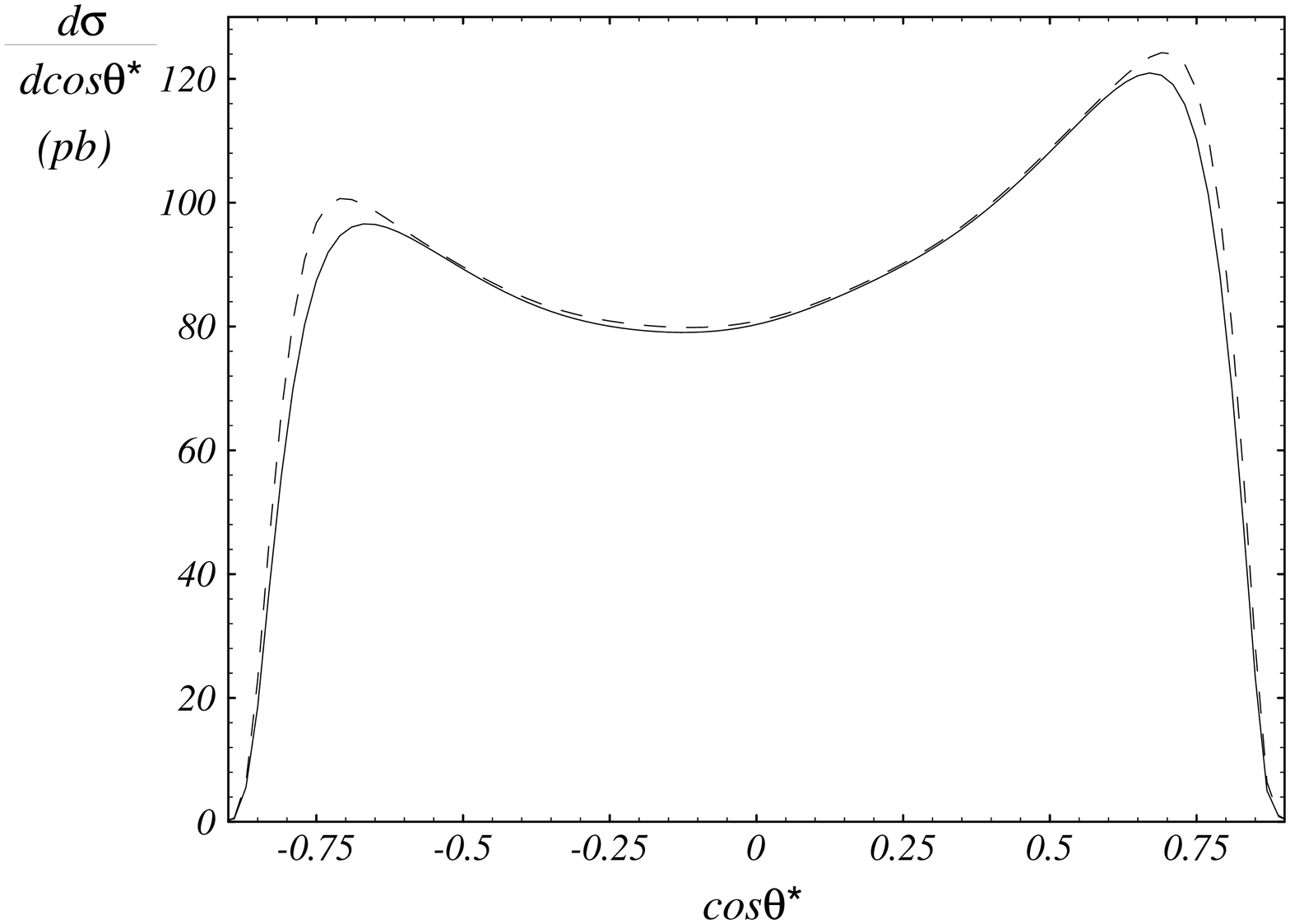} } 
\fi & 
\end{tabular}
\end{center}
\caption{Distribution of the $e^{+}$ polar angle 
$\cos (\theta ^{*})$ in the Collins-Soper frame from $Z^0$ 
decays at the Tevatron with cuts indicated in the text.}
\label{fig:Z_CTS}
\end{figure*}

Another interesting angular variable is the lepton polar angle
distribution $\cos \theta ^{*\ell }$ in the Collins-Soper frame. It can be
calculated for the $Z^0$ decay and used to extract $\sin ^2\theta _w$ at the
Tevatron~\cite{CDFAFB}. % >>> Relation to A_{FB} ? 
The asymmetry in the polar angle distribution is essentially the same as the
forward-backward asymmetry $A_{FB}$ measured at LEP. Since $A_{FB}$ depends
on the invariant mass $Q$ and around the energy of the $Z^0$ peak $A_{FB}$
happens to be very small, the measurement is quite challenging. At the hadron
collider, on the other hand, the invariant mass of the incoming partons is
distributed over a range so the asymmetry is enhanced~\cite{Rosner}. 
The potentials of the
measurement deserve a separated study. In Fig.~\ref{fig:Z_CTS} we show the
distributions of $\cos \theta ^{*\ell }$ predicted from the NLO and the
resummed results. 
% FFFFFFFFFFFFFFFFFFFFFFFFFFFFFFFFFFFFFFFFFFFFFFFFFFFFFFFFFFFFFFFFFFFFFFF
% ffffffffffffffffffffffffffffffffffffffffffffffffffffffffffffffffffffff

% sssssssssssssssssssssssssssssssssssssssssssssssssssssssssssssssssssss

\subsection{Vector Boson Longitudinal Distributions}

\begin{figure*}[t]
\begin{center}
\begin{tabular}{cc} 
\ifx\nopictures Y 
\else{ \epsfysize=6.0cm 
\epsffile{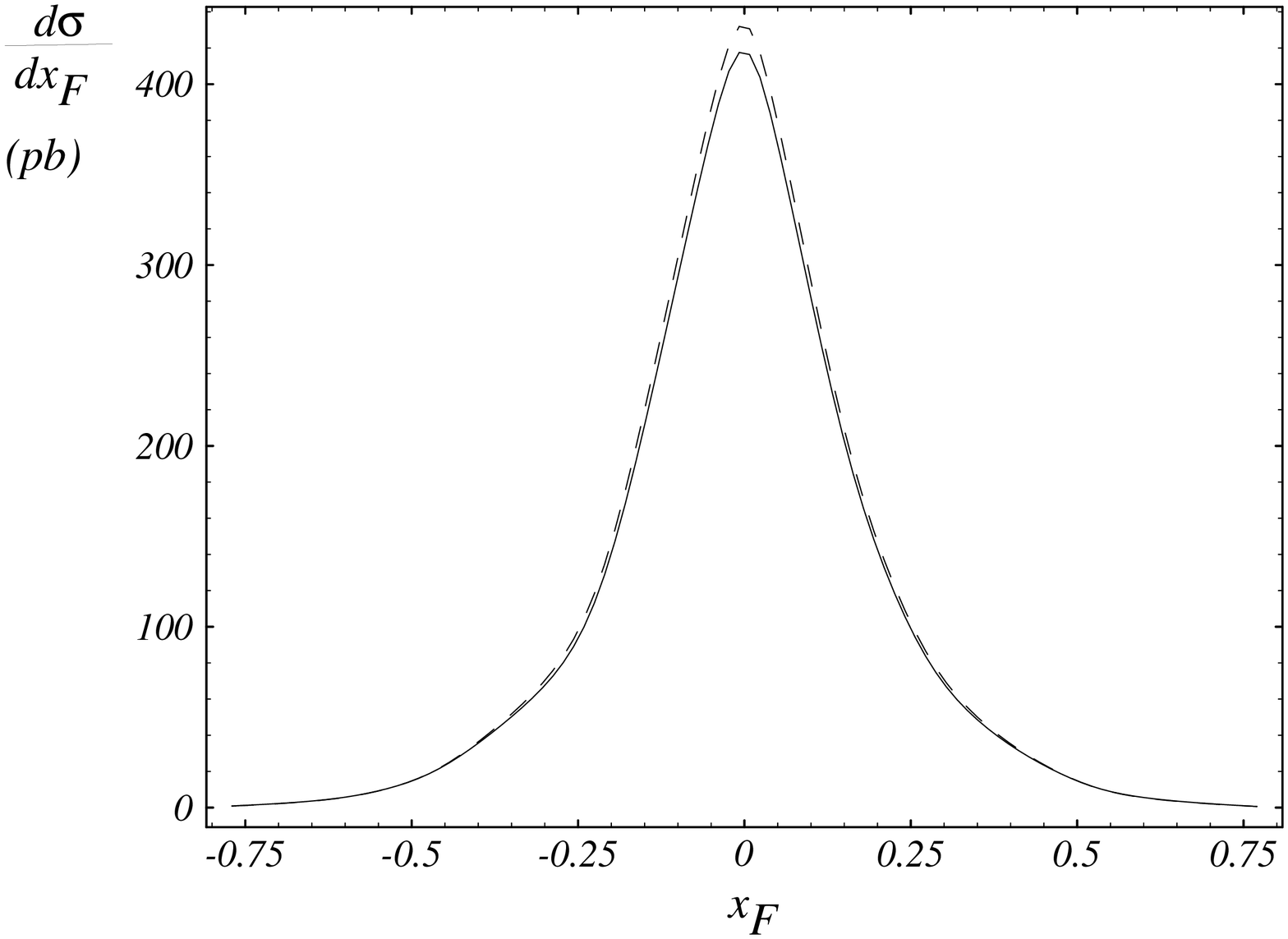} } 
\fi & 
\end{tabular}
\end{center}
\caption{
Longitudinal $x_F$
%=2q^3/\sqrt S $ 
distributions of $Z^0$ bosons produced
at the Tevatron. The NLO (dashed) curves overestimate the rate compared
to the resummed (solid) ones, because kinematic cuts enhance the low
$Q_T$ region where the NLO and resummed distributions are qualitatively
different. Without cuts, the NLO and the resummed $x_F$ distributions are the 
same.
}
\label{fig:Z_qZ_Z}
\end{figure*}

The resummation of the logs involving the transverse momentum of the vector
boson does not directly affect the shape of the longitudinal distributions
of the vector bosons. % >>> for $Z^0$: x_F = $2p_z/\sqrt{S}$
A good example of this is the distribution of the longitudinal
momentum of the $Z^0$ boson which can be measured at the Tevatron with high 
precision, and can be used to extract information on the parton 
distributions. It is customary to plot the rescaled quantity 
$x_F=2q^3/\sqrt{S}$, where $q^3 = \sinh (y) \sqrt{Q^2 + Q_T^2}$ 
is the longitudinal momentum of the $Z^0$ boson measured in the
laboratory frame. In Fig.~\ref{fig:Z_qZ_Z}, we plot 
the distributions predicted in the resummed and the NLO calculations.
As shown, their total event rates are different in the presence of kinematic 
cuts. (Although they are the same if no kinematic cuts imposed.)
This conclusion is similar to that of the $y^\ell $ distributions,
as discussed in Sections~\ref{subsec:Total} and~\ref{subsec:LCA}. 
% FFFFFFFFFFFFFFFFFFFFFFFFFFFFFFFFFFFFFFFFFFFFFFFFFFFFFFFFFFFFFFFFFFFFFFF
% ffffffffffffffffffffffffffffffffffffffffffffffffffffffffffffffffffffff

\begin{figure*}[tbp]
\begin{center}
\begin{tabular}{cc}
\ifx\nopictures Y 
\else{ \epsfysize=6.0cm 
\epsffile{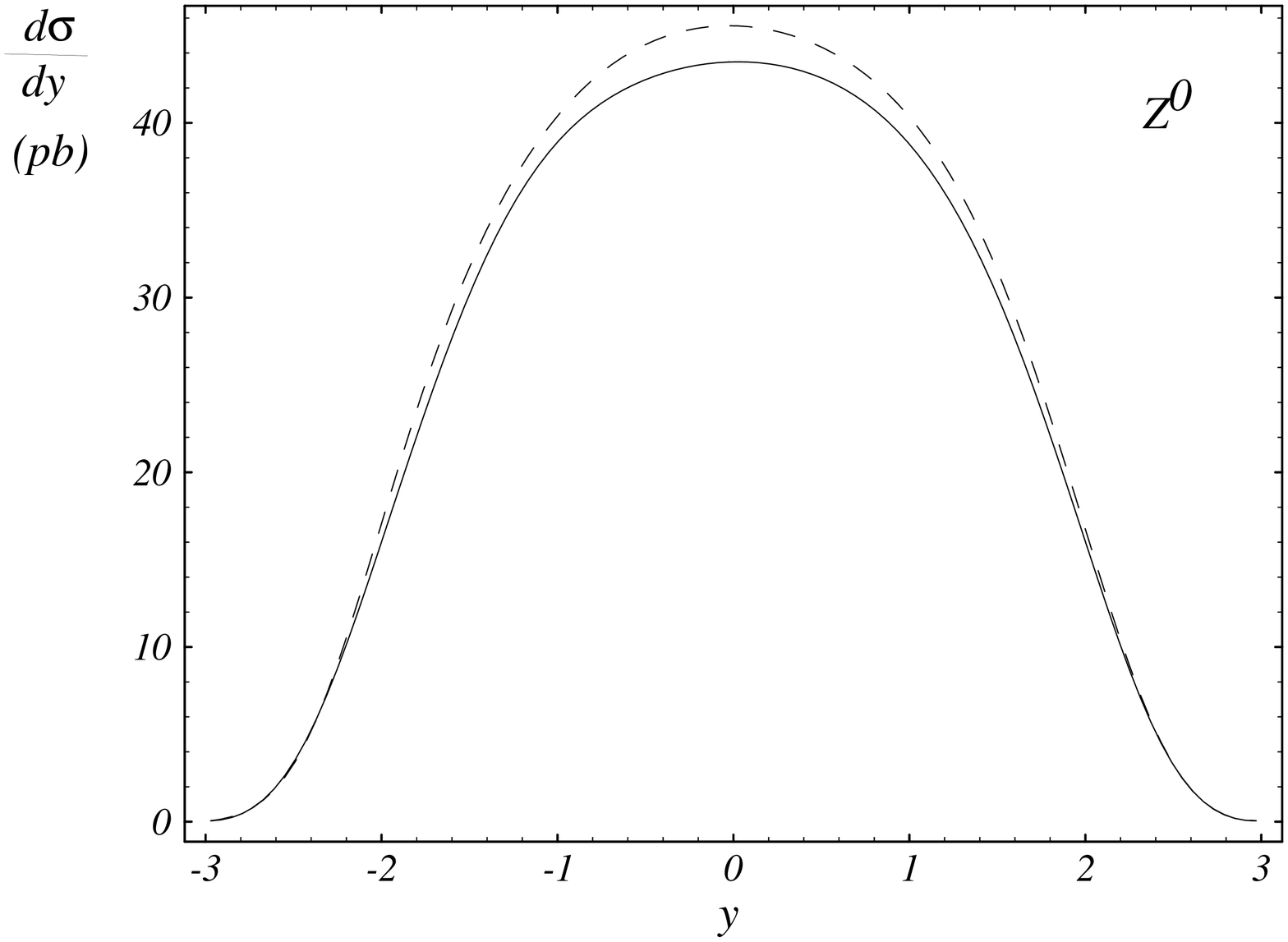} } 
\fi & 
\end{tabular}
\end{center}
\caption{Rapidity distributions (resummed: solid, NLO: dashed) of $Z^0$ 
bosons produced at the Tevatron with the kinematic cuts given in the text.}
\label{fig:Z_y_Z}
\end{figure*}

Without any kinematic cuts, the vector boson rapidity distributions are
also the same in the
resummed and the NLO calculations. This is so because when calculating
the $y$ distribution the transverse momentum $Q_T$ is integrated out so that
the integral has the same value in the  NLO and the resummed calculations.
On the other hand, experimental cuts on the final state leptons restrict the
phase space, so the difference between the NLO and the resummed 
$Q_T$ distributions affects the vector boson rapidity distributions. This shape
difference is very small at the vector boson level, as shown in 
Fig.~\ref{fig:Z_y_Z}. 
% FFFFFFFFFFFFFFFFFFFFFFFFFFFFFFFFFFFFFFFFFFFFFFFFFFFFFFFFFFFFFFFFFFFFFFF
% ffffffffffffffffffffffffffffffffffffffffffffffffffffffffffffffffffffff

% sssssssssssssssssssssssssssssssssssssssssssssssssssssssssssssssssssssss

\section{Discussion and Conclusions}

With a $100\,{\rm pb}^{-1}$ luminosity at the Tevatron, around $2 \times
10^6$ $W^\pm$ and $6 \times 10^5$ $Z^0$ bosons are produced, and the
data sample will increase by a factor of 20 in the Run 2 era. In view
of this large event rate, a careful study of the distributions of 
leptons from the decay of the vector bosons can provide a stringent test
of the rich dynamics of the multiple soft gluon emission predicted by
the QCD theory. Since an accurate determination of the mass of the $W^\pm$ 
boson and the test of parton distribution functions demand a highly precise
knowledge of the kinematical acceptance and the detection efficiency of
$W^\pm$ or $Z^0$ bosons, the effects of the multiple gluon radiation have
to be taken into account. In this work, we have extended the formalism
introduced by Collins, Soper and Sterman for calculating an on-shell
vector boson to include the effects of the polarization
and the decay width of the vector boson on the distributions of the
decay leptons. Our resummation formalism can be applied to any vector
boson $V$ where $V=\gamma^*, \, W^\pm, \, Z^0, \, W', \, Z'$, etc., with
either vector or axial-vector couplings to fermions (leptons or quarks).
To illustrate how the multiple gluon radiation can affect the
distributions of the decay leptons, we studied in detail various
distributions for the production and the decay of the vector bosons at
the Tevatron.

One of the methods to test the rich dynamics of the multiple soft gluon
radiation predicted by the QCD theory is to measure the 
ratio $R_{CSS} \equiv \frac{\sigma (Q_T>Q_T^{\min })}{\sigma _{Total}}$
for the $W^\pm$ and $Z^0$ bosons. We found that, for the vector boson
transverse momentum less than about 30\,GeV, the
difference between the resummed and the
fixed order predictions (either at the $\alpha_S$ or $\alpha^2_S$ order)
can be distinguished by experimental data. 
This suggests that in this kinematic region, the effects of the multiple soft 
gluon radiation are important, hence, the $Q_T$ distribution of the vector 
boson provides an ideal opportunity to test this aspect of the QCD dynamics.
For $Q_T$ less than about 10 GeV, the distribution of $Q_T$ is largely
determined by the non-perturbative sector of QCD. At the
Tevatron this non-perturbative physics,
when parametrized by Eq.~(\ref{eq:WNPLY}) for $W^\pm$ and $Z^0$ production, 
is dominated by the parameter $g_2$ which was shown to be related to 
properties of the QCD vacuum~\cite{Korchemsky-Sterman}. 
Therefore, precisely measuring the $Q_T$ distribution of the vector boson 
in the low $Q_T$ region, e.g. from the ample $Z^0$ events, can advance our 
knowledge of the non-perturbative QCD physics.

Although the rapidity distributions of the leptons are not directly
related to the transverse momentum of the vector boson, they are
predicted to be different in the resummed and the fixed order calculations.
This is because to compare the theoretical predictions with the
experimental data, some kinematic cuts have to be imposed so that the
signal events can be observed over the backgrounds. We showed that the
difference is the largest when the rapidity of the lepton is near the
boundary of the phase space (i.e. in the large rapidity
region), and the difference diminishes when no kinematic cuts are
imposed. 
When kinematic cuts are imposed another important difference between the 
results of the resummed and the NLO calculations is the prediction of 
the event rate. These two calculations predict different normalizations of 
various distributions.
For example, the rapidity distributions of charged leptons 
($y^{\ell^\pm}$) from the decays of $W^\pm$ bosons are different. 
They even differ in the central 
rapidity region in which the lepton charge asymmetry distributions  
are about the same (cf. Figs.~\ref{fig:W_y_e_Asy} and~\ref{fig:W_y_e}).
As noted in Ref.~\cite{Stirling-Martin}, with kinematic cuts, the measurement 
of $M_W$ is correlated to that of the rapidity and its asymmetry 
through the transverse momentum of the decay lepton. 
Since the resummed and the NLO results are different and the former includes 
the multiple soft gluon emission dynamics, the resummed calculation should be 
used for a precision measurement of $M_W$.

In addition to the rapidity distribution, we have also shown
various distributions of the leptons which are either directly or
indirectly related to the transverse momentum of the vector boson. For
those which are directly related to the transverse momentum of the
vector boson, such as the transverse momentum of the lepton and the
azimuthal correlation of the leptons, our resummation formalism
predicts significant differences from the fixed order perturbation
calculations in some kinematic regions. The details were discussed in
Section III.

\begin{figure*}[t]
\begin{center}
\begin{tabular}{cc}
\ifx\nopictures Y \else{ 
\epsfysize=6.0cm 
\epsffile{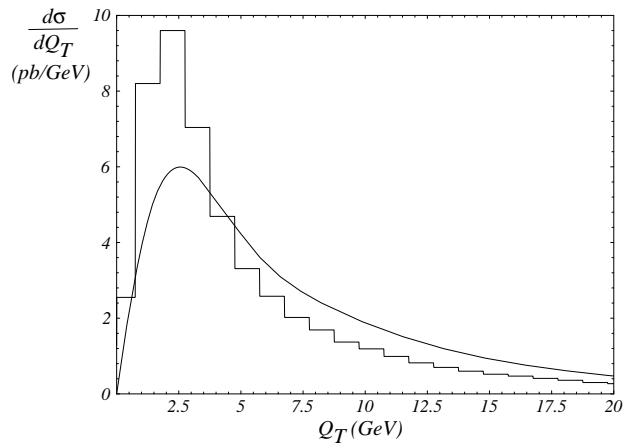}}
\fi & 
\end{tabular}
\end{center}
\caption{ 
Transverse momentum distribution of virtual photons in $p {\bar p}
\rightarrow \gamma^* \rightarrow e^+ e^-$ events predicted by ResBos 
(solid curve) and ISAJET (histogram), calculated for the invariant mass range 
30 GeV $< Q <$ 60 GeV at the 1.8 TeV Tevatron. 
}
\label{fig:QTph}
\end{figure*}

As noted in the Introduction, a full event generator, such as ISAJET,
can predict a reasonable shape for various distributions because it
contains the backward radiation algorithm~\cite{Sjostrand}, which effectively
includes part of the Sudakov factor, i.e. effects of the multiple gluon 
radiation.
However, the total event rate predicted by the full event generator is
usually only accurate at the tree level, as the short distance part
of the virtual corrections cannot yet be consistently implemented in
this type of Monte Carlo program. To illustrate the effects of the high
order corrections coming from the virtual corrections, which contribute to the 
Wilson coefficients $C$ in our resummation formalism, we showed in 
Fig.~\ref{fig:QTph} the
predicted distributions of the transverse momentum of the Drell-Yan
pairs by ISAJET and by ResBos (our resummed calculation). In this
figure we have rescaled the ISAJET prediction to have the same total
rate as the ResBos result, so that the shape of the distributions can be
directly compared. 
We restrict the invariant mass of the virtual photons $Q$ to be between 
30 and 60 GeV without any kinematic cuts on the leptons.
If additional kinematic cuts on the leptons are imposed, then the
difference is expected to be enhanced, as discussed in 
Section~\ref{subsec:LCA}.
As clearly shown, with a large data sample in the future, 
it will be possible to experimentally distinguish between these two
predictions, and, more interestingly, to start probing the
non-perturbative sector of the QCD physics.
% FFFFFFFFFFFFFFFFFFFFFFFFFFFFFFFFFFFFFFFFFFFFFFFFFFFFFFFFFFFFFFFFFFFFFFF
% ffffffffffffffffffffffffffffffffffffffffffffffffffffffffffffffffffffff

% aaaaaaaaaaaaaaaaaaaaaaaaaaaaaaaaaaaaaaaaaaaaaaaaaaaaaaaaaaaaaaaaaaaaaa

\acknowledgments
We thank G.A. Ladinsky and  J.W. Qiu 
for their vital collaboration in this project, R. Brock, S. Mrenna and 
W.K. Tung for numerous discussions and suggestions, and to the CTEQ 
collaboration for discussions on resummation and related topics. 
This work was supported in part by NSF under
grant PHY-9507683.

\newpage

% AAAAAAAAAAAAAAAAAAAAAAAAAAAAAAAAAAAAAAAAAAAAAAAAAAAAAAAAAAAAAAAAAAAAAA

\appendix

\section{Kinematics}
\label{app:Kin}

Here we summarize some details of the kinematics for the lepton pair
production process $h_1h_2\rightarrow V(\rightarrow \ell _1{\bar \ell_2})X$.
The laboratory ($lab$) frame is the center-of-mass frame of the colliding
hadrons $h_1$ and $h_2$. In the $lab$ frame, the cartesian coordinates of
the hadrons are: $p_{h_1,h_2}^\mu (lab)=\sqrt{S}/2\,(1,0,0,\pm 1)$, where $%
\sqrt{S}$ is the center-of-mass energy of the collider. Transverse momentum
resummation is performed in the Collins-Soper ($CS$) frame~\cite{CSFrame}.
This is the special rest frame of the vector boson in which the $z$ axis
bisects the angle between the $h_1$ hadron momentum $p_{h_1}(CS)$ and the
negative $h_2$ hadron momentum $-p_{h_2}(CS)$~\cite{lamtung}.

% >>> Formulae relating momenta in the $CS$ and lab frames
To derive the Lorentz transformation $\Lambda _{~\nu} ^\mu (lab\rightarrow CS)$
that connects the $lab$ and $CS$ frames (in the active view point): $p^\mu
(CS)=\Lambda _{~\nu} ^\mu (lab\rightarrow CS)\;p^\nu (lab)$, we follow the
definition of the $CS$ frame. First, we find the rotation which makes the
azimuthal angle $\phi _V$ of the vector boson vanish. Then, we find the
boost into a vector boson rest frame. Finally, in the vector boson rest
frame we find the rotation which brings the hadron momentum $p_{h_1}(CS)$
and negative hadron momentum $-p_{h_2}(CS)$ into the desired directions.

The transformation that takes $q^\mu (lab)$=$(q^0,Q_T\cos \phi _V,Q_T\sin
\phi _V,q^3)$ from the $lab$ frame into a longitudinally co-moving frame ($%
long$), in which $q^\mu (long)=(q^0,Q_T,0,q^3)$, is a rotation around the $z$
axis, which is 
%% EEEEEEEEEEEEEEEEEEEEEEEEEEEEEEEEEEEEEEEEEEEEEEEEEEEEEEEEEEEEEEEEEEEEEE
\[
\Lambda _{~\nu} ^\mu (lab\rightarrow long)=\left( \text{%
\begin{tabular}{rrrr}
${1}$ \quad & $0$ \quad & $0$ \quad & $0$ \\ 
$0$ \quad & $\;\;{\cos \phi _V}$ \quad & $\sin {\phi _V}$ \quad & ${0}$ \\ 
$0$ \quad & --$\sin {\phi _V}$ \quad & ${\cos \phi _V}$ \quad & $0$ \\ 
$0$ \quad & ${0}$\quad & $0$ \quad & ${1}$%
\end{tabular}
}\right) . 
\]
%% eeeeeeeeeeeeeeeeeeeeeeeeeeeeeeeeeeeeeeeeeeeeeeeeeeeeeeeeeeeeeeeeeeeeee
A boost by $\vec{\beta}=-\vec{q}(long)/q^0$ brings four vectors from the
longitudinal ($long$) frame into a vector boson rest frame ($rest$). The
matrix of the Lorentz boost from the $long$ frame to the $rest$ frame,
expressed explicitly in terms of $q^\mu $ is 
%% EEEEEEEEEEEEEEEEEEEEEEEEEEEEEEEEEEEEEEEEEEEEEEEEEEEEEEEEEEEEEEEEEEEEEE
\[
\Lambda _{~\nu} ^\mu (long\rightarrow rest)=\left( \text{%
\begin{tabular}{rrrr}
${\frac{q^0}Q}$ \quad & $-{\frac{Q_T}Q}$ \quad & $0$ \quad & $-{\frac{q^3}Q}$
\\ 
$-{\frac{Q_T}Q}$ \quad & ${\frac{Qq^0+M_T^2}{Q(q^0+Q)}}$ \quad & $0$ \quad & 
${\frac{Q_Tq^3}{Q(q^0+Q)}}$ \\ 
$0$ \quad & $0$ \quad & $1$ \quad & $0$ \\ 
$-{\frac{q^3}Q}$ \quad & ${\frac{Q_Tq^3}{Q(q^0+Q)}}$\quad & $0$ \quad & ${%
\frac{Q(q^0+Q)+(q^3)^2}{Q(q^0+Q)}}$%
\end{tabular}
}\right) , 
\]
%% eeeeeeeeeeeeeeeeeeeeeeeeeeeeeeeeeeeeeeeeeeeeeeeeeeeeeeeeeeeeeeeeeeeeee
where $Q=\sqrt{(q^0)^2-Q_T^2-(q^3)^2}$ is the vector boson invariant mass,
and the transverse mass is defined as $M_T=\sqrt{Q^2+Q_T^2}$. The
transformation from the $lab$ frame to the $rest$ frame is then the product
of the above boost and rotation: $\Lambda _{~\nu} ^\mu (lab\rightarrow
rest)=\Lambda _{~\lambda} ^\mu (long\rightarrow rest)\Lambda _{~\nu} ^\lambda
(lab\rightarrow long)$. Had we used only the boost $\vec{\beta}^{\prime }=-%
\vec{q}(lab)/q^0$, we would have obtained the same result for $\Lambda _{~\nu}
^\mu(lab\rightarrow rest)$ in one step: 
%% EEEEEEEEEEEEEEEEEEEEEEEEEEEEEEEEEEEEEEEEEEEEEEEEEEEEEEEEEEEEEEEEEEEEEE
\[
\Lambda _{~\nu} ^\mu (lab\rightarrow rest)=\left( \text{%
\begin{tabular}{rrrr}
${\frac{q^0}Q}$ \quad & $-{\frac{Q_T}Q\cos \phi _V}$ \quad & $-{\frac{Q_T}Q%
\sin \phi _V}$ \quad & $-{\frac{q^3}Q}$ \\ 
$-{\frac{Q_T}Q}$ \quad & ${\frac{Qq^0+M_T^2}{Q(q^0+Q)}\cos \phi _V}$ \quad & 
${\frac{Qq^0+M_T^2}{Q(q^0+Q)}\sin \phi _V}$ \quad & ${\frac{Q_Tq^3}{Q(q^0+Q)}%
}$ \\ 
$0$ \quad & $-{\sin \phi _V}$ \quad & ${\cos \phi _V}$ \quad & $0$ \\ 
$-{\frac{q^3}Q}$ \quad & ${\frac{Q_Tq^3}{Q(q^0+Q)}\cos \phi _V}$ \quad & ${%
\frac{Q_Tq^3}{Q(q^0+Q)}\sin \phi _V}$ \quad & ${\frac{Q(q^0+Q)+(q^3)^2}{%
Q(q^0+Q)}}$%
\end{tabular}
}\right) . 
\]
%% eeeeeeeeeeeeeeeeeeeeeeeeeeeeeeeeeeeeeeeeeeeeeeeeeeeeeeeeeeeeeeeeeeeeee

After boosting the lab frame hadron momenta into this rest frame, we obtain 
%% EEEEEEEEEEEEEEEEEEEEEEEEEEEEEEEEEEEEEEEEEEEEEEEEEEEEEEEEEEEEEEEEEEEEEE
\begin{eqnarray*}
p_{h_1,h_2}^\mu (rest) &=&\Lambda _{~\nu} ^\mu (lab\rightarrow
rest)\;p_{h_1,h_2}^\nu (lab)= \\
&&\frac{\sqrt{S}}2\left( \frac{q^0\mp q^3}Q,-\frac{Q_T}Q\frac{q^0+Q\pm q^3}{%
q^0+Q},0,\frac{(\pm Q-q^3)(q^0+Q)\pm (q^3)^2}{Q(q^0+Q)}\right) ,
\end{eqnarray*}
%% eeeeeeeeeeeeeeeeeeeeeeeeeeeeeeeeeeeeeeeeeeeeeeeeeeeeeeeeeeeeeeeeeeeeee
and the polar angles of $p_{h_1}^\mu (rest)$ and $-p_{h_2}^\mu (rest)$ are
not equal unless $Q_T=0$. (In the above expressions the upper signs refers
to $h_1$ and the lower signs to $h_2$.) In the general $Q_T\neq 0$ case we
have to apply an additional rotation in the $rest$ frame so that the $z$-axis
bisects the angle between the hadron momentum $p_{h_1}(CS)$ and the negative
hadron momentum $-p_{h_2}(CS)$. 
It is easy to see that to keep $\vec{p}_{h_1,h_2}$ in the $xz$ plane, 
this rotation should be a rotation around the $y$
axis by an angle $\alpha =\arccos {[Q(q^0+M_T)/(M_T(q^0+Q))]}$.

Thus the Lorentz transformation from the $lab$ frame to the $CS$ frame is 
$\Lambda _{~\nu} ^\mu (lab\rightarrow CS)=\Lambda _{~\lambda} ^\mu 
(rest\rightarrow CS)\;\Lambda _{~\nu} ^\lambda (lab\rightarrow rest)$. 
Indeed, this
transformation results in equal polar angles $\theta _{h_1,-h_2}=\arctan
(Q_T/Q)$. The inverse of this transformation takes vectors from the $CS$
frame to the $lab$ frame is: 
%% EEEEEEEEEEEEEEEEEEEEEEEEEEEEEEEEEEEEEEEEEEEEEEEEEEEEEEEEEEEEEEEEEEEEEE
\begin{eqnarray*}
\Lambda _{~\nu} ^\mu (CS &\rightarrow& lab) = \left( \Lambda _{~\nu} ^\mu
(lab\rightarrow CS)\right) ^{-1} = \\
&& \left( \text{%
\begin{tabular}{rrrr}
${\frac{q^0}Q}$ \quad & ${\frac{Q_Tq^0}{QM_T}}$ \quad & $0$ \quad & $-{\frac{%
q^3}{M_T}}$ \\ 
$-{\frac{Q_T}Q\cos \phi _V}$ \quad & ${\frac{M_T}Q\cos \phi _V}$ \quad & ${%
-\sin \phi _V}$ \quad & ${0}$ \\ 
$-{\frac{Q_T}Q\sin \phi _V}$ \quad & ${\frac{M_T}Q\sin \phi _V}$ \quad & ${%
\cos \phi _V}$ \quad & $0$ \\ 
${\frac{q^3}Q}$ \quad & ${\frac{Q_Tq^3}{QM_T}}$ \quad & ${0}$ \quad & ${%
\frac{Q_Tq^0}{QM_T}}$%
\end{tabular}
}\right) .
\end{eqnarray*}
%% eeeeeeeeeeeeeeeeeeeeeeeeeeeeeeeeeeeeeeeeeeeeeeeeeeeeeeeeeeeeeeeeeeeeee

The kinematics of the leptons from the decay of the vector boson can be
described by the polar angle $\theta $ and the azimuthal angle $\phi $,
defined in the Collins-Soper frame. The above transformation formulae lead
to the four-momentum of the decay product fermion (and anti-fermion) in the
lab frame as 
% EEEEEEEEEEEEEEEEEEEEEEEEEEEEEEEEEEEEEEEEEEEEEEEEEEEEEEEEEEEEEEEEEEEEEE
\begin{eqnarray*}
p^\mu &=&{\frac Q2}\left( {\frac{q^\mu }Q}+\sin {\theta }\cos {\phi }~X^\mu
+\sin {\theta }\sin {\phi }~Y^\mu +\cos {\theta }~Z^\mu \right) , \\
\overline{p}^\mu &=&q^\mu -p^\mu ,
\end{eqnarray*}
%% eeeeeeeeeeeeeeeeeeeeeeeeeeeeeeeeeeeeeeeeeeeeeeeeeeeeeeeeeeeeeeeeeeeeee
where 
% EEEEEEEEEEEEEEEEEEEEEEEEEEEEEEEEEEEEEEEEEEEEEEEEEEEEEEEEEEEEEEEEEEEEEE
\begin{eqnarray}
&&q^\mu =(M_T\cosh {y},\,Q_T\cos {\phi _V},\,Q_T\sin {\phi _V},\,M_T\sinh {y}%
),  \nonumber \\
\ &&X^\mu =-{\frac Q{Q_TM_T}}\left( q_{+}n^\mu +q_{-}{\bar{n}}^\mu -{\frac{%
M_T^2}{Q^2}}q^\mu \right) ,  \nonumber \\
\ &&Z^\mu ={\frac 1{M_T}}\left( q_{+}n^\mu -q_{-}{\bar{n}}^\mu \right) , 
\nonumber \\
\ &&Y^\mu =\varepsilon ^{\mu \nu \alpha \beta }{\frac{q_\nu }Q}Z_\alpha
X_\beta .  \label{eq:QXYZ}
\end{eqnarray}
%% eeeeeeeeeeeeeeeeeeeeeeeeeeeeeeeeeeeeeeeeeeeeeeeeeeeeeeeeeeeeeeeeeeeeee
Here, $q_{\pm }={\frac 1{\sqrt{2}}}(q^0\pm q^3)$, $y=\frac 12\ln
(q_{+}/q_{-})$, $n^\nu =\frac 1{\sqrt{2}}(1,0,0,1)$, ${\bar{n}}^\nu ={\frac 1%
{\sqrt{2}}}(1,0,0,-1)$ and the totally anti-symmetric tensor is defined as $%
\varepsilon ^{0123}=-1$.

% ssssssssssssssssssssssssssssssssssssssssssssssssssssssssssssssssssssss

\section{${\cal O}(\alpha _S)$ Results}
\label{app:OaS}

To correctly extract the distributions of the leptons, we have to calculate 
the production and the decay of a polarized vector boson. The ${\cal O}(\alpha
_S)$ QCD corrections to the production and decay of a polarized vector boson 
can be found in the literature~\cite{Aurenche}, in which both the symmetric 
and the anti-symmetric parts of the hadronic tensor were calculated. Such a
calculation was, as usual, carried out in general number ($D$) of space-time
dimensions, and dimensional regularization scheme was used to regulate
infrared (IR) divergences because it preserves the gauge and the Lorentz
invariances. Since the anti-symmetric part of the hadronic tensor contains
traces with an odd number of $\gamma _5$'s, one has to choose a definition
(prescription) of $\gamma _5$ in $D$ dimensions. It was shown in a series of
papers~\cite{gamma5c} that in $D \neq 4$ dimension, the consistent $\gamma_5$
prescription to use is the t'Hooft-Veltman prescription. Since in Ref.~\cite
{Aurenche} a different prescription was used, we give below the results of
our calculation in the t'Hooft-Veltman $\gamma_5$ prescription.

For calculating the virtual corrections, we follow the argument of Ref.~\cite
{Korner-Schuler} and impose the chiral invariance relation, which is
necessary to eliminate ultraviolet anomalies of the one loop axial vector
current when calculating the structure function. Applying this relation for
the virtual corrections we obtain the same result as that in Refs.~\cite
{Aurenche} and~\cite{AEM79}. The final result of the virtual corrections
gives 
% EEEEEEEEEEEEEEEEEEEEEEEEEEEEEEEEEEEEEEEEEEEEEEEEEEEEEEEEEEEEEEEEEEEEEE
\begin{eqnarray}
{\cal M}_{Born}^{\dagger }{\cal M}_{Virt}&+&{\cal M}_{Virt}^{\dagger }
{\cal M}_{Born}=  \nonumber \\
&& C_F\frac{\alpha _S}{2\pi }\left( \frac{4\pi \mu ^2}{Q^2}\right) ^\epsilon 
\frac 1{\Gamma (1-\epsilon )}\left( -\frac 2{\epsilon ^2}-\frac 3 \epsilon
+\pi ^2-8\right) \left| {\cal M}_{Born}\right| ^2,  \label{Eq:VirtCorr}
\end{eqnarray}
% eeeeeeeeeeeeeeeeeeeeeeeeeeeeeeeeeeeeeeeeeeeeeeeeeeeeeeeeeeeeeeeeeeeeee
where $\epsilon =(4-D)/2$, $\mu $ is the t'Hooft mass scale, and $C_F=4/3$
in QCD. The four dimensional Born level amplitude is 
% EEEEEEEEEEEEEEEEEEEEEEEEEEEEEEEEEEEEEEEEEEEEEEEEEEEEEEEEEEEEEEEEEEEEEE
\begin{eqnarray}
\left| {\cal M}_{Born}\right| ^2 &=&
{\frac{16Q^4}{(Q^2-M_V^2)^2+Q^4\Gamma_V^2/M_V^2}} 
%\delta ^2({\vec{Q}}_T)\delta (1-z_1)\delta (1-z_2) 
\nonumber \\ && \times 
\left[ (g_L^2+g_R^2)(f_L^2+f_R^2){\cal L}_0+(g_L^2-g_R^2)(f_L^2-f_R^2)%
{\cal A}_3\right],  
\label{eq:MBorn}
\end{eqnarray}
% eeeeeeeeeeeeeeeeeeeeeeeeeeeeeeeeeeeeeeeeeeeeeeeeeeeeeeeeeeeeeeeeeeeeee
where we have used the LEP prescription for the vector boson resonance with
mass $M_V$ and width $\Gamma_V$. The angular functions are ${\cal L}_0 = 1 +
\cos^2{\theta}$ and ${\cal A}_3 = 2 \cos{\theta}$. The initial state
spin average (1/4), and color average (1/9) factors are not yet included 
in Eq.~(\ref{eq:MBorn}).

When calculating the real emission diagrams, we use the same
(t'Hooft-Veltman) $\gamma _5$ prescription. It is customary to organize the $%
{\cal O}(\alpha _S^n)$ corrections by separating the lepton degrees of
freedom from the hadronic ones, so that 
% EEEEEEEEEEEEEEEEEEEEEEEEEEEEEEEEEEEEEEEEEEEEEEEEEEEEEEEEEEEEEEEEEEEEEE
\begin{eqnarray*}
&&{\left( {\frac{d\sigma (h_1h_2\rightarrow V(\rightarrow \ell _1{\bar{\ell}%
_2})X)}{dQ^2dydQ_T^2d\cos {\theta }d\phi }}\right) }_{{\cal O}(\alpha _S)}^{%
{\rm real~emission}}=
{\frac{\alpha _S(Q) C_F}{(2\pi )^3S}}{\frac{Q^2}{%
(Q^2-M_V^2)^2+Q^4\Gamma _V^2/M_V^2}} \\
&&~~~\times \sum_{a,b,i}\int_{x_1}^1{\frac{d\xi _1}{\xi _1}}\int_{x_2}^1{%
\frac{d\xi _2}{\xi _2}}~{\cal G}^{i\;}{\cal L}_{ab}(\xi _1,\xi _2,Q^2)\;%
{\cal T}_{ab}^i(Q_T,Q,z_1,z_2)\ {\cal A}_i(\theta ,\phi ),
\end{eqnarray*}
% eeeeeeeeeeeeeeeeeeeeeeeeeeeeeeeeeeeeeeeeeeeeeeeeeeeeeeeeeeeeeeeeeeeeee
with $z_1 = x_1/\xi_1$ and $z_2 = x_2/\xi_2$.
The dependence on the lepton kinematics is carried by the angular functions 
% EEEEEEEEEEEEEEEEEEEEEEEEEEEEEEEEEEEEEEEEEEEEEEEEEEEEEEEEEEEEEEEEEEEEEE
\begin{eqnarray*}
\ &&{\cal L}_0=1+\cos ^2{\theta },~{\cal A}_0={\frac 12}(1-3\cos ^2{\theta }%
),~{\cal A}_1=\sin {2\theta }\cos {\phi },~{\cal A}_2={\frac 12}\sin ^2{%
\theta }\cos {2\phi },~ \\
\ &&{\cal A}_3=2\cos {\theta },~{\cal A}_4=\sin {\theta }\cos {\phi }.
\end{eqnarray*}
% eeeeeeeeeeeeeeeeeeeeeeeeeeeeeeeeeeeeeeeeeeeeeeeeeeeeeeeeeeeeeeeeeeeeee
In the above differential cross section, $i=-1,...,4$ with ${\cal A}%
_{-1}\equiv {\cal L}_0$; and ${\cal G}^i=(g_L^2+g_R^2)(f_L^2+f_R^2)$ for $%
i=-1,0,1,2;$ ${\cal G}^i=(g_L^2-g_R^2)(f_L^2-f_R^2)$ for $i=3,4$. The parton
level helicity cross sections are summed for the parton indices $a$, $b$ in
the following fashion 
% EEEEEEEEEEEEEEEEEEEEEEEEEEEEEEEEEEEEEEEEEEEEEEEEEEEEEEEEEEEEEEEEEEEEEE
\[
\ \sum_{a,b}{\cal L}_{ab}\;{\cal T}_{ab}^i=\sum_{q=u,d,s,c,b}\left( 
{\cal L}_{q\bar{q}}\;{\cal T}_{q\bar{q}}^i+{\cal L}_{\bar{q}q}\;{\cal T}_{%
\bar{q}q}^i+{\cal L}_{qG}\;{\cal T}_{qG}^i+{\cal L}_{\bar{q}G}\;{\cal T}_{%
\bar{q}G}^i+{\cal L}_{Gq}\;{\cal T}_{Gq}^i+{\cal L}_{G\bar{q}}\;{\cal T}%
_{G\bar{q}}^i\right) . 
\]
% eeeeeeeeeeeeeeeeeeeeeeeeeeeeeeeeeeeeeeeeeeeeeeeeeeeeeeeeeeeeeeeeeeeeee
The partonic luminosity functions ${\cal L}_{ab}$\ are defined as 
% EEEEEEEEEEEEEEEEEEEEEEEEEEEEEEEEEEEEEEEEEEEEEEEEEEEEEEEEEEEEEEEEEEEEEE
\[
{\cal L}_{ab}(\xi_1,\xi_2,Q^2)=
%\sum_{a,b}
\;f_{a/h_1}(\xi_1,Q^2) \;f_{b/h_2}(\xi_2,Q^2), 
\]
% eeeeeeeeeeeeeeeeeeeeeeeeeeeeeeeeeeeeeeeeeeeeeeeeeeeeeeeeeeeeeeeeeeeeee
where $f_{a/h_1}$ is the parton probability density of parton $a$ in hadron 
$h_1$, etc. The squared matrix elements for the annihilation sub-process 
$q{\bar{q}}\rightarrow VG$ in the $CS$ frame, including the $\epsilon$ 
dependent terms, are as follows: 
% eeeeeeeeeeeeeeeeeeeeeeeeeeeeeeeeeeeeeeeeeeeeeeeeeeeeeeeeeeeeeeeeeeeeee
\begin{eqnarray*}
\ {\cal T}_{q\bar{q}}^{-1} &=&\frac 1{ut}\left( T_{+}(u,t)-(t+u)^2\epsilon
\right) ,\;\; \\
{\cal T}_{q\bar{q}}^0\;\; &=&{\cal T}_{q\bar{q}}^2=\frac 1{ut}\frac{Q_T^2}{%
M_T^2}\left( T_{+}(u,t)-(Q^2+s)^2\epsilon \right) ,\; \\
{\cal T}_{q\bar{q}}^1\;\; &=&\frac 1{ut}\frac{Q_TQ}{M_T^2}%
T_{-}(u,t)(1-\epsilon ),\; \\
{\cal T}_{q\bar{q}}^3\;\; &=&\frac 1{ut}\frac Q{M_T}\left( T_{+}(u,t)-{\frac{%
(Q^2-u)t^2+(Q^2-t)u^2}{Q^2}}\epsilon \right) ,\; \\
{\cal T}_{q\bar{q}}^4\;\; &=&\frac 2{ut}\frac{Q_T}{M_T}\left(
T_{-}(u,t)+Q^2(u-t)\epsilon \right) .
\end{eqnarray*}
% eeeeeeeeeeeeeeeeeeeeeeeeeeeeeeeeeeeeeeeeeeeeeeeeeeeeeeeeeeeeeeeeeeeeee
For the Compton sub-process $qG\rightarrow Vq$, we obtain 
% eeeeeeeeeeeeeeeeeeeeeeeeeeeeeeeeeeeeeeeeeeeeeeeeeeeeeeeeeeeeeeeeeeeeee
\begin{eqnarray*}
\ {\cal T}_{qG}^{-1} &=&\frac{-1}{su}\left( T_{+}(s,u)-{(s+u)^2}\epsilon
\right) ,\;\; \\
{\cal T}_{qG}^0\;\; &=&{\cal T}_{qG}^2=\frac{-1}{su}\frac{Q_T^2}{M_T^2}%
\left( (Q^2-u)^2+(Q^2+s)^2-{(s+u)^2}\epsilon \right) , \\
\;{\cal T}_{qG}^1\;\; &=&\frac{-1}{su}\frac{Q_TQ}{M_T^2}\left(
2(Q^2-u)^2-(Q^2-t)^2+{(s+u)^2}\epsilon \right) ,\; \\
{\cal T}_{qG}^3\;\; &=&\frac{-1}{su}\frac Q{M_T}\left( T_{+}(s,u)-2u(Q^2-s)+%
\frac{(Q^2-t)(Q^2s-su-u^2)}{Q^2}\epsilon \right) , \\
{\cal T}_{qG}^4\;\; &=&\frac{-2}{su}\frac{Q_T}{M_T}\left(
2s(Q^2-s)+T_{+}(s,u)-(Q^2-t)u\epsilon \right) .
\end{eqnarray*}
% eeeeeeeeeeeeeeeeeeeeeeeeeeeeeeeeeeeeeeeeeeeeeeeeeeeeeeeeeeeeeeeeeeeeee
In the above equations, the Mandelstam variables: $s=(k+l)^2$, $t=(k-q)^2$,
and $u=(l-q)^2$ where $k$, $l$ and $q$ are the four momenta of the partons
from hadrons $h_1$, $h_2$ and that of the vector boson, respectively. 
%The ${\cal T}_{qG}^i$ functions of the 
%Compton supprocess cannot simply be obtained by 
%crossing from ${\cal T}_{q\bar q}^i$ since they are 
%already expressed in the specific ($CS$) frame.
% where we used the relations (\ref{Eq:kl12}) to express 
%${\calT}^i$ as the function of the lepton angles $\theta $ and $\phi $
%On the other hand, all other relevant parton level 
%cross sections can be obtained from 
%the aboves, summarized by the following rules 
All other relevant parton level cross sections can be obtained from the
above, and summarized by the following rules: 
% EEEEEEEEEEEEEEEEEEEEEEEEEEEEEEEEEEEEEEEEEEEEEEEEEEEEEEEEEEEEEEEEEEEEEE
\vspace{.3cm}
%\begin{eqnarray}
\begin{center}
\begin{tabular}{ll}
\hline\hline
$i=-1,0,1,2$ & $i=3,4$ \\ \hline
${\cal T}_{\bar{q}q}^i={\cal T}_{q\bar{q}}^i$ & ${\cal T}_{\bar{q}q}^i=
-{\cal T}_{q\bar{q}}^i$ \\ 
${\cal T}_{Gq}^i={\cal T}_{qG}^i\;(u\leftrightarrow t)\;\;\;\;\;\;\;$ & $%
{\cal T}_{Gq}^i=-{\cal T}_{qG}^i\;(u\leftrightarrow t)$ \\ 
${\cal T}_{\bar{q}G}^i={\cal T}_{qG}^i$ & ${\cal T}_{\bar{q}G}^i=-{\cal T}%
_{qG}^i$ \\ 
${\cal T}_{G\bar{q}}^i={\cal T}_{Gq}$ & ${\cal T}_{G\bar{q}}^i=-{\cal T}%
_{Gq}^i,$ \\ \hline\hline
\end{tabular}
\end{center}
%\end{eqnarray}
\vspace{.3cm} 
% eeeeeeeeeeeeeeeeeeeeeeeeeeeeeeeeeeeeeeeeeeeeeeeeeeeeeeeeeeeeeeeeeeeeee
with the only exceptions that ${\cal T}_{Gq}^1=-{\cal T}_{qG}^1\;(u%
\leftrightarrow t)$\ and ${\cal T}_{Gq}^4={\cal T}_{qG}^4\;(u\leftrightarrow
t)$. These results are consistent with the regular pieces of the $Y$ term
given in Appendix E and with~those in Ref. \cite{Mirkes}.

In the above matrix elements, only the coefficients of ${\cal L}_0$ and $%
{\cal A}_3$ are not suppressed by $Q_T$ or $Q_T^2$, so they contribute to
the singular pieces which are resummed in the CSS formalism. By definition
we call a term singular if it diverges as $Q_T^{-2}\times [1$ or $\ln
(Q^2/Q_T^2)]$ as $Q_T \rightarrow 0$. Using the t'Hooft-Veltman prescription
of $\gamma _5$ we conclude that the singular pieces of the symmetric (${\cal %
L}_0$) and anti-symmetric (${\cal A}_3$) parts are the same, and 
%% EEEEEEEEEEEEEEEEEEEEEEEEEEEEEEEEEEEEEEEEEEEEEEEEEEEEEEEEEEEEEEEEEEEEEE
\begin{eqnarray*}
\lim\limits_{Q_T\rightarrow 0}\ {\cal T}_{q\bar{q}}^{-1}\;\delta (s+t+u-Q^2)
&=&\lim\limits_{Q_T\rightarrow 0}\ {\cal T}_{q\bar{q}}^3\;\delta (s+t+u-Q^2)=%
{s}_{q\bar{q}}, \\
\lim\limits_{Q_T\rightarrow 0}\ {\cal T}_{Gq}^{-1}\;\delta (s+t+u-Q^2)
&=&\lim\limits_{Q_T\rightarrow 0}\ {\cal T}_{Gq}^3\;\delta (s+t+u-Q^2)={s}%
_{Gq},
\end{eqnarray*}
% eeeeeeeeeeeeeeeeeeeeeeeeeeeeeeeeeeeeeeeeeeeeeeeeeeeeeeeeeeeeeeeeeeeeee
where 
%% EEEEEEEEEEEEEEEEEEEEEEEEEEEEEEEEEEEEEEEEEEEEEEEEEEEEEEEEEEEEEEEEEEEEEE
\begin{eqnarray*}
{s}_{q\bar{q}} &=&\frac 1{Q_T^2}\left[ 2\;\delta (1-z_1)\;\delta
(1-z_2)\left( \ln \frac{Q^2}{Q_T^2}-\frac 32\right) \right.  \nonumber \\
&&\left. +\delta (1-z_1)\left( \frac{1+z_2{}^2}{1-z_2}\right) _{+}+\delta
(1-z_2)\left( \frac{1+z_1{}^2}{1-z_1}\right) _{+}\right.  \nonumber \\
&&\left. -\left( (1-z_1)\;\delta (1-z_2)+(1-z_2)\;\delta (1-z_1)\right)
\epsilon \right] +{\cal O}\left( {\frac 1{Q_T}}\right) , \\
{s}_{Gq} &=&\frac 1{Q_T^2}\left[ (z_1^2+\ (1-z_1^2))\;\delta (1-z_2)-\delta
(1-z_2)\epsilon \right] +{\cal O}\left( {\frac 1{Q_T}}\right) .
%\label{Eq:sqG}
\end{eqnarray*}
% eeeeeeeeeeeeeeeeeeeeeeeeeeeeeeeeeeeeeeeeeeeeeeeeeeeeeeeeeeeeeeeeeeeeee
As $Q_T\rightarrow 0$, only the ${\cal L}_0$ and ${\cal A}_3$ helicity cross
sections survive as expected, since the ${\cal O}(\alpha _S^0)$ differential
cross section contains only these angular functions 
[cf. Eq.~(\ref{eq:MBorn})]. 

% ssssssssssssssssssssssssssssssssssssssssssssssssssssssssssssssssssssss

\section{Expansion of the Resummation Formula} 
\label{app:ExpResFor}

In this section we expand the resummation formula, as given in 
Eq.~(\ref{eq:ResFor}), up to ${\cal O}(\alpha _S)$, and calculate the ${Q_T}$ 
singular piece as well as the
integral of the ${\cal O}(\alpha _S)$ corrections from 0 to ${P_T}$. These
are the ingredients, together with the regular pieces to be given in 
Appendix~\ref{app:RegPcs}, needed to construct our NLO calculation.

First we calculate the singular part at the ${\cal O}(\alpha _S)$. By
definition, this consist of terms which are at least as singular as 
$Q_T^{-2}\times$ [1 or $\ln {(Q_T^2/Q^2)}$]. We use the perturbative
expansion of the $A,\;B$ and $C$ functions in the strong coupling constant
$\alpha _S$ as: 
% EEEEEEEEEEEEEEEEEEEEEEEEEEEEEEEEEEEEEEEEEEEEEEEEEEEEEEEEEEEEEEEEEEEEEE
\begin{eqnarray}
A\left( \alpha _S({\bar{\mu}}),C_1\right) &=&
\sum_{n=1}^\infty 
\left( \frac{\alpha _S({\bar{\mu}})}\pi \right) ^nA^{(n)}(C_1),
\nonumber \\
B\left( \alpha _S({\bar{\mu}}),C_1,C_2\right) &=&\sum_{n=1}^\infty \left( 
\frac{\alpha _S({\bar{\mu}})}\pi \right) ^nB^{(n)}(C_1,C_2), 
\label{eq:ABCExp} \\
C_{ja}(z,b,\mu ,C_1,C_2) &=&
\sum_{n=0}^\infty 
\left( \frac{\alpha _S({\mu })}\pi \right)^n
C_{ja}^{(n)}(z,b,\mu ,\frac{C_1}{C_2}). 
\nonumber
\end{eqnarray}
% eeeeeeeeeeeeeeeeeeeeeeeeeeeeeeeeeeeeeeeeeeeeeeeeeeeeeeeeeeeeeeeeeeeeee
The explicit expressions of the $A^{(n)},\;B^{(n)}$ and $C^{(n)}$
coefficients are given in Appendix \ref{app:ABC}. 
After integrating over the lepton variables and the angle between ${\vec{b}}$ 
and ${\vec{Q}_T}$, and dropping the regular ($Y$) piece in 
Eq.~(\ref{eq:ResFor}), we obtain 
% EEEEEEEEEEEEEEEEEEEEEEEEEEEEEEEEEEEEEEEEEEEEEEEEEEEEEEEEEEEEEEEEEEEEEE
\begin{eqnarray*}
\left. \frac{d\sigma }{dQ^2dydQ_T^2}\right| _{Q_T\rightarrow 0} &=&\frac{%
\sigma _0}S\delta (Q^2-M_V^2)\left\{ \frac 1{2\pi Q_T^2}\int_0^\infty
d\eta \;\eta \;J_0(\eta )\;\text{e}^{-{\cal S}(\eta /Q{%
_T,Q,C}_{{1}},{C}_{{2}})}\right. \\
&&\ \ \left. \times \;f_{j/h_1}\left( x_1,\frac{C_3^2Q_T^2}{\eta ^2}%
\right) \;f_{\overline{k}/h_2}\left( x_2,\frac{C_3^2Q_T^2}{\eta ^2}%
\right) +\;j\leftrightarrow \overline{k}\right\} +{\cal O(}Q_T^{-1}),
\end{eqnarray*}
% eeeeeeeeeeeeeeeeeeeeeeeeeeeeeeeeeeeeeeeeeeeeeeeeeeeeeeeeeeeeeeeeeeeeee
where we have substituted the resonance behavior by a fixed mass for 
simplicity, and defined $\sigma _0$ as\footnote{%
For our numerical calculation (inside the ResBos Monte Carlo package), we
have consistently used the on-shell scheme for all the electroweak parameters 
in the improved Born level formula for including large electroweak radiative
corrections. In the $V=Z^0$ case, they are the same as those used in
studying the $Z^0$-pole physics at LEP~\cite{Peccei}.}~\cite{CSS} 
% EEEEEEEEEEEEEEEEEEEEEEEEEEEEEEEEEEEEEEEEEEEEEEEEEEEEEEEEEEEEEEEEEEEEEE
\begin{eqnarray*}
\sigma _0 &=&\frac{4\pi \alpha ^2}{9Q^2},~~~\text{for\ }V=\gamma ^{*}, \\
\sigma _0 &=&\frac{\pi ^2\alpha }{3 s_W^2}\sum_{jk}|V_{jk}|^2,~~~%
\text{for\ }V=W^{\pm }, \\
\sigma _0 &=&\frac{\pi ^2\alpha }{12 s_W^2 c_W^2}
\sum_{jk}[(1-4|Q_j| s_W^2)^2+1]|V_{jk}|^2,~~~\text{for\ }V=Z^0.
\end{eqnarray*}
% eeeeeeeeeeeeeeeeeeeeeeeeeeeeeeeeeeeeeeeeeeeeeeeeeeeeeeeeeeeeeeeeeeeeee
Here $\alpha $ is the fine structure constant, $s_W$ ($c_W$) is the sine 
(cosine) of the weak mixing angle $\theta _W$, 
$Q_j$ is the electric charge of the incoming quark in the
units of the charge of the positron (e.g. $Q_{up}=2/3$, $Q_{down}=-1/3$,
etc.), and $V_{jk}$ is defined by Eq.~(\ref{eq:Vjk}). To evaluate the
integral over $\eta =bQ_T$, we use the following property of the Bessel
functions: 
% EEEEEEEEEEEEEEEEEEEEEEEEEEEEEEEEEEEEEEEEEEEEEEEEEEEEEEEEEEEEEEEEEEEEEE
\[
\int_0^\infty d\eta \;\eta \;J_0(\eta )\;F(\eta
)=-\int_0^\infty d\eta \;\eta \;J_1(\eta )\;\frac{%
dF(\eta )}{d\eta }, 
\]
% eeeeeeeeeeeeeeeeeeeeeeeeeeeeeeeeeeeeeeeeeeeeeeeeeeeeeeeeeeeeeeeeeeeeee
which holds for any function $F(\eta )$ satisfying $\left[ \eta
\;J_1(\eta )\;F(\eta )\right] _0^\infty =0$. Using the expansion
of the Sudakov exponent ${\cal S}(b,Q,C_1,C_2)={\cal S}^{(1)}(b,Q,C_1,C_2)+%
{\cal O}(\alpha _S^2)$ with%
% EEEEEEEEEEEEEEEEEEEEEEEEEEEEEEEEEEEEEEEEEEEEEEEEEEEEEEEEEEEEEEEEEEEEEE
\[
{\cal S}^{(1)}(b,Q,C_1,C_2)=\frac{\alpha _S(Q^2)}\pi \left[ \frac 12%
A^{(1)}(C_1)\ln ^2\left( \frac{C_2^2Q^2}{C_1^2/b^2}\right)
+B^{(1)}(C_1,C_2)\ln \left( \frac{C_2^2Q^2}{C_1^2/b^2}\right) \right] , 
\]
and the evolution equation of the parton distribution functions
% EEEEEEEEEEEEEEEEEEEEEEEEEEEEEEEEEEEEEEEEEEEEEEEEEEEEEEEEEEEEEEEEEEEEEE
\[
\frac{d\;f_{j/h}(x,\mu ^2)}{d\ln \mu ^2}=\frac{\alpha _S(\mu ^2)}{2\pi }%
\left( P_{j\leftarrow a}^{(1)}\otimes f_{a/h}\right) (x,\mu ^2)+{\cal O}%
(\alpha _S^2), 
\]
% eeeeeeeeeeeeeeeeeeeeeeeeeeeeeeeeeeeeeeeeeeeeeeeeeeeeeeeeeeeeeeeeeeeeee
we can calculate the derivatives of the Sudakov factor and the parton
distributions with respect to $\eta $:
% EEEEEEEEEEEEEEEEEEEEEEEEEEEEEEEEEEEEEEEEEEEEEEEEEEEEEEEEEEEEEEEEEEEEEE
\begin{eqnarray*}
\frac d{d\eta }\text{e}^{-{\cal S}(\eta /Q_T,Q,C_1,C_2)} &=&
\frac{-2}\eta \frac{\alpha _S(Q^2)}\pi \left[ A^{(1)}(C_1)\ln \left( 
\frac{C_2^2Q^2\eta ^2}{C_1^2Q_T^2}\right) +B^{(1)}(C_1,C_2)\right] +%
{\cal O}(\alpha _S^2)\;\;\text{and} \\
\frac d{d\eta }f_{j/h}(x,\frac{C_3^2Q_T^2}{\eta ^2}) &=&\frac{-2}%
\eta \frac{\alpha _S(Q^2)}{2\pi }\left( P_{j\leftarrow a}^{(1)}\otimes
f_{a/h}\right) (x,Q^2)+{\cal O}(\alpha _S^2).
\end{eqnarray*}
% eeeeeeeeeeeeeeeeeeeeeeeeeeeeeeeeeeeeeeeeeeeeeeeeeeeeeeeeeeeeeeeeeeeeee
Note that $\alpha _S$ itself is expanded as 
% EEEEEEEEEEEEEEEEEEEEEEEEEEEEEEEEEEEEEEEEEEEEEEEEEEEEEEEEEEEEEEEEEEEEEE
\[
\frac{\alpha _S(\mu ^2)}{2\pi} =\frac{\alpha _S(Q^2)}{2\pi} -\beta _0
\left( \frac{\alpha _S(Q^2)}{2\pi} \right) ^2\ln \left( \frac{\mu ^2}{Q^2}%
\right) +{\cal O}(\alpha _S^3(Q^2)),
\]
% eeeeeeeeeeeeeeeeeeeeeeeeeeeeeeeeeeeeeeeeeeeeeeeeeeeeeeeeeeeeeeeeeeeeee
with $\beta _0=(11N_C-2N_f)/6$, where $N_C$ is the number of colors (3 in
QCD) and $N_f$ is the number of light quark flavors with masses less than $Q$.
In the evolution equation of the parton distributions,
% EEEEEEEEEEEEEEEEEEEEEEEEEEEEEEEEEEEEEEEEEEEEEEEEEEEEEEEEEEEEEEEEEEEEEE
\begin{eqnarray}
P_{{j\leftarrow k}}^{(1)}(z) &=&C_F\left( {\frac{1+z^2}{{1-z}}}\right) _{+}~~~%
{\rm and}  \nonumber \\
P_{{j\leftarrow G}}^{(1)}(z) &=&{\frac 12}\left[ z^2+(1-z)^2\right]
\label{eq:DGLAP} 
\end{eqnarray}
% eeeeeeeeeeeeeeeeeeeeeeeeeeeeeeeeeeeeeeeeeeeeeeeeeeeeeeeeeeeeeeeeeeeeee
are the leading order DGLAP splitting kernels~\cite{DGLAP}, 
and $\otimes $ denotes the convolution defined by 
% EEEEEEEEEEEEEEEEEEEEEEEEEEEEEEEEEEEEEEEEEEEEEEEEEEEEEEEEEEEEEEEEEEEEEE
\begin{eqnarray*}
\left( P_{j\leftarrow a}^{(1)}\otimes f_{a/h}\right) (x,\mu ^2)=\int_x^1{%
\frac{d\xi }\xi }\,P_{j\leftarrow a}^{(1)}\left( {\frac x\xi }\right)
\;f_{a/h}\left( \xi ,\mu ^2\right) ,
\end{eqnarray*}
% eeeeeeeeeeeeeeeeeeeeeeeeeeeeeeeeeeeeeeeeeeeeeeeeeeeeeeeeeeeeeeeeeeeeee
and the double parton index $a$ is running over all light quark flavors
and the gluon. In Eq.~(\ref{eq:DGLAP}), the ``+'' prescription is defined as
% EEEEEEEEEEEEEEEEEEEEEEEEEEEEEEEEEEEEEEEEEEEEEEEEEEEEEEEEEEEEEEEEEEEEEE
\[
\int_{x}^{1} dz \left( G(z) \right)_+ F(z) = 
\int_{0}^{1} dz G(z) \left[ F(z) \Theta(z - x) - F(1) \right] ,
\]
% eeeeeeeeeeeeeeeeeeeeeeeeeeeeeeeeeeeeeeeeeeeeeeeeeeeeeeeeeeeeeeeeeeeeee
where 
% EEEEEEEEEEEEEEEEEEEEEEEEEEEEEEEEEEEEEEEEEEEEEEEEEEEEEEEEEEEEEEEEEEEEEE
\[
\Theta(x) = \cases{~0, & {\rm if} $x < 0$, \cr ~1, & {\rm if} $x \geq 0$}
\]
% eeeeeeeeeeeeeeeeeeeeeeeeeeeeeeeeeeeeeeeeeeeeeeeeeeeeeeeeeeeeeeeeeeeeee
is the unit step function and $F(z)$ is an arbitrary function.

After utilizing the Bessel function property and substituting the
derivatives into the resummation formula above, the integral over $\eta 
$ can be evaluated using
% EEEEEEEEEEEEEEEEEEEEEEEEEEEEEEEEEEEEEEEEEEEEEEEEEEEEEEEEEEEEEEEEEEEEEE
\begin{equation}
\int_0^\infty d\eta \;J_1(\eta )\;\ln ^m\left( \frac \eta {b_0}\right) =
\cases{~1, & {\rm if} $m = 0$, \cr ~0, & {\rm if} $m=1,2$ {\rm
and} $b_0=2e^{-\gamma_E}$, }  \label{eq:IntBess}
\end{equation}
% eeeeeeeeeeeeeeeeeeeeeeeeeeeeeeeeeeeeeeeeeeeeeeeeeeeeeeeeeeeeeeeeeeeeee
where $\gamma _E$ is the Euler constant.
Fixing the renormalization constants 
$C_1 = C_2 b_0 = C_3 = b_0 = 2e^{-\gamma_E}$, 
we obtain the singular piece up to ${\cal O}(\alpha_S)$ as
% EEEEEEEEEEEEEEEEEEEEEEEEEEEEEEEEEEEEEEEEEEEEEEEEEEEEEEEEEEEEEEEEEEEEEE
\begin{eqnarray}
\left. \frac{d\sigma }{dQ^2dydQ_T^2}\right| _{Q_T\rightarrow 0} &=
&\frac{\sigma _0}S\delta (Q^2-M_V^2)\frac 1{2\pi Q_T^2}\frac{\alpha _S (Q^2)}
\pi \left\{
\left[ \;f_{j/h_1}(x_1,Q^2)\left( P_{\overline{k}\leftarrow b}\otimes
f_{b/h_2}\right) (x_2,Q^2)\right. \right.   \nonumber \\
&&+\left. \left( P_{j\leftarrow a}\otimes f_{a/h_1}\right) (x_1,Q^2)\;f_{%
\overline{k}/h_2}(x_2,Q^2)\right]   \nonumber \\
&&+\left. \left[ A^{(1)}\ln \left( \frac{Q^2}{Q_T^2}\right) +B^{(1)}\right]
\;f_{j/h_1}(x_1,Q^2)\;f_{\overline{k}/h_2}(x_2,Q^2)+\;j\leftrightarrow 
\overline{k}\right\}   \nonumber \\
&&+{\cal O}(\alpha _S^2,Q_T^{-1}).  
\label{eq:SingPart}
\end{eqnarray}
% eeeeeeeeeeeeeeeeeeeeeeeeeeeeeeeeeeeeeeeeeeeeeeeeeeeeeeeeeeeeeeeeeeeeee

To derive the integral of the ${\cal O}(\alpha _S)$ corrections over $Q_T$,
we start again from the resummation formula [Eq.~(\ref{eq:ResFor})] and the
expansion of the $A,\;B$ and $C$ functions [Eq.~(\ref{eq:ABCExp})]. 
This time the evolution of parton distributions is expressed as 
% EEEEEEEEEEEEEEEEEEEEEEEEEEEEEEEEEEEEEEEEEEEEEEEEEEEEEEEEEEEEEEEEEEEEEE
\begin{eqnarray*}
f_{j/h}(x,\mu ^2) &=&f_{j/h}(x,Q^2)+f_{j/h}^{(1)}(x,\mu ^2)+{\cal O}(\alpha
_S^2),~\text{ with} \\
f_{j/h}^{(1)}(x,\mu ^2) &=&\frac{\alpha _S(Q^2)}{2\pi }\ln \left( \frac{\mu
^2}{Q^2}\right) \left( P_{j\leftarrow a}^{(1)}\otimes f_{a/h}\right) (x,Q^2),
\end{eqnarray*}
% eeeeeeeeeeeeeeeeeeeeeeeeeeeeeeeeeeeeeeeeeeeeeeeeeeeeeeeeeeeeeeeeeeeeee
where summation over the partonic index $a$ is implied. Substituting these
expansions in the resummation formula Eq.~(\ref{eq:ResFor})
and integrating over both sides with respect to $Q_T^2$. 
We use the integral formula, valid for an arbitrary function $F(b)$:
% EEEEEEEEEEEEEEEEEEEEEEEEEEEEEEEEEEEEEEEEEEEEEEEEEEEEEEEEEEEEEEEEEEEEEE
\[
\frac 1{(2\pi )^2}\int_0^{P_T^2}dQ_T^2\int d^2b\;\text{e}^{i{\vec{Q}_T}\cdot 
{\vec{b}}}F(b)=\frac 1{2\pi }\int_0^\infty db\;P_TJ_1(bP_T)F(b),
\]
% eeeeeeeeeeeeeeeeeeeeeeeeeeeeeeeeeeeeeeeeeeeeeeeeeeeeeeeeeeeeeeeeeeeeee
together with Eq.~(\ref{eq:IntBess}) to derive 
% EEEEEEEEEEEEEEEEEEEEEEEEEEEEEEEEEEEEEEEEEEEEEEEEEEEEEEEEEEEEEEEEEEEEEE
\begin{eqnarray}
&& \int_0^{P_T^2}dQ_T^2\;\frac{d\sigma }{dQ^2dydQ_T^2}=\frac{\sigma _0}S
\delta (Q^2-M_V^2)  \nonumber \\
&&\times \left\{ \left( 1-\frac{\alpha _S (Q^2)}\pi \left[ \frac 12A^{(1)}\ln
^2\left( \frac{Q^2}{P_T^2}\right) +B^{(1)}\ln \left( \frac{Q^2}{P_T^2}
\right) \right] \right) \;f_{j/h_1}(x_1,Q^2)\;f_{\overline{k}%
/h_2}(x_2,Q^2)\right. \nonumber \\
&&-\frac{\alpha _S (Q^2)}{2\pi }\ln \left( \frac{Q^2}{P_T^2} \right)   
 \left[ \left( P_{j\leftarrow a}\otimes f_{a/h_1}\right)
(x_1,Q^2)\;f_{\overline{k}/h_2}(x_2,Q^2) \right. \nonumber \\
&& ~~~~~~~~~~~~~~~~~~~~~~~~~~ \left.
- f_{j/h_1}(x_1,Q^2)\left( P_{
\overline{k}\leftarrow b}\otimes f_{b/h_2}\right) (x_2,Q^2)\right]  
\nonumber \\
&&+\frac{\alpha _S (Q^2)}\pi \left[ \left( C_{ja}^{(1)}\otimes f_{a/h_1}\right)
(x_1,Q^2)\;\;f_{\overline{k}/h_2}(x_2,Q^2)+\;f_{j/h_1}(x_1,Q^2)\;\left( C_{
\overline{k}b}^{(1)}\otimes f_{b/h_2}\right) (x_2,Q^2)\right]   \nonumber \\
&&\left. +j\leftrightarrow \overline{k}+\int_0^{P_T^2}dQ_T^2%
\;Y(Q_T,Q,x_1,x_2)\right\}, 
\label{Eq:DelSig}
\end{eqnarray}
% eeeeeeeeeeeeeeeeeeeeeeeeeeeeeeeeeeeeeeeeeeeeeeeeeeeeeeeeeeeeeeeeeeeeee
where $x_1 = {\rm e}^y Q/\sqrt{S}$ and $x_2 = {\rm e}^{-y} Q/\sqrt{S}$.
Equations (\ref{eq:SingPart}) and (\ref{Eq:DelSig}) (together with the
regular pieces, discussed in Appendix \ref{app:RegPcs}) are used to program
the ${\cal O}(\alpha _S)$ results as discussed in the beginning of
Section \ref{sec:Phenom}.

% ssssssssssssssssssssssssssssssssssssssssssssssssssssssssssssssssssssss

\section{$A$, $B$ and $C$ functions}
\label{app:ABC}

For completeness, we give here the coefficients $A$, $B$ and $C$ utilized
in our numerical calculations. The coefficients in the Sudakov exponent are~%
\cite{CSS,Davies}. 
% EEEEEEEEEEEEEEEEEEEEEEEEEEEEEEEEEEEEEEEEEEEEEEEEEEEEEEEEEEEEEEEEEEEEEE
\begin{eqnarray*}
A^{(1)}(C_1) &=&C_F, \\
A^{(2)}(C_1) &=&C_F\left[ \left( {\frac{67}{36}}-{\frac{\pi ^2}{12}}\right)
N_C-{\frac 5{18}}N_f-\beta _0\ln \left( {\frac{b_0}{C_1}}\right) \right] , \\
B^{(1)}(C_1,C_2) &=&C_F\left[ -{\frac 32}-2\ln \left( {\frac{C_2b_0}{C_1}}%
\right) \right] , \\
B^{(2)}(C_1,C_2) &=&C_F\left\{ C_F\left( {\frac{\pi ^2}4}-{\frac 3{16}}%
-3\zeta (3)\right) +N_C\left( {\frac{11}{36}}\pi ^2-{\frac{193}{48}}+{\frac 3%
2}\zeta (3)\right) \right. + \\
&&\frac{N_F}{2}\left( -{\frac 19}\pi ^2+{\frac{17}{12}}\right) -
\left[ \left( {\frac{67}{18}}-{\frac{\pi ^2}6}\right) 
N_C-{\frac{5}{9}}N_f\right] \ln \left( {\frac{C_2b_0}{C_1}}\right) + \\
&&\left. \beta _0\left[ \ln ^2\left( {\frac{b_0}{C_1}}\right) -\ln ^2(C_2)-{%
\frac 32}\ln (C_2)\right] \right\} ,
\end{eqnarray*}
% eeeeeeeeeeeeeeeeeeeeeeeeeeeeeeeeeeeeeeeeeeeeeeeeeeeeeeeeeeeeeeeeeeeeee
where $N_f$ is the number of light quark flavors ($m_q<Q_V$, e.g. $N_f=5$ for 
$W^{\pm }$ or $Z^0$ production), $C_F={\rm tr}(t_at_a)$ is the second order 
Casimir of the quark representation (with $t_a$ being the SU(N$_C$) generators
in the fundamental representation), $\beta _0=(11N_C-2N_f)/6$ and $\zeta (x)$
is the Riemann zeta function, and $\zeta (3) \approx 1.202$. 
For QCD, $N_C = 3$ and $C_F = 4/3$. 

The $C_{jk}^{(n)}$ coefficients up to $n=1$ are: 
% EEEEEEEEEEEEEEEEEEEEEEEEEEEEEEEEEEEEEEEEEEEEEEEEEEEEEEEEEEEEEEEEEEEEEE
\begin{eqnarray*}
C_{jk}^{(0)}(z,b,\mu ,{\frac{C_1}{C_2}}) &=&\delta _{jk}\delta ({1-z}), \\
C_{jG}^{(0)}(z,b,\mu ,{\frac{C_1}{C_2}}) &=&0, \\
C_{jk}^{(1)}(z,b,\mu ,{\frac{C_1}{C_2}}) &=&\delta _{jk}C_F\left\{ {\frac 12}%
(1-z)-{\frac 1{C_F}}\ln \left( {\frac{\mu b}{b_0}}\right) P_{j\leftarrow
k}^{(1)}(z)\right. \\
&&\left. +\delta (1-z)\left[ -\ln ^2\left( {\frac{C_1}{{b_0C_2}}}%
e^{-3/4}\right) +{\frac{\pi ^2}4}-{\frac{23}{16}}\right] \right\} , \\
C_{jG}^{(1)}(z,b,\mu ,{\frac{C_1}{C_2}}) &=&{\frac 12}z(1-z)-\ln \left( {%
\frac{\mu b}{b_0}}\right) P_{j\leftarrow G}^{(1)}(z),
\end{eqnarray*}
% eeeeeeeeeeeeeeeeeeeeeeeeeeeeeeeeeeeeeeeeeeeeeeeeeeeeeeeeeeeeeeeeeeeeee
where $P_{j\leftarrow a}^{(1)}(z)$ are the leading order DGLAP splitting
kernels~\cite{DGLAP} given in Appendix C, and $j$ and $k$ represent quark 
or anti-quark flavors.

The constants $C_1,\;C_2$ and $C_3\equiv \mu b$ were introduced when solving
the renormalization group equation for $\widetilde{W}_{jk}$. $C_1$ enters
the lower limit $\overline{\mu }=C_1/b$ in the integral of the Sudakov
exponent [cf. Eq.~(\ref{eq:SudExp})], and determines the onset of the
non-perturbative physics. The renormalization constant $C_2,$ in the upper
limit $\overline{\mu }=C_2 Q$ of the Sudakov integral, specifies the scale
of the hard scattering process. The scale $\mu = C_3/b$ is the scale at
which the $C$ functions are evaluated. The canonical choice of these
renormalization constants is $C_1=C_3=2e^{-\gamma _E}\equiv b_0$ and 
$C_2=C_4=1$ \cite{CSS}. We adopt these choices of the renormalization
constants in the numerical results of this work, because they eliminate
large constant factors inside the $A,\;B$ and $C$ functions.

After fixing the renormalization constants to the canonical values, we
obtain much simpler expressions of $A^{(1)}$, $B^{(1)}$, $A^{(2)}$ and $%
B^{(2)}$. The first order coefficients in the Sudakov exponent become 
% EEEEEEEEEEEEEEEEEEEEEEEEEEEEEEEEEEEEEEEEEEEEEEEEEEEEEEEEEEEEEEEEEEEEEE
\[
A^{(1)}(C_1)=C_F,~~~{\rm and}~~~B^{(1)}(C_1=b_0,C_2=1)=-3C_F/2. 
\]
% eeeeeeeeeeeeeeeeeeeeeeeeeeeeeeeeeeeeeeeeeeeeeeeeeeeeeeeeeeeeeeeeeeeeee
The second order coefficients in the Sudakov exponent simplify to 
% EEEEEEEEEEEEEEEEEEEEEEEEEEEEEEEEEEEEEEEEEEEEEEEEEEEEEEEEEEEEEEEEEEEEEE
\begin{eqnarray*}
\ A^{(2)}(C_1=b_0) &=&C_F\left[ \left( {\frac{67}{36}}-{\frac{\pi ^2}{12}}%
\right) N_C-{\frac 5{18}}N_f\right] , \\
B^{(2)}(C_1=b_0,C_2=1) &=&C_F^2\left( {\frac{\pi ^2}4}-{\frac 3{16}}-3\zeta
(3)\right) +C_FN_C\left( {\frac{11}{36}}\pi ^2-{\frac{193}{48}}+{\frac 32}%
\zeta (3)\right) \\
&&+C_FN_f\left( -{\frac 1{18}}\pi ^2+{\frac{17}{24}}\right).
\end{eqnarray*}
% eeeeeeeeeeeeeeeeeeeeeeeeeeeeeeeeeeeeeeeeeeeeeeeeeeeeeeeeeeeeeeeeeeeeee

The Wilson coefficients $C_{ja}^{(i)}$ for the parity-conserving part of the
resummed result are also greatly simplified under the canonical definition
of the renormalization constants. Their explicit forms are 
% EEEEEEEEEEEEEEEEEEEEEEEEEEEEEEEEEEEEEEEEEEEEEEEEEEEEEEEEEEEEEEEEEEEEEE
\begin{eqnarray*}
C_{jk}^{(1)}(z,b,\mu =\frac{b_0}b,\frac{C_1}{C_2}=b_0) &=&\delta
_{jk}\left\{ {\frac 23}(1-z)+{\frac 13}\,\delta (1-z)(\pi ^2-8)\right\} 
~~~{\rm and}~~~ \\
C_{jG}^{(1)}(z,b,\mu =\frac{b_0}b,\frac{C_1}{C_2}=b_0) &=&{\frac 12}z(1-z).
\end{eqnarray*}
% eeeeeeeeeeeeeeeeeeeeeeeeeeeeeeeeeeeeeeeeeeeeeeeeeeeeeeeeeeeeeeeeeeeeee
As noted in Appendix~\ref{app:ExpResFor}, the same Wilson coefficient
functions $C_{ja}$ also apply to the parity violating part which is
multiplied by the angular function ${\cal A}_3 = 2 \cos{\theta}$.

% sssssssssssssssssssssssssssssssssssssssssssssssssssssssssssssssssssss

\section{Regular Pieces} 
\label{app:RegPcs}

The $Y$ piece in Eq.~(\ref{eq:ResFor}), which is the difference of the fixed
order perturbative result and their singular part, is given by the
expression 
% EEEEEEEEEEEEEEEEEEEEEEEEEEEEEEEEEEEEEEEEEEEEEEEEEEEEEEEEEEEEEEEEEEEEEE
\begin{eqnarray}
Y(Q_T,Q,x_1,x_2,\theta ,\phi ,C_4) &=&\int_{x_1}^1{\frac{d\xi _1}{\xi _1}}%
\int_{x_2}^1{\frac{d\xi _2}{\xi _2}}\sum_{n=1}^\infty \left[ {\frac{\alpha
_S(C_4Q)}\pi }\right] ^n \nonumber \\
&&\times f_{a/h_1}(\xi _1,C_4Q)\,R_{ab}^{(n)}(Q_T,Q,z_1,z_2,\theta ,\phi)
\,f_{b/h_2}(\xi _2,C_4Q),
\label{eq:YPc}
\end{eqnarray}
% eeeeeeeeeeeeeeeeeeeeeeeeeeeeeeeeeeeeeeeeeeeeeeeeeeeeeeeeeeeeeeeeeeeeee
where $z_i=x_i/\xi _i$ $(i=1,2)$. The regular functions $R_{ab}^{(n)}$ only
contain contributions which are less singular than $Q_T^{-2}\times $[1 or $\ln
(Q_T^2/Q^2)]$ as $Q_T\rightarrow 0$. Their explicit expressions for 
$h_1h_2\rightarrow V(\rightarrow \ell _1 {\bar \ell _2})X$ are given below.
The scale for evaluating the regular pieces is $C_4 Q$. To minimize the
contribution of large logarithmic terms from higher order corrections, we
choose $C_4=1$ when calculating the $Y$ piece.

We define the $q\bar{q^{\prime }}V$ and the $\ell _1{\bar \ell _2}V$
vertices, respectively, as 
% EEEEEEEEEEEEEEEEEEEEEEEEEEEEEEEEEEEEEEEEEEEEEEEEEEEEEEEEEEEEEEEEEEEEEE
\[
i\gamma _\mu \left[ g_L(1-\gamma _5)+g_R(1+\gamma _5)\right] 
~~~{\rm and}~~~
i\gamma _\mu \left[ f_L(1-\gamma _5)+f_R(1+\gamma _5)\right] . 
\]
% eeeeeeeeeeeeeeeeeeeeeeeeeeeeeeeeeeeeeeeeeeeeeeeeeeeeeeeeeeeeeeeeeeeeee
For example, for $V=W^{+},q=u$, ${\bar{q}^{\prime }}={\bar{d}}$, $\ell
_1=\nu _e$, and ${\bar \ell_2}=e^{+}$, the couplings $g_L^2=f_L^2=G_FM_W^2/%
\sqrt{2} $ and $g_R^2=f_R^2=0$, where $G_F$ is the Fermi constant. Table \ref
{tbl:parameters} shows all the couplings for the general case. In 
Eq.~(\ref{eq:YPc}), 
% EEEEEEEEEEEEEEEEEEEEEEEEEEEEEEEEEEEEEEEEEEEEEEEEEEEEEEEEEEEEEEEEEEEEEE
\[
R_{ab}^{(1)}={\frac{16\mid V_{ab}\mid ^2}{\pi Q^2}}\left[
(g_L^2+g_R^2)(f_L^2+f_R^2)R_1^{ab}+(g_L^2-g_R^2)(f_L^2-f_R^2)R_2^{ab}%
\right], 
\]
% eeeeeeeeeeeeeeeeeeeeeeeeeeeeeeeeeeeeeeeeeeeeeeeeeeeeeeeeeeeeeeeeeeeeee
where the coefficient functions $R_i^{ab}$ are given as follows\footnote{%
Note that in Ref.~\cite{Balazs-Qui-Yuan} there were typos in $R_1^{j{\bar{k}}%
}$ and $R_2^{Gj}$.}: 
% EEEEEEEEEEEEEEEEEEEEEEEEEEEEEEEEEEEEEEEEEEEEEEEEEEEEEEEEEEEEEEEEEEEEEE
\begin{eqnarray*}
&&R_1^{j{\bar{k}}}=r^{j{\bar{k}}}{\cal L}_0+{\frac{{\cal R_{+}}(t,u)}s}%
\,\delta (s+t+u-Q^2)\left[ {\cal A}_0+{\cal A}_2+{\frac Q{Q_T}}{\frac{{\cal R%
}_{-}(u,t)}{{\cal R}_{+}(t,u)}}{\cal A}_1\right] {\frac{Q^2}{M_T^2}}, \\
&&R_2^{j{\bar{k}}}=r^{j{\bar{k}}}{\cal A}_3+{\frac{{\cal R_{+}}(t,u)}s}%
\,\delta (s+t+u-Q^2) \\
&&~~~~~~~~~~~~\times \left\{ {\frac{Q^2}{Q_T^2}}\left( {\frac Q{M_T}}%
-1\right) {\cal A}_3-{\frac{2Q^2}{Q_TM_T}}{\frac{{\cal R_{-}}(t,u)}{{\cal %
R_{+}}(t,u)}}{\cal A}_4\right\} , \\
&&R_1^{Gj}=r^{Gj}{\cal L}_0-{\frac{Q^2Q_T^2}{uM_T^2}}\,{\frac{{\cal R_{+}}%
(u,s)}s}\,\delta (s+t+u-Q^2) \\
&&~~~~~~~~~~~~\times \left\{ {\frac{{\cal R_{+}}(u,-s)}{{\cal R_{+}}(u,s)}}%
\left[ {\cal A}_0+{\cal A}_2\right] +{\frac Q{Q_T}}{\frac{(Q^2-u)^2+{\cal %
R_{-}}(u,t)}{{\cal R_{+}}(u,s)}}{\cal A}_1\right\} , \\
&&R_2^{Gj}=-r^{Gj}{\cal A}_3-{\frac{Q_T^2}u}~{\frac{{\cal R_{+}}(u,s)}s}%
\,\delta (s+t+u-Q^2) \\
&&~~~~~~~~~~~~\times \left\{ {\frac{Q^2}{Q_T^2}}\left[ {\frac Q{M_T}}\left( {%
\frac{2u(Q^2-s)}{{\cal R_{+}}(u,s)}}-1\right) +1\right] {\cal A}_3\right. \\
&&~~~~~~~~~~~~~~~~~~~~\left. -{\frac{2Q^2}{Q_TM_T}}\left[ {\frac{2s(Q^2-s)}{%
{\cal R_{+}}(u,s)}}+1\right] {\cal A}_4\right\} ,
\end{eqnarray*}
with 
% EEEEEEEEEEEEEEEEEEEEEEEEEEEEEEEEEEEEEEEEEEEEEEEEEEEEEEEEEEEEEEEEEEEEEE
\begin{eqnarray*}
&&r^{j{\bar{k}}}={\frac{Q^2}{Q_T^2}}\left\{ {\frac{{\cal R}_{+}(t,u)}s}%
~\delta (s+t+u-Q^2)-2~\delta (1-z_1)~\delta (1-z_2)\left[ \ln {\left( {\frac{%
Q^2}{Q_T^2}}\right) }-{\frac 32}\right] \right. \\
&&~~~~~~~~~~~~~~~\left. -\delta (1-z_1)\left( {\frac{1+z_2^2}{1-z_2}}\right)
_{+}-\delta (1-z_2)\left( {\frac{1+z_1^2}{1-z_1}}\right) _{+}\right\} ,
\end{eqnarray*}
and 
% EEEEEEEEEEEEEEEEEEEEEEEEEEEEEEEEEEEEEEEEEEEEEEEEEEEEEEEEEEEEEEEEEEEEEE
\[
r^{Gj}={\frac{Q^2}{Q_T^2}}\left\{ -{\frac{Q_T^2}u}~{\frac{{\cal R_{+}}(u,s)}s%
}\,\delta (s+t+u-Q^2)-\left[ z_1^2+(1-z_1)^2\right] \,\delta (1-z_2)\right\}
, 
\]
% eeeeeeeeeeeeeeeeeeeeeeeeeeeeeeeeeeeeeeeeeeeeeeeeeeeeeeeeeeeeeeeeeeeeee
where ${\cal R_{\pm }}(t,u)=(Q^2-t)^2\pm (Q^2-u)^2$. The Mandelstam
variables $s,t,u$ and the angular functions ${\cal L}_0,{\cal A}_i$ are
defined in Appendix B. The $V_{jk}$ coefficients are defined by Eq.~(\ref
{eq:Vjk}). For $a=j$ and $b = G$: $\displaystyle |V_{jG}|^2 = \sum_k
|V_{jk}|^2$ where $j$ and $k$ are light quark flavors with opposite weak
isospin quantum numbers. Up to this order, there is no contribution from
gluon-gluon initial state, i.e. $R_{GG}^{(1)} = 0$. 
The remaining coefficient functions with all
possible combinations of the quark and gluon indices (for example $R^{{\bar k%
}j}$, $R^{G{\bar j}}$ or $R^{jG}$, etc.) are obtained by the same crossing
rules summarized in Appendix~\ref{app:OaS}.
%Eq.~(\ref{eq:CrossRule}).

Having both the singular and the regular pieces expanded up to ${\cal O}%
(\alpha_S)$, we can construct the NLO Monte Carlo calculation by first 
including the contribution from Eq.~(\ref{Eq:DelSig}), with $P_T=Q_T^{Sep}$, 
for $Q_T<Q_T^{Sep}$. Second, for $Q_T>Q_T^{Sep}$, we
include the ${\cal O}(\alpha_S)$ perturbative results, which is equal to the
sum of the singular [Eq.~(\ref{eq:SingPart})] and the regular 
[Eq.~(\ref{eq:YPc})] pieces up to ${\cal O}(\alpha_S)$.
(Needless to say that the relevant angular functions for using 
Eqs.(\ref{eq:SingPart}) and (\ref{Eq:DelSig}) are 
${\cal L}_0=1 + \cos^2{\theta}$ and ${\cal A}_3=2 \cos{\theta}$, 
cf. Eq.~(\ref{eq:MBorn}).)
Hence, the NLO total rate is given by the sum of the contributions from both 
the $Q_T<Q_T^{Sep}$ and the $Q_T>Q_T^{Sep}$ regions.

% aaaaaaaaaaaaaaaaaaaaaaaaaaaaaaaaaaaaaaaaaaaaaaaaaaaaaaaaaaaaaaaaaaaaaaaaa


\begin{thebibliography}{99}

\bibitem{Bj} J.D. Bjorken, SLAC-PUB-7361, 1996.

\bibitem{qcdhand}  CTEQ Collaboration, R. Brock et al., Rev. Mod.
Phys. {\bf 67} (1995) 157.

\bibitem{Mueller} A.H. Mueller (editor), 
{\it Perturbative Quantum Chromodynamics},
(Singapore, World Scientific, Advanced Series on Directions in High
Energy Physics v5, 1989).

\bibitem{isajet} F.E. Paige, S.D. Protopopescu, 
in {\it Physics of the Superconducting Supercollider} 
(Proceedings of Snowmass Summer Study, 
Edited by R. Donaldson and J. Marx, Am. Phys. Soc., 1988, p320).

\bibitem{pythia}  T. Sj\"ostrand, Comp. Phys. Com. {\bf 82} (1994) 74.

\bibitem{herwig}  G. Marchesini, B.R. Webber, G. Abbiendi, I.G. Knowles,
M.H. Seymour, L. Stanco, 
%hepph-9607393, 
Comput. Phys. Commun. {\bf 67} (1992) 465.

\bibitem{CSS}  J. Collins, D. Soper, G. Sterman, Nucl. Phys. {\bf B250}
(1985) 199.

\bibitem{Collins}  J. Collins, D. Soper, Nucl. Phys. {\bf B193} (1981)
381; Erratum {\bf B213} (1983) 545; {\bf B197} (1982) 446.

\bibitem{DDT}  Y.I. Dokshitser, D.I. D'Yakonov, S.I. Troyan, Phys. Lett. 
{\bf B79} (1978) 269.

\bibitem{Parisi}  G. Parisi, R. Petronzio, Nucl. Phys. {\bf B154} (1979)
427.

\bibitem{AEGM}  G. Altarelli, R.K. Ellis, M. Greco, G. Martinelli, 
Nucl. Phys. {\bf B246} (1984) 12.

\bibitem{Arnold-Kauffman}  P.B. Arnold, R.P. Kauffman, Nucl. Phys. 
{\bf B349} (1991) 381.

\bibitem{Balazs-Qui-Yuan}  C. Bal{\'{a}}zs, J.W. Qui, C.--P. Yuan, Phys.
Lett. {\bf B355} (1995) 548.

\bibitem{CSFrame}  J. Collins, D. Soper, Phys. Rev. {\bf D16} (1977) 2219.

\bibitem{Rosner}  J.L. Rosner, Phys. Rev. {\bf D54} (1996) 1078.

\bibitem{gamma5c}  G. t'Hooft, M. Veltman, Nucl. Phys. {\bf B44} (1972)
189; \newline
P. Breitenlohner, D. Maison, Comm. Math. Phys. {\bf 52} (1977) 11; 
\newline
J.G. K{\"{o}}rner, G. Schuler, G. Kramer, B. Lampe, Phys. Lett. {\bf B164}
(1985) 136; \newline
M. Veltman, Nucl. Phys. {\bf B319} (1989) 253;

\bibitem{twoloopf3}  J.G. K\"{o}rner, G. Schuler, G. Kramer, B. Lampe, Z.
Phys. {\bf C32} (1986) 181.

\bibitem{Korner-Mirkes}  J.G. K\"{o}rner, E. Mirkes, G. Schuler, 
Internat. J. of Mod. Phys. {\bf A4} (1989) 1781.

\bibitem{lamtung}  C.S. Lam, W.K. Tung, Phys. Rev. {\bf D18} (1978) 2447.

\bibitem{Korchemsky-Sterman}  G.P. Korchemsky, G. Sterman, Nucl. Phys. 
{\bf B437} (1995) 415.

\bibitem{Davies}  C. Davies, Ph.D. Thesis, Churchill College (1984); \newline
C. Davies, W. Stirling, Nucl. Phys. {\bf B244} (1984) 337; \newline
C. Davies, B. Webber, W. Stirling, Nucl. Phys. {\bf B256} (1985) 413.

\bibitem{Ladinsky-Yuan}  G.A. Ladinsky, C.--P. Yuan, Phys. Rev. {\bf D50}
(1994) 4239.

\bibitem{Sjostrand} T. Sj\"ostrand, Phys. Lett. {\bf B157} (1985) 321. 

\bibitem{Giele}  W. Giele, E. Glover, D.A. Kosover, Nucl. Phys. 
{\bf B403} (1993) 633.

\bibitem{Reno}  M.H. Reno, Phys. Rev. {\bf D49} (1994) 4326.

\bibitem{D0alphaS}  \D0 collaboration, S. Abachi et al., Phys. Rev. Lett. 
{\bf 75} (1995) 3226.

\bibitem{AEM85}  G. Altarelli, R.K. Ellis, G. Martinelli, Z. Phys. 
{\bf C27} (1985) 617.

\bibitem{Arnold-Reno} P.B. Arnold, M.H. Reno, Nucl. Phys. {\bf B319} (1989) 37;
Erratum {\bf B330} (1990) 284.

\bibitem{CDFLCA} M. Dickson, Ph.D. Thesis, Rochester URochester University 
(1994); 
CDF Collaboration, F. Abe et al., Phys. Rev. Lett. {\bf 74}, 850 (1995).

\bibitem{TEV2000} TeV-2000 Study Group (D. Amidei, R. Brock ed.), {\it Future
ElecroWeak Physics at the Fermilab Tevatron}, FERMILAB-Pub-96/082.

\bibitem{Stirling-Martin}  W.J. Stirling, A.D. Martin, Phys. Lett. 
{\bf B237} (1990) 551.

\bibitem{Barger}  V. Barger, A.D. Martin, R.J.N. Phillips, Z. Phys. 
{\bf C21} (1983) 99.

\bibitem{Smith}  J. Smith, W.L. van Neerven, J.A.M. Vermaseren, Phys. Rev.
Lett. {\bf 50} (1983) 1738.

\bibitem{TEV33}  P.P. Bagley, et al., FERMILAB-Conf-96/392 (1996).

\bibitem{Adam} I. Adam, Ph.D. Thesis, Columbia University (1997).

\bibitem{Flattum} E. Flattum, Ph.D. Thesis, Michigan State University (1996).

\bibitem{SM96EW} U. Baur, M. Demarteau, FERMILAB-Conf-96/423 (1996). 

\bibitem{Giele-Keller}  W.T. Giele, S. Keller, FERMILAB-Conf-96/307-T (1996).

\bibitem{CDFAFB}  CDF Collaboration, F. Abe et al., Phys. Rev. Lett. 
{\bf 67} (1991) 1502.

\bibitem{Aurenche}  P. Aurenche, J. Lindfors, Nucl. Phys. {\bf B185}
(1981) 301.

\bibitem{Korner-Schuler}  J.G. K\"{o}rner, G. Schuler, G. Kramer, B. Lampe,
Z. Phys. {\bf C32} (1986) 181.

\bibitem{AEM79}  G. Altarelli, R.K. Ellis, G. Martinelli, Nucl. Phys. 
{\bf B157} (1979) 461.

\bibitem{Mirkes}  E. Mirkes, Nucl. Phys. {\bf B387} (1992) 3.

\bibitem{gamma5n}  M. Chanowitz, M. Furman, I. Hinchliffe, Phys. Lett. 
{\bf B78} (1978) 285; Nucl. Phys. {\bf B159} (1979) 225.

\bibitem{CDFTotal}  CDF Collaboration, F. Abe et al., Phys. Rev. Lett. 
{\bf 76} (1996) 3070.

\bibitem{CDFMW}  CDF Collaboration, F. Abe et al., Phys. Rev. 
{\bf D52} (1995) 4784.

\bibitem{Peccei} R.D. Peccei, UCLA-96-TEP-35 (1996).

\bibitem{DGLAP} Yu.L. Dokshitzer, JETP {\bf 46} (1977) 641; 
V.N. Gribov, L.N. Lipatov, Sov. Journ. Nucl. Phys. {\bf 15} (1972) 78;
G. Altarelli, G. Parisi, Nucl. Phys. {\bf B126} (1977) 298.

\end{thebibliography}
\end{document}